\documentclass{jfm}

\usepackage{graphicx}
\usepackage{natbib}
\font\smallfont=cmsy10 at 10truept
\mathchardef\bigCircle="280D

\font\bigfont=cmsy10 at 14.4truept
\mathchardef\tiMes="2902        %

\font\Bigfont=cmsy10 at 17.28truept
\mathchardef\DiaMond="2A05        %
\mathchardef\cirCle="2A0E
\mathchardef\BigCircle="2A0D

\font\Bbigfont=cmsy10 at 24.88truept
\mathchardef\buLLet="2B0F

\def\bigCirc{\raise 0.3ex\hbox{$\bigCircle$}\nobreak$\,$}
\def\Bullet{\raise-0.35ex\hbox{$\buLLet$}\nobreak$\,$}

\def\minisquare{\hbox{${\vcenter{
               \hrule height 0.3pt \kern-0.4pt
               \hbox{\vrule width  0.3pt height 3.0pt \kern 2.6pt
               \vrule width  0.3pt height 3.0pt} \kern-0.4pt
               \hrule height 0.3pt}}$}}
\def\ssquare{\raise 0.175ex\hbox{${\vcenter{
               \hrule height 0.5truept       \kern-0.25truept
               \hbox{\vrule width 0.5truept height 3.0truept \kern 2.75truept
                     \vrule width 0.5truept height 3.0truept} \kern-0.25truept
               \hrule height 0.5truept}}$}\nobreak$\,$}
\def\squarex{\raise 0.175ex\hbox{${\vcenter{
               \hrule height 0.8truept       \kern-1.80truept
          \hbox{\vrule width 0.8truept height 8.0truept \kern-1.95truept
                \raise 0.8truept\hbox{$\tiMes$}     \kern-6.70truept
                \vrule width 0.8truept height 8.0truept} \kern-0.80truept
               \hrule height 0.8truept}}$}\nobreak$\,$}

\def\sqbull{\raise0.175ex\hbox{\vrule height 1.4ex width 1.6ex depth 0.2ex}\nobreak$\,$}
\def\smsqbull{\raise0.175ex\hbox{\vrule height 0.8ex width 0.9ex depth 0.2ex}\nobreak$\,$}

\def\Diamondplus{${\vcenter{\vcenter{\DiaMond} \kern-10truept
                            \hbox{\vrule width .4truept}\kern -3truept
                            \hrule height .4truept}}$\nobreak$\,$}
\newcount\ndots

\def\drawline#1#2{\raise 2.5truept\vbox{\hrule width #1truept height #2truept}}
\def\moonspace#1{\hskip #1truept}

\def\shortchain{\drawline{6.0}{0.75}}

\def\shortchainspace{\shortchain\moonspace{2}}

\def\dashy{\drawline{4.00}{0.75}}     
\def\thindashy{\drawline{4.00}{0.25}}     
\def\dashyspace{\dashy\moonspace{2}}
\def\thindashyspace{\thindashy\moonspace{2}}
\def\longdashy{\drawline{8.00}{0.75}} 
\def\thinlongdashy{\drawline{8.00}{0.25}} 
\def\longdashyspace{\longdashy\moonspace{2}}
\def\thinlongdashyspace{\thinlongdashy\moonspace{2}}
\def\dotty{\drawline{1.00}{0.75}}     
     
\def\dottyspace{\dotty\moonspace{2}}

\def\solid{\drawline{24}{0.75}\nobreak$\,$}

\def\dotbox{\hbox{\dottyspace}}
\def\dotted{\hbox{\leaders\dotbox\hskip 24truept}\nobreak$\,$}  
\def\dashbox{\hbox{\dashyspace}}  
\def\dashed{\hbox {\ndots=0 \loop\ifnum\ndots<3\advance\ndots by 1
        \dashbox\repeat\dashy}\nobreak$\,$}       
\def\thindashbox{\hbox{\thindashyspace}}  
\def\thindashed{\hbox {\ndots=0 \loop\ifnum\ndots<3\advance\ndots by 1
        \thindashbox\repeat\thindashy}\nobreak$\,$}       
\def\thindash{\hbox {\ndots=0 \loop\ifnum\ndots<3\advance\ndots by 1
        \thindashbox\repeat\thindashy}\nobreak$\,$}       
\def\longdashbox{\hbox{\longdashyspace}}  
\def\thinlongdashbox{\hbox{\thinlongdashyspace}}  
\def\longdash{\hbox {\ndots=0 \loop\ifnum\ndots<3\advance\ndots by 1
        \longdashbox\repeat\longdashy}\nobreak$\,$}       
\def\thinlongdash{\hbox {\ndots=0 \loop\ifnum\ndots<3\advance\ndots by 1
        \thinlongdashbox\repeat\thinlongdashy}\nobreak$\,$}       
\def\dotdashed{\hbox{\shortchainspace\dottyspace\shortchain}\nobreak$\,$}      
\usepackage{color}
\usepackage{comment}
\usepackage{amssymb,wasysym}
\definecolor{light-gray}{gray}{0.75}
\ifCUPmtlplainloaded \else
  \checkfont{eurm10}
  \iffontfound
    \IfFileExists{upmath.sty}
      {\typeout{^^JFound AMS Euler Roman fonts on the system,
                   using the 'upmath' package.^^J}%
       \usepackage{upmath}}
      {\typeout{^^JFound AMS Euler Roman fonts on the system, but you
                   dont seem to have the}%
       \typeout{'upmath' package installed. JFM.cls can take advantage
                 of these fonts,^^Jif you use 'upmath' package.^^J}%
      }
  \else
  \fi
\fi

\ifCUPmtlplainloaded \else
  \checkfont{msam10}
  \iffontfound
    \IfFileExists{amssymb.sty}
      {\typeout{^^JFound AMS Symbol fonts on the system, using the
                'amssymb' package.^^J}%
       \usepackage{amssymb}%
         \let\leq=\leqslant
         
      }{}
  \fi
\fi

\ifCUPmtlplainloaded \else
  \IfFileExists{amsbsy.sty}
    {\typeout{^^JFound the 'amsbsy' package on the system, using it.^^J}%
     \usepackage{amsbsy}}
    {\providecommand\boldsymbol[1]{\mbox{\boldmath $##1$}}}
\fi

\providecommand\bnabla{\boldsymbol{\nabla}}
\providecommand\bcdot{\boldsymbol{\cdot}}

\newsavebox{\astrutbox}
\sbox{\astrutbox}{\rule[-5pt]{0pt}{20pt}}

\usepackage[normalem]{ulem}

\title{Particle transport in turbulent curved pipe flow}

\author[A. Noorani, G. Sardina, L. Brandt, P. Schlatter]{Azad Noorani \thanks{Email address for correspondence: azad@mech.kth.se},\ns
Gaetano Sardina,\break Luca Brandt and Philipp Schlatter}

\affiliation{Swedish e-Science Research Centre (SeRC), Linn\'e FLOW Centre,\\ KTH Mechanics, SE-100 44 Stockholm, Sweden}

\date{\today}%?; revised ?; accepted ?. - To be entered by editorial office}
\begin{document}

\maketitle

\begin{abstract}
Direct numerical simulations (DNS) of particle-laden turbulent flow in straight, mildly curved and strongly bent pipes are performed in which the solid phase is modelled as small heavy spherical particles. A total of seven populations of dilute particles with different Stokes numbers, one-way coupled with their carrier phase, are simulated. The objective is to examine the effect of the curvature on micro-particle transport and accumulation. It is shown that even a slight non-zero curvature in the flow configuration strongly impact the particle concentration map such that the concentration of  inertial particles with bulk Stokes number $0.45$ (based on bulk velocity and pipe radius) at the inner-bend wall of mildly curved pipe becomes $12.8$ times larger than that in the viscous sublayer of the straight pipe. Near-wall helicoidal particle streaks are observed in the curved configurations with their inclination varying with the strength of the secondary motion of the carrier phase. A reflection layer, as previously observed in particle laden turbulent S-shaped channels, is also apparent in the strongly curved pipe with heavy particles. In addition, depending on the curvature, the central regions of the mean Dean vortices appear to be completely depleted of particles, as observed also in the partially re-laminarised region at the inner bend. The turbophoretic drift of the particles is shown to be affected by weak and strong secondary motions of the carrier phase and geometry-induced centrifugal forces. The first and second-order moments of the velocity and acceleration of the particulate phase in the same configurations are addressed in a companion paper by the same authors. The current data-set will be useful for modelling particles advected in wall-bounded turbulent flows where the effects of the curvature are not negligible.
\end{abstract}

%\begin{keywords}
%Keywords: \highlight{Add keywords from the JFM homepage from the list!}
%\end{keywords}

\section{Introduction}\label{sec:introduction}

Most of the flows observed in nature and industrial applications are turbulent, and are often carrying dispersed particulate phases. Some typical examples are the atmospheric boundary layer transporting volcanic ash and pollen particles, or the flow in engine combustion chambers with suspended droplets. In the context of turbulent flows, the dispersion process of heavier-than-fluid particles predominantly results in two peculiar phenomena: \emph{small-scale clustering} and \emph{turbophoresis}. Small-scale clustering affects the particle distribution resulting in the loss of spatial homogeneity since the particles aggregate in specific regions characterised by higher values of the turbulent kinetic energy dissipation rate; this phenomenon appears in all kinds of turbulent flow laden with heavy particles. On the contrary, turbophoresis is typical of turbulent wall-bounded flows and occurs as a pronounced particle accumulation in the near-wall region. These two phenomena are not completely independent but it has been shown that small-scale clustering, induces turbophoresis in wall flows \citep{sardina_schlatter_brandt_picano_casciola_2012}. Recent publications by \citet{balachandar_eaton_2010} and \citet{toschi_bodenschatz_2009} provide a review of the research activities in the area of dispersed phase in turbulent flows.

The origin of the preferential particle concentration resulting in small-scale clustering is still being studied. It is usually explained in terms of the combination of particle inertia and turbulent vortical structures: Inertia prevents particles from following the fluid trajectories leading to preferential concentration outside of the vortical regions \citep{squires_eaton_1991,wang_maxey_1993}. Other explanations have also been proposed for small-scale clustering since this peculiar phenomenon is also observed in random flows without the coherent vortical structures that usually appear in turbulence \citep[see][]{mehlig_etal_2005}. Among these, the sweep-stick mechanism proposed by \citet{goto_vassilicos_2008} and \citet{coleman_vassilicos_2009} suggests that the inertial particles tend to preferentially accumulate in regions characterised by a stagnation point of the flow acceleration field. 

On the other hand, turbophoresis is essentially an inertial particle drift towards the walls driven by turbulent fluctuations \citep{caporaloni_tampieri_trombetti_vittori_1975}. This process has been theoretically studied by \citet{reeks_1983,young_leeming_1997}, and  several experiments have been conducted to observe and quantify this near-wall accumulation \citep{kaftori_etal_1995_1,kaftori_etal_1995_2,ninto_garcia_1996}. Today, the most common way to address the problem is by means of direct numerical simulation (DNS) in classical channel or pipe-flow configurations at moderate Reynolds number (see \emph{e.g.}\ \citet{soldati_marchioli_2009} for a review). In particular, it is assumed that turbophoresis is strongly linked to the vortical structures of near-wall turbulence. Sweep and ejection events affect the particle transfer towards the wall as described by \citet{rouson_eaton_2001}. \citet{marchioli_soldati_2002} observed that the inertial particles tend to preferentially oversample the low streamwise velocity regions in the buffer layer, \emph{i.e.}\  ejection events. \citet{sardina_schlatter_brandt_picano_casciola_2012} further showed a link between this preferential particle localisation in the low speed streaks and the steady-state concentration profiles of the particles. According to \citet{milici_etal_2014} turbophoresis may be completely suppressed when the streaky patterns of wall-turbulence are destroyed \emph{e.g.}\ by adding wall roughness to a channel flow. The turbophoretic motion of the particles is also present in boundary-layer flows since the turbulent structures close to the wall are essentially similar to those of the canonical channel flow \citep[see][]{sardina_schlatter_picano_casciola_brandt_henningson_2012}. A recent study by \citet{sikovsky_2014} investigates the stochastic behaviour of inertial particles near the wall by means of matched asymptotic expansions. This study suggests that the particle concentration in the viscous sublayer of wall-bounded turbulent flows exhibits a power-law singularity and the corresponding exponent depends on the particle relaxation time.  

Internal flows relevant to industrial applications often contain parts with geometrical complexities such as curved sections or converging/diverging compartments, which in general produce secondary flows, separation and other complex flow phenomena. Despite some previous efforts, particle motion has mostly been studied in the context of canonical wall-bounded flows. Given the strong link between turbophoresis and turbulent near-wall structures, it is fundamental to investigate the particulate phase behaviour in complex geometries where a slight modification of the geometrical characteristics of the flow can induce a significant change in the vortical wall-dominated structures and consequently in the turbophoretic drift and particle clustering. The current research, therefore, will examine the effects of  curvature on particulate dispersion in bent pipes in comparison to the canonical (and more studied) straight pipe. More specifically, turbulent flow in curved pipes is occurring in many typical engineering applications such as heat exchangers, chemical reactors and pipeline systems. In these typical heat and mass transfer systems the flow is commonly laden with a dispersed phase \emph{e.g.}\ membrane ultrafiltration hydraulic systems or mixing devices applied in pharmaceutical industries. 

The imbalance between the cross-stream pressure gradient and geometry induced centrifugal forces causes a secondary motion orthogonal to the main flow direction, which is characterised by the formation of a pair of counter-rotating vortices, the so-called Dean vortices. These vortices force the streamwise velocity to be distributed non-uniformly in the cross section of the pipe such that its maximum is deflected towards the outer wall.  Secondary motion that is geometry or skew-induced appears both in laminar and turbulent flows unlike the stress-induced secondary flow that is solely formed as a result of the local variation of the Reynolds stresses in turbulent regimes. The former type is usually referred to as Prandtl's secondary flow of first kind and the latter is called Prandtl's secondary flow of second kind which can be observed, for instance, in the turbulent flow in straight ducts with non-circular cross-section \citep[see][]{bradshaw_1987}.

In general, curved pipe geometries can be classified as: \emph{(a)} Helically coiled tubes where the flow is fully developed inside the curved geometry; \emph{(b)} Spatially developing bends such as \emph{U}-bends or elbows (90$^\circ$ bends) where a fully developed flow from a straight pipe enters the bend. Generally the pitch of the resulting coil in category \emph{(a)} causes an additional force which is the torsion acting alongside the centrifugal force on the fluid flow. However, when the coil pitch is small compared to the coil diameter, which is the case in most practical applications, the influence of the torsion is negligible with respect to the effect of the curvature \citep[see][]{manlapaz_churchill_1981, germano_1982}. This reduces the geometry to that of a toroidal shape. The latter category \emph{(b)} and in particular bends with small curvature where the entry flow region is short compared to the length of the pipe in the curved section can also be modelled as toroidal pipes. This abstraction -- infinitely long bent pipe configuration -- thus provides a unique opportunity to isolate the effect of the curvature on the turbulent characteristics and also to study the influence of the centrifugal forces and the secondary motion on the near-wall features.

Due to the complexity of the flow field in curved pipe configurations very limited efforts have been made in the past to unfold the intrinsic dynamics related to the turbulent flow in toroidal pipes. \citet{boersma_nieuwstadt_1996} employed large-eddy simulation (LES) to examine the effect of curvature on the mean flow and fluctuations. \citet{huttl_friedrich_2000} performed the first DNS in the framework of Germano's coordinates \citep{germano_1982} to test the curvature and torsion effect on turbulence at a friction Reynolds number $Re_{\tau} = 230$, based on the azimuthally averaged friction velocity and pipe section radii. Only recently, \citet{noorani_etal_2013} performed DNSs with high enough Reynolds numbers for the flow to remain turbulent even in strongly curved pipes. These authors computed the full Reynolds stress budget at various curvature configurations for the first time and have shown that the turbulent flow in the inner side of the strongly curved pipe is highly damped while the outer bend remained fully turbulent with essentially unchanged near-wall dynamics. 

A dispersed phase transported in turbulent curved pipes is subjected to the wall-dominated turbulence of the carrier phase, its secondary motion due to the Dean vortices and volumetric geometry-induced centrifugal forces. These effects may have a large impact on the turbophoretic motion of the particles. Investigations of a dispersed phase in turbulent flow in toroidal pipes are difficult to perform. Therefore, numerical or experimental data are very rare in the literature \citep[see][]{vashisth_kumar_nigam_2008}. Examining membrane filtration modules, as an example, shows the reduction of near-wall concentration of the particles in curved geometries compared to the straight configuration \citep[see][]{winzeler_belfort_1993}. Recently, \citet{wu_young_2012} conducted a set of experimental measurements and theoretical calculations considering the inertial particle dispersion rate in spatially developing mildly curved ducts (with rectangular cross-section). In comparison to the straight configuration, they reported a dramatic increase in the accumulation rate of particles at the outer bend surface. Their study showed that large heavy inertial particles are mostly driven by centrifugal forces rather than turbophoresis even in a mildly curved configuration. From a numerical point of view, some attempts to study particle transport in $90^\circ$ pipe bends have been recently conducted by means of LES \citep{breuer_etal_2006,berrouk_laurence_2008}. Applying an artificial swirl in a spatially developing turbulent straight pipe simulation \citet{zonta_marchioli_soldati_2013} studied the effect of centrifugal forces along with turbophoretic drift on near-wall particle accumulation. 

In another study to examine the particles behaviour in the presence of geometrical curvature, \citet{huang_durbin_2010, huang_durbin_2012} performed DNS of particulate dispersion in a strongly curved S-shaped channel. They examined the erosive effect of wall--particle collisions at the channel walls and reported that the maximum erosion rate is dramatically increased for heavier particles. They also observed a plume of particles bouncing back from the outer wall of the curved section and building up a high-particle concentration region adjacent to the wall, which is named as \textit{'reflection layer'}. \citet{huang_durbin_2012} further presented a simple model to explain the oscillatory behaviour of the wall collision for heavy inertial particles. %As the number of studies in this realm is growing still the presence of a fully developed particle dispersion that can be suitable for theoretical analysis is missing.  

The current study is devoted to the complex dynamics of particle dispersion in the presence of the curvature where a stably secondary motion and centrifugal forces acting on the particles co-exist. Fully developed turbulent pipe flows at three different curvatures laden with inertial small heavy particles are simulated. In these configurations, the transport and accumulation of these dilute micro-size particles are investigated and addressed in the current paper. The first and second-order moments of the velocity and acceleration of the particulate phase in turbulent bent pipes are addressed in a companion paper by \citet{noorani_sardina_brandt_schlatter_FTC_2015}. In the pages that follow, the computational methodology for simulating the carrier and dispersed phase is explained in \S \ref{sec:method} along with a discussion of the basic flow features in \S \ref{sec:flow}. The instantaneous particle distribution in the pipe, particle trajectories, and Eulerian statistics of the Lagrangian phase, in particular concentration maps, are discussed in \S \ref{sec:part}. Conclusions and an outlook are given in the final section \S \ref{sec:conclusions}.      

\section{Computational Methodology}\label{sec:method}
\subsection{Flow configuration and governing equations}
\begin{figure}
  \centerline{\includegraphics*[width=10cm]{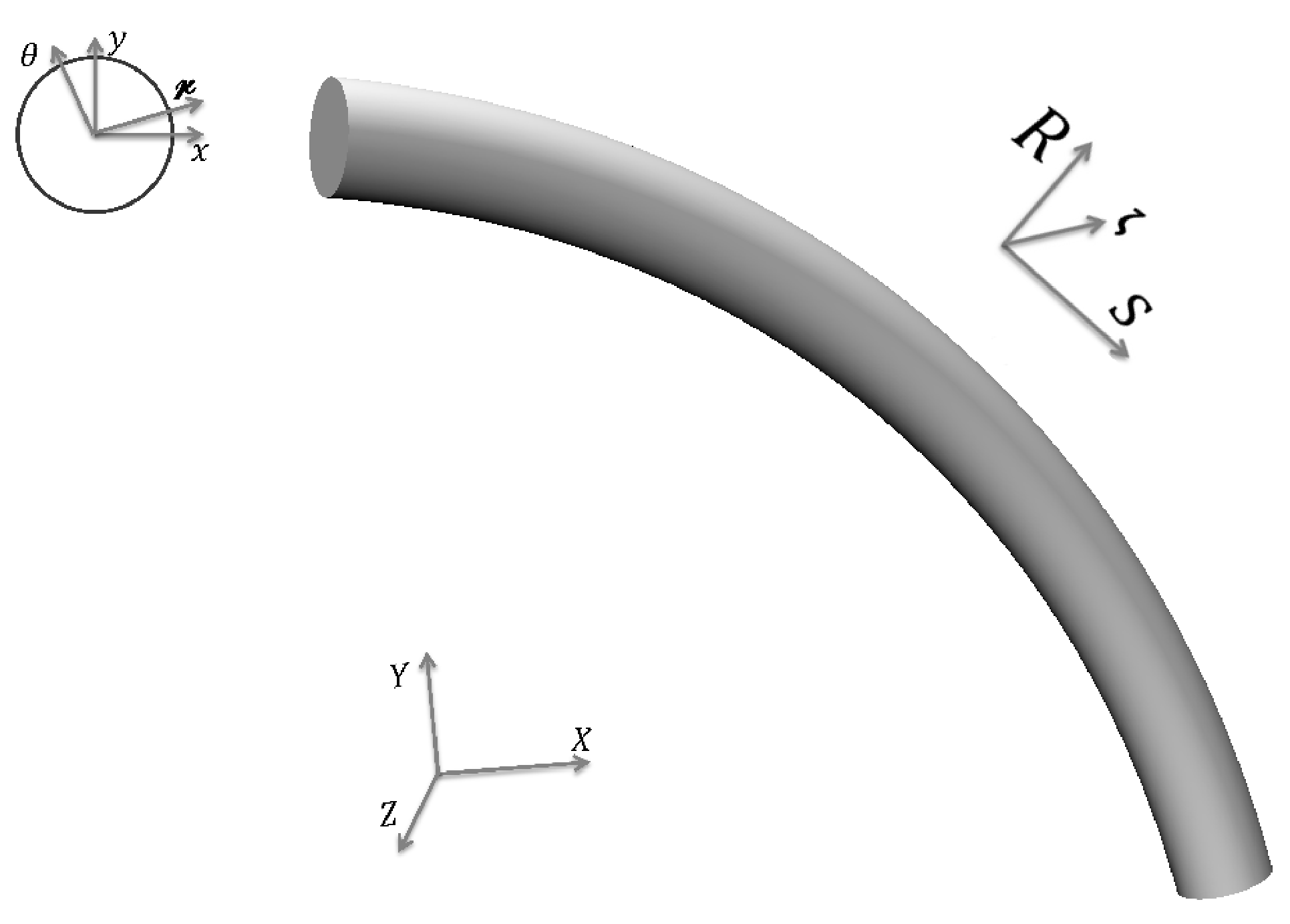}
  \put(-320,140){$(a)$}}
  \centerline{\includegraphics*[width=4cm]{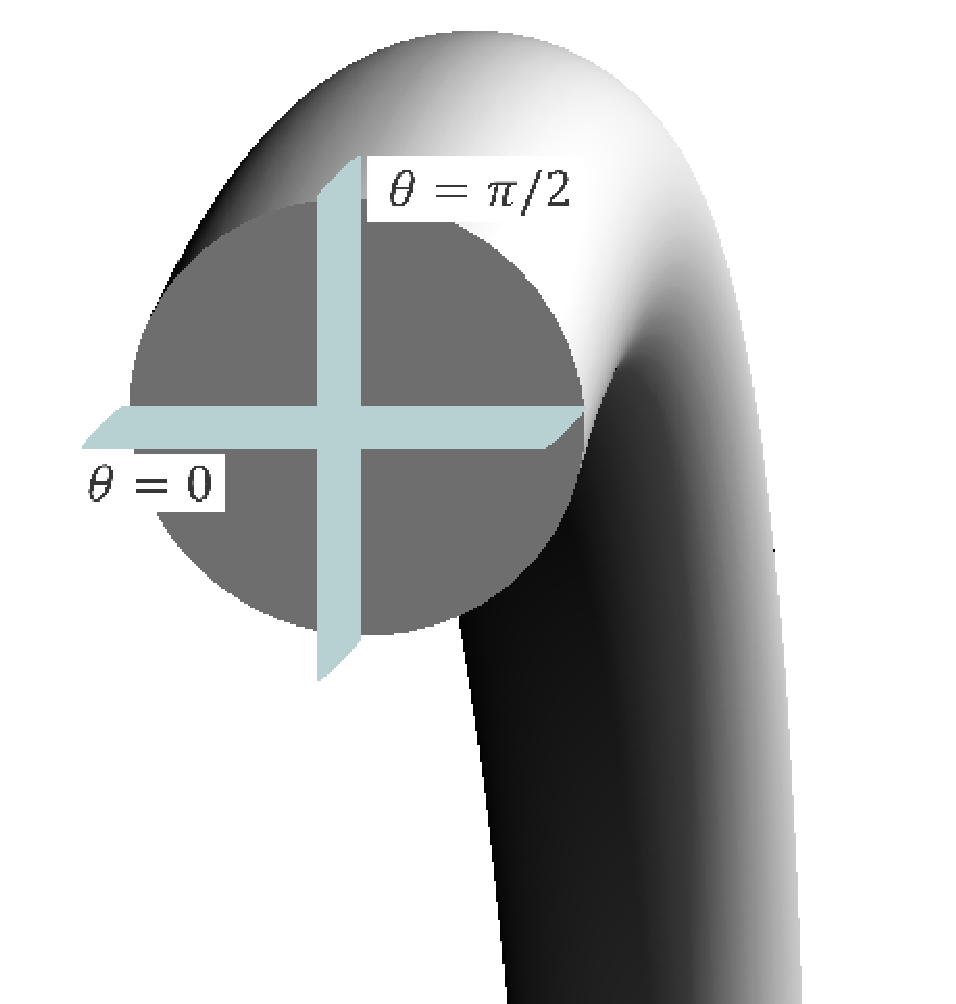}\put(-150,140){$(b)$}
              \includegraphics*[width=6cm]{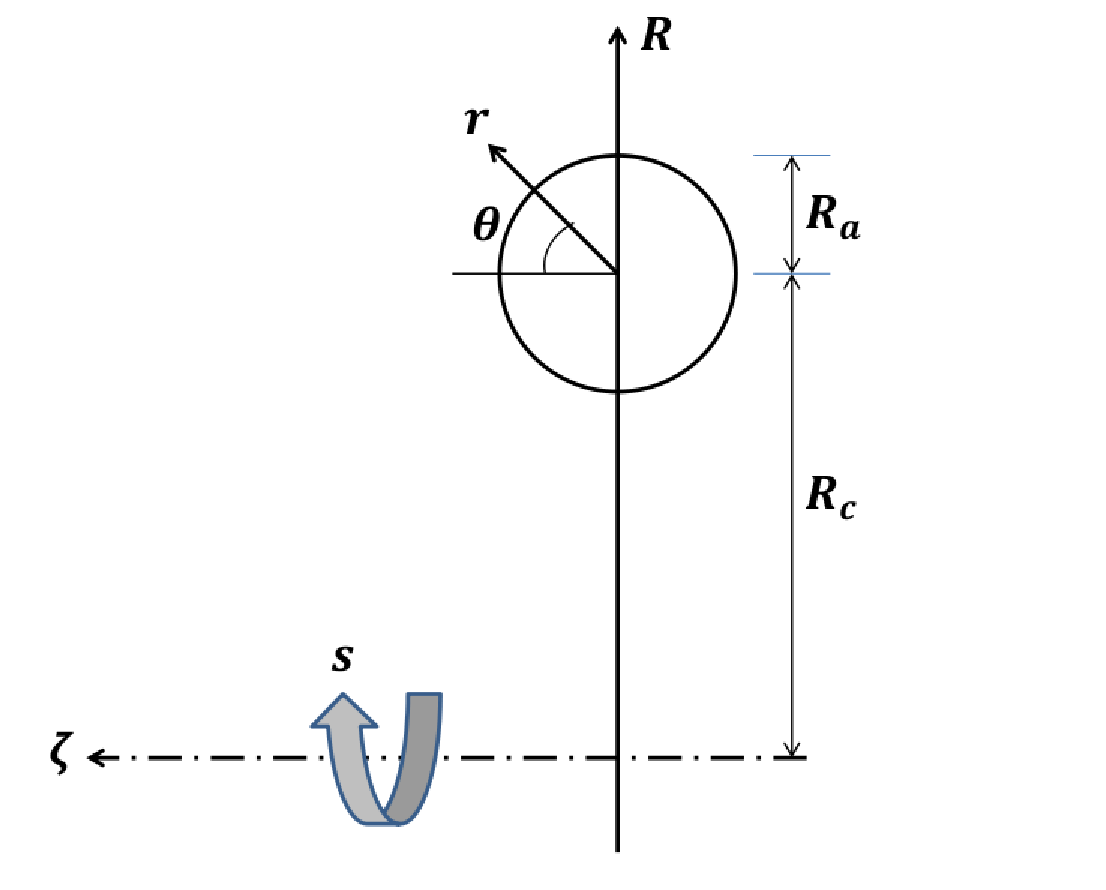}\put(-150,140){$(c)$}}
  \caption{\emph{(a)} Side view of curved pipe geometry and its associated Cartesian coordinates, $(X,Y,Z)$, with embedded toroidal coordinate system, $(R,s,\zeta)$. The in-plane polar coordinates, $(r,\theta)$, relative to the cross-section Cartesian triads, $(x,y)$, of the pipe-section are also illustrated at the \emph{left} panel. \emph{(b)} Cut-away view of a curved pipe, with locations of the horizontal and vertical cross-sections displayed at $\theta=0$ and $\theta=\pi/2$ respectively. \emph{(c)} The schematic front view of the curved configuration with toroidal and in-plane poloidal coordinates.}
\label{fig:fig1}
\end{figure}

In order to provide a clear overview of the computational domain employed for the DNSs a schematic of a curved pipe is given in figure \ref{fig:fig1}. The pipe consists of a part of a torus whose centre is located at the origin of the Cartesian system $(X,Y,Z)$. Relative to this system the toroidal coordinates $(R,s,\zeta)$ of the pipe together with local (in-plane) poloidal coordinates $(r,\theta)$ are shown in figure \ref{fig:fig1} \emph{(a)}. The equatorial mid-plane of the pipe which is also the plane of symmetry is indicated as vertical cut of the cross-section of the pipe at $\theta=\pi/2$. The horizontal cut of the pipe section at $\theta=0$ is displayed in figure  \ref{fig:fig1} \emph{(b)}. Of particular relevance in this configuration is the curvature parameter $\kappa$ that can be defined as $R_a/R_c$ where $R_a$ is the radius of the pipe cross-section and $R_c$ is the major radius of the torus at the pipe centreline (shown in figure \ref{fig:fig1} \emph{c}). Generally $\kappa$ distinguishes between mildly (weakly) curved pipes $(\kappa \approx 0.01)$ and strongly curved configurations $(\kappa \gtrsim 0.1)$.
%%%%%%%%%%%%%%%%%%%%%%%%%%%%%%%%%%%%%%%%%%%%%%%%%%%%%%%%%%%%%%%%%%%%%%%%%%%%%%%%%%%%%%
%%%%%%%%%%%%%%%%%%%%%%%%%%%%%%%%%%%%%%%%%%%%%%%%%%%%%%%%%%%%%%%%%%%%%%%%%%%%%%%%%%%%%%

All  simulations are performed in Cartesian coordinates where the incompressible Navier--Stokes equations for the carrier phase read in non-dimensional form
%%%%%%%%%%%%%%%%%%%%%%%%%%%%%%%%%%%%%%%%%%%%%%%%%%%%%%%%%%%%%%%%%%%%%%%%%%%%%%%%%%%%%%
\begin{eqnarray}
\bnabla\bcdot\boldsymbol{u} & = & 0, \label{eq:Cont} \\
\frac{\mathrm{\partial} \boldsymbol{u}}{\mathrm{\partial} t} + (\boldsymbol{u}\bcdot \bnabla)\boldsymbol{u} & = & -\bnabla p +  \frac{1}{Re_{b}}\nabla^{2}\boldsymbol{u} \label{eq:moment} \ .
\end{eqnarray}
%%%%%%%%%%%%%%%%%%%%%%%%%%%%%%%%%%%%%%%%%%%%%%%%%%%%%%%%%%%%%%%%%%%%%%%%%%%%%%%%%%%%%%
Here, $\boldsymbol{u}$ indicates the velocity vector of the fluid flow, and $p$ is the pressure. $Re_b$ stands for the bulk Reynolds number defined as $2R_au_b/\nu$; $u_b$ is the bulk velocity, and $\nu$ is the kinematic viscosity. The Dean number  is often used in the literature to characterise the secondary motion and is defined as $De_b=Re_b\sqrt{\kappa}$ for the current flow configurations \citep{berger_talbot_yao_1983}.

\subsection{Numerical approach for the carrier phase}
The present DNSs are performed using the spectral-element code \texttt{nek5000}. This massively parallel code is developed by \citet{nek5000-web-page} at Argonne National Laboratory (ANL). The physical domain is divided into a number of hexahedral local elements on which the incompressible Navier--Stokes equations are solved by means of local approximations based on high-order orthogonal polynomial basis on Gauss--Lobatto--Legendre (GLL) nodes. This provides spectral accuracy with geometrical flexibility applicable to problems with moderately complex geometries. The spatial discretisation is obtained by means of the Galerkin projection, applying the $\mathbb{P}_N-\mathbb{P}_N$ scheme where the velocity and pressure spaces are represented by the same polynomial order \citep{tomboulides_lee_orszag_1997}. The temporal discretisation uses a $3^{rd}$ order mixed Backward Difference/Extrapolation (BDF3/EXT3) semi-implicit scheme. The walls are set to no-slip boundary condition while periodicity is applied for the end sections of the pipe domains in the axial direction as the flow is assumed to be homogenous in this direction. The initial condition for all the simulations are set to be the laminar Poiseuille profile of the straight pipe. This initial flow is perturbed with low-amplitude pseudo-random noise which is then evolved in time until the turbulent state is statistically fully developed in the pipe; this corresponds to approximately $200$ integral (convective) time units from the start. The nondimensional convective time unit is defined as $tu_b/R_a$. The driving force of the flow is adapted in such a way to keep the mass flow rate fixed throughout the simulations. This is performed implicitly in the time-integration scheme.

\begin{figure}
  \centerline{\includegraphics*[width=4.6cm]{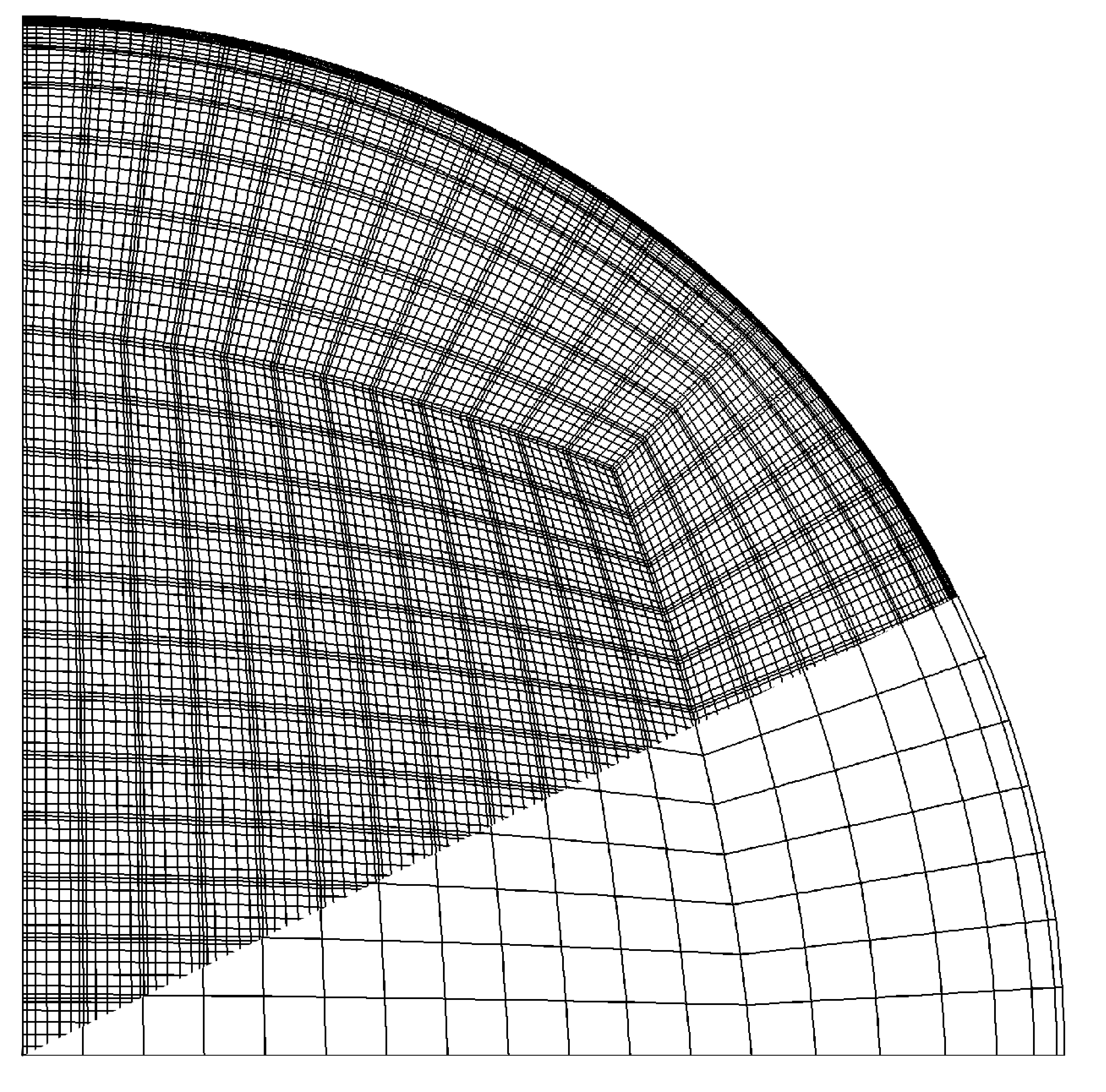}
              \includegraphics*[width=5.6cm]{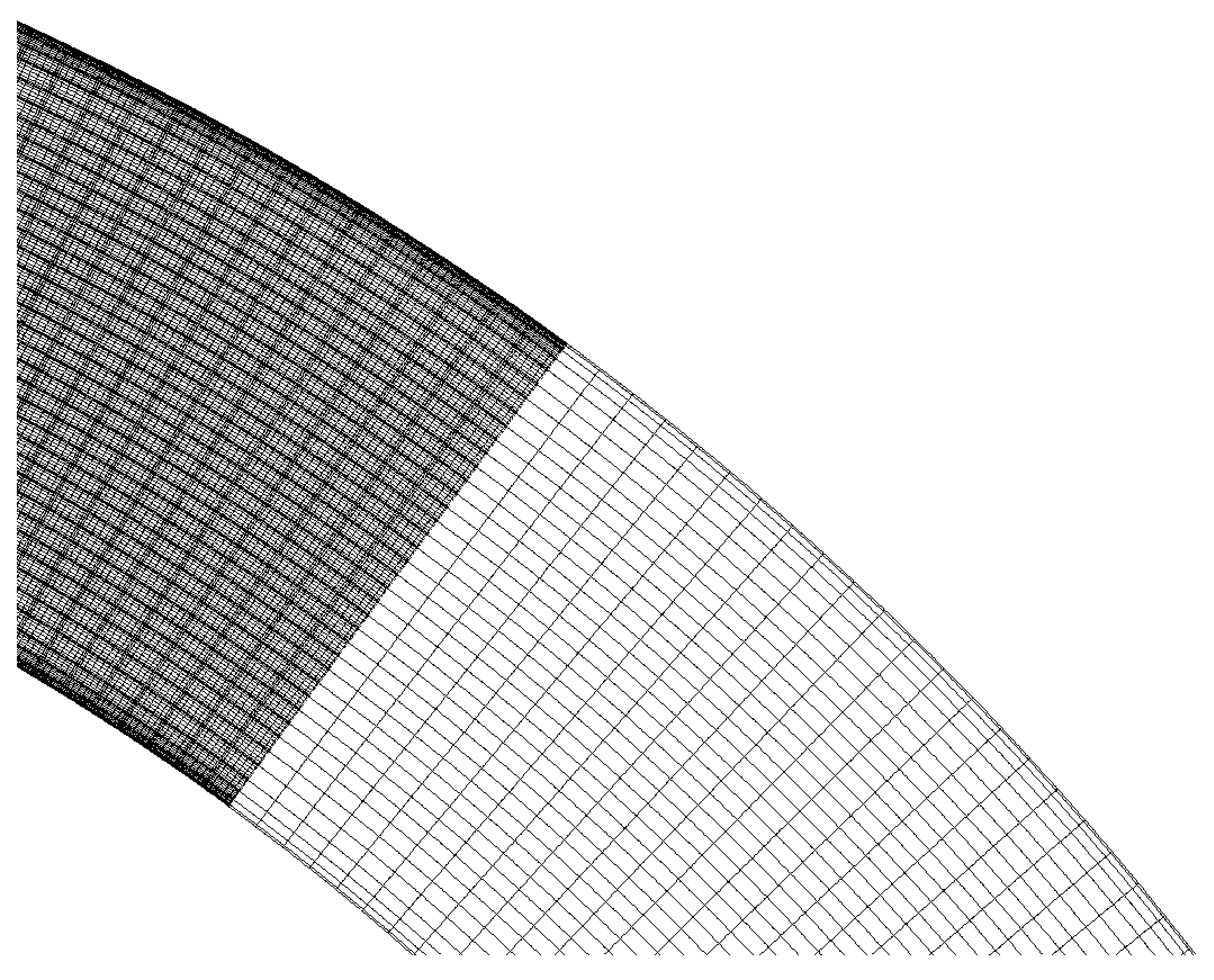}
              \put(-320,120){$(a)$}
              \put(-180,120){$(b)$}}
  \caption{\emph{(a)} A quarter of a cross-sectional plane pertaining to the pipe mesh. \emph{(b)} Grid representation in the equatorial mid-plane of the toroidal pipe. Here, element boundaries and Gauss--Lobatto--Legendre (GLL) points are shown in the darker region.}
\label{fig:fig2}
\end{figure}

Note that the numerical setup for the carrier phase is similar to the one used by \citet{noorani_etal_2013} for their highest Reynolds number; in particular the mesh topology and resolution are the same. Nevertheless, we provide a short overview of the main characteristics here for completeness. Figure \ref{fig:fig2} \emph{(a)} displays a quarter of the cross-sectional grid that is generated for the pipe section. The two dimensional mesh is uniformly extruded in the axial direction to generate the hexahedral elements of the 3D straight pipe mesh. The required grid for curved/toroidal configurations is obtained using an analytical morphing of the straight pipe grid. For all the cases, $8^3$ GLL points (polynomial order $7$) resolve the grid inside each individual element. The resulting mesh in the equatorial section of the bent pipe is sketched in figure \ref{fig:fig2} \emph{(b)}.
%%%%%%%%%%%%%%%%%%%%%%%%%%%%%%%%%%%%%%%%%%%%%%%%%%%%%%%%%%%%%%%%%%%%%%%%%%%%%%%%%%%%%
%%%%%%%%%%%%%%%%%%%%%%%%%%%%%%%%%%%%%%%%%%%%%%%%%%%%%%%%%%%%%%%%%%%%%%%%%%%%%%%%%%%%%%

It is known that the critical Reynolds number for transition to turbulence is generally higher in curved pipes compared to the straight configuration. The present simulations are carried out at a fixed $Re_b=11700$ such that the flow in all curved configurations is above the critical value given by \citet{ito_1959}. The domain length of all configurations is fixed to be $25R_a$ along the pipe centreline. \cite{chin_2010} studied the effect of the streamwise periodic length on the convergence of turbulence statistics in the straight pipe. Their study concludes that a pipe length of $8\pi R_a$ is sufficient for the flow statistics to be independent of the periodic domain length; our choice of the domain length is based on these findings. In the straight configuration the grid spacing is designed to have 14 grid points placed below $y^+=(1-r/R_a)^+=10$, measured from the wall, and $\Delta(R_a\theta)_{max}^+\leq5$ and $\Delta s_{max}^+\leq10$. These are measured in wall units where the viscous length scale is based on the straight pipe friction Reynolds number (the superscript $^+$ indicates the viscous scaling). The resolution of the grid for the curved pipes is chosen to be the same as that of the straight configuration. The characteristics of the grid are summarised in table \ref{tab:tab1}. 
%%%%%%%%%%%%%%%%%%%%%%%%%%%%%%%%%%%%%%%%%%%%%%%%%%%%%%%%%%%%%%%%%%%%%%%%%%%%%%%%%%%%%%
\begin{table}
\begin{center}
\begin{tabular}{lccccc}
%\hline
$Re_b$  & \# of elements & \# Grid points & $\Delta r^+$  &  $\Delta (R_a\theta)^+$ & $\Delta s^+$\\ [3pt]
\hline
$11700$ & $237120$ & $ 121.4  \times 10^{6}$ & $(0.16,4.70)$ & $(1.49,4.93)$  & $(3.03,9.91)$ \\
%\hline
\end{tabular}
\caption{Resolution details for the present pipe flow meshes. All quantities are based on the straight pipe configuration.}
\label{tab:tab1}
\end{center}
\end{table} 
%%%%%%%%%%%%%%%%%%%%%%%%%%%%%%%%%%%%%%%%%%%%%%%%%%%%%%%%%%%%%%%%%%%%%%%%%%%%%%%%%%%%%%

After the fluid flow reached the statistically stationary state, statistics of the turbulent flow are computed based on averages (denoted by $\langle \cdot \rangle$) in time ($t$), and along the axial direction ($s$) which is the homogeneous direction of the flow. The statistical sampling is started approximately after $200 R_a/u_b$ to make sure the flow is fully developed and settled in the turbulent state. The averaging time frame and other relevant simulation parameters are listed in table \ref{tab:tab2}. Note that for all cases 3000 convective time units were included in the temporal average, which proved to be necessary for obtaining converged statistics for the Lagrangian phase; for the flow statistics a shorter time frame would have been sufficient. Further details on how to obtain statistical averages via tensor transformations under rotation from Cartesian coordinates to toroidal/poloidal coordinates are given in \citet{noorani_etal_2013}. The verification of the method and the validations of the simulation setup along with statistical analysis for the carrier phase is also performed by \citet{noorani_etal_2013}.
%%%%%%%%%%%%%%%%%%%%%%%%%%%%%%%%%%%%%%%%%%%%%%%%%%%%%%%%%%%%%%%%%%%%%%%%%%
\begin{table}
   \begin{center}
   \def~{\hphantom{0}}
   \begin{tabular}{lccc}
%\hline
 $Re_{b,D}$ & ${\kappa}$ & $Re_{\tau}$ & Averaging time frame ($tu_b/R_a$) \\[3pt]
   \hline
 $11700$   &  $0.00$    & $360$       &   $3000$\\
 $11700$   &  $0.01$    & $368$       &   $3000$\\
 $11700$   &  $0.10$    & $410$       &   $3000$\\[3pt]
%   \hline
   \end{tabular}
   \caption{Simulation parameters of the present study. The time $t$ is normalised by $R_a/u_b$. The non-dimensional friction Reynolds number $Re_\tau$ is defined as $\overline{u_{\tau}}R_a/\nu$.}   
   \label{tab:tab2}
   \end{center}
\end{table}
%%%%%%%%%%%%%%%%%%%%%%%%%%%%%%%%%%%%%%%%%%%%%%%%%%%%%%%%%%%%%%%%%%%%%%%%%%%%

In the curved configurations the mean friction velocity $\overline{u_{\tau}}$ is defined as $\sqrt{\overline{\tau_{w,tot}}/\rho_f}$ where $\overline{\tau_{w,tot}}$ indicates the total mean shear stress over the wall and $\rho_f$ is the fluid density. The mean shear stress can be obtained as $\surd{(\overline{\langle \tau_{w,s} \rangle}^2+ \overline{\langle \tau_{w,\theta}\rangle}^2)}$ where overbar denotes averaging along the circumference of the pipe section of the streamwise mean shear stress $\langle \tau_{w,s}\rangle$ and its azimuthal component $\langle \tau_{w,\theta}\rangle$. Note that for the straight pipe, due to symmetry $\langle \tau_{w,\theta}\rangle$ vanishes. The resulting $\overline{u_{\tau}}$ is used to compute the viscous scaling.

\subsection{Numerical approach for the particulate phase}
In order to characterise the dispersed phase one commonly defines the volume fraction of particles with respect to the fluid as $\Phi_v = N_p\mathbb{V}_p/\mathbb{V}$ where $N_p$ is the number of particles, $\mathbb{V}_p$ is the volume of one particle and $\mathbb{V}$ is the global volume occupied by the carrier phase. The density ratio is defined as $\rho_p/\rho_f$; $\rho_p$ being the particle density. According to \citet{elghobashi_1991, elghobashi_1994} these two parameters ($\Phi_v$ and $\rho_p/\rho_f$) determine the level of interaction between the dispersed and carrier phase. Considering sufficiently small volume fraction $\Phi_v \lesssim 10^{-6}$ and large enough density ratio (around $ \sim 10^3$) the dynamics of the particles is essentially governed by the turbulent carrier phase. In fact, for dilute small particles the feedback force acting on the fluid phase and the inter-particle collisions can be considered negligible \citep{balachandar_eaton_2010}. In this case the inertial particles are one-way coupled with the carrier (Eulerian) field. \citet{maxey_riley_1983} rigorously derived the equation of motion for a spherical rigid particle immersed in a non-uniform flow. The resulting equation of motion includes various forces acting on the individual particles. These derivations have been subjected to a number of corrections and sometimes additional terms are included \citep{crowe_schwarzkopf_sommerfeld_tsuji_2011}. Among the different forces the only significant contributions to drive a heavy point particle are the aerodynamic Stokes drag and the gravitational force \citep[see][]{elghobashi_truesdell_1992}. %Recently  \citet{olivieri_picano_sardina_iudicone_brandt_2014} have shown that the Basset history force becomes also important for inertial particles with $\rho_p/\rho_f \sim 1 - 10$. 

The current research is concerned with micron-sized dilute heavier-than-fluid particles that are assumed to be smaller than the smallest spatial scale of the flow. With these assumptions, the particle equation of motion reads
\begin{eqnarray}
\frac{\mathrm{d} \boldsymbol{v}_p}{\mathrm{d} t} & = & \frac{\boldsymbol{u}(\boldsymbol{x}_p,t)-\boldsymbol{v}_p}{St_b} \left( 1+0.15Re_p^{0.687}  \right), \label{eq:part1}\\
\frac{\mathrm{d} \boldsymbol{x}_p}{\mathrm{d} t} & = & \boldsymbol{v}_p, \label{eq:part2}
\end{eqnarray}
where $\boldsymbol{x}_p$ is the particle position,  $\boldsymbol{v}_p$ and $\boldsymbol{u}(\boldsymbol{x}_p,t)$ are the particle velocity and the fluid velocity at the particle position, respectively. The non-dimensional bulk Stokes number $St_b$ is defined as the ratio of particle relaxation time $\tau_p$ to bulk flow residence time ($R_a/u_b$), and characterises the particle response to the fluid. It can be computed as $\tau_pu_b/R_a$ where $\tau_p=\rho_p d_p^2/(18\rho_f \nu)$; $d_p$ is the particle diameter. The Stokes number as dimensionless form of the particle relaxation time can also be expressed considering the viscous (turbulence) time units such that $St^+=\tau_p\overline{u_{\tau}}^2/\nu$. The Stokes number based on the friction velocity ($St^+$) is more meaningful in the straight pipe/channel where near-wall turbulence is responsible for nonuniform inertial particles dispersion.

Equation (\ref{eq:part1}) essentially expresses the particle acceleration ($\boldsymbol{a}_p$) due to a steady-state drag force acting on the spherical solid particle in a uniform velocity field. The last term on the right-hand side is the \emph{non-linear} correction due to particle finite Reynolds number effects \citep{schiller_naumann_1933}. The particle Reynolds number $Re_p$ is defined as $|\boldsymbol{v}_p-\boldsymbol{u}(\boldsymbol{x}_p,t)|d_p/\nu$. For the current investigation, the effect of gravitational acceleration will be neglected to avoid losing generality and to be able to isolate the effect of the centrifugal force and the secondary motion of the Dean vortices on the turbophoretic behaviour of particles. 

Unlike the carrier phase which is simulated on a set of fixed grid points (Eulerian framework), the particles are tracked freely inside the computational domain (Lagrangian framework). The accuracy of the interpolation method is significant for the correct evaluation of the hydrodynamic forces and for instance it has direct consequences for the acceleration spectrum of the fluid particles \citep[see][]{vanhinsberg_tenThijeBoonkkamp_toschi_clercx_2013}. A spectrally accurate interpolation scheme, hence, is applied in the current study to evaluate fluid velocity at the centre of the spherical particle. Since in each time step the particle tracking is performed after updating the fluid phase, the Legendre basis functions and the corresponding expansion coefficients are already available for each element. Therefore, without losing the accuracy, the fluid properties (velocity, pressure, \emph{etc.}\ ) can be determined in arbitrary positions inside each element. The order of accuracy of this  interpolation, thus, is equal to the order of spectral element method ($7$ in the current simulations). The time stepper implemented for the particle tracking is the classical $3^{rd}$ order multistep Adams--Bashforth (\emph{AB3}) scheme. The time step of the carrier phase solver ($\Delta t$) is fixed such that the convective CFL condition is always below $0.5$. The time step of the particle solver is also set to be equal to the flow solver time step. As for the velocity field, re-starts of the simulations are based on storing all necessary previous times in order to avoid reducing the temporal order.

The particle interaction with solid walls is treated as \emph{elastic} collisions. A restitution coefficient can be defined as $e=|v_{n,2}/v_{n,1}|$; $v_{n,2}$ and $v_{n,1}$ are the wall-normal pre- and post-collisional particle velocities. In a \emph{purely elastic} collision (as used in the current study), $e$ equals unity and the total kinetic energy is conserved in the collision process. It has been shown previously that \emph{inelastic} reflection from the wall has a negligibly small effect on the dynamics of small inertial particles in turbulent flows \citep[see][]{li_mcLaughlin_kontomaris_portela_2001}. However, the analysis of \citet{huang_durbin_2012} suggested that in the presence of centrifugal forces the amount of collisions and their significance will increase especially for particles with larger $St$. Applying an inelastic collision scheme might therefore slightly alter the turbulent particle structures. % implementation details (e.g. velocity switching / curved boundaries)?
Similar to the carrier phase, periodic boundary conditions are used for the dispersed phase in the axial direction. 

\begin{figure}
  \centerline{\includegraphics*[width=10cm]{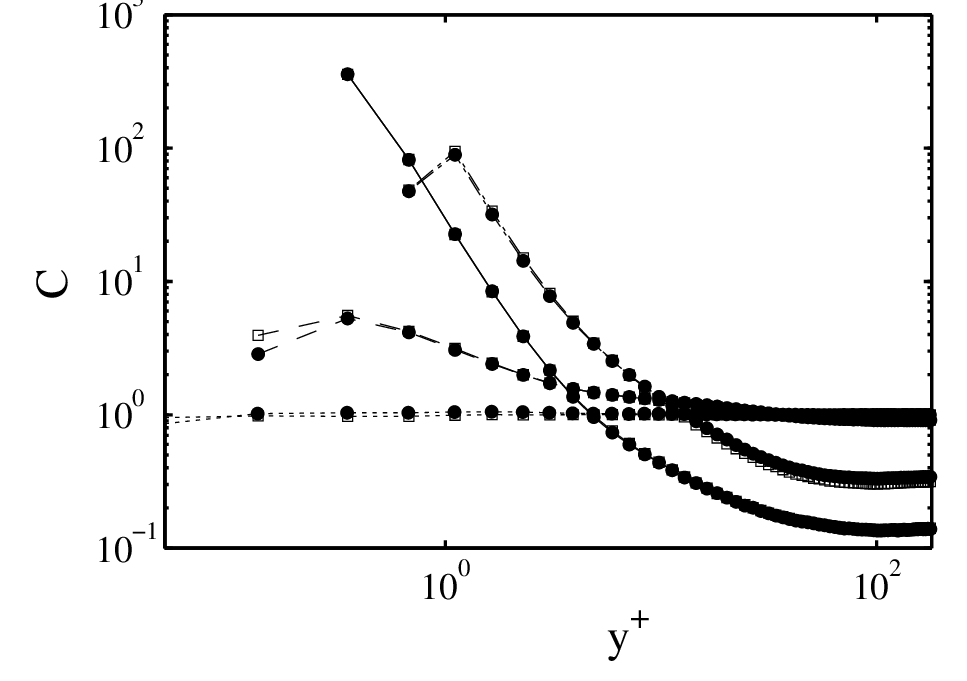}}
  \caption{Normalised statistically steady-state particle concentration distributed in the wall-normal direction of the turbulent channel. \dotted $St^+=0$, \dashed $St^+=1$, \solid $St^+=10$, \dotdashed $St^+=100$; $\square$ data by \citet{sardina_schlatter_brandt_picano_casciola_2012} with the pseudo-spectral code SIMSON \citep{chevalier_schlatter_lundbladh_henningson_2007}; \textbullet~present \texttt{nek5000} simulation.}
\label{fig:fig3}
\end{figure}

The implemented particle tracking scheme is validated for a classical particle-laden turbulent channel flow. For that purpose, DNS at $Re_\tau=180$ (based on the half-height of the channel) in a standard domain ($4\pi \times 2 \times 4/3\pi$) was performed and compared with the existing data of \citet{sardina_schlatter_brandt_picano_casciola_2012}. The latter simulation was performed using the fully spectral code SIMSON \citep{chevalier_schlatter_lundbladh_henningson_2007}.  A total of four particle populations is considered with Stokes number ($St^+ =0, 1, 10, 100$) where $\rho_p/\rho_f$ is set to $770$. The two simulations share similar set-ups in terms of initial conditions, start  and end time of the particle tracking and statistical sampling time frames. The normalised particle concentrations in the wall-normal direction, shown in figure \ref{fig:fig3}, show an excellent agreement between the data obtained with the two different codes.
 %%%%%%%%%%%%%%%%%%%%%%%%%%%%%%%%%%%%%%%%%%%%%%%%%%%%%%%%%%%%%%%%%%%%%%%%%%%%%%%%%%%%%%
%%%%%%%%%%%%%%%%%%%%%%%%%%%%%%%%%%%%%%%%%%%%%%%%%%%%%%%%%%%%%%%%%%%%%%%%%%%%%%%%%%%%%%

\section{Results and analysis}

In this section, results from the simulations detailed in \S \ref{sec:method} are discussed. In a first step in \S \ref{sec:flow}, the carrier phase is discussed with some detail, however the main focus is on characterising the particle phase in \S \ref{sec:part}.

\subsection{Carrier phase}\label{sec:flow}
It is crucial to have a clear overview of the carrier phase turbulence before analysing the complex dynamics of a particulate phase in bent pipes. To this end the main features of the turbulent flow in these configurations are initially described and discussed in this section. A more detailed analysis of the turbulent curved pipe flows can be found in \citet{noorani_etal_2013} which uses the same numerical setup.

\begin{figure}
   \centerline{\includegraphics*[width=3.6cm]{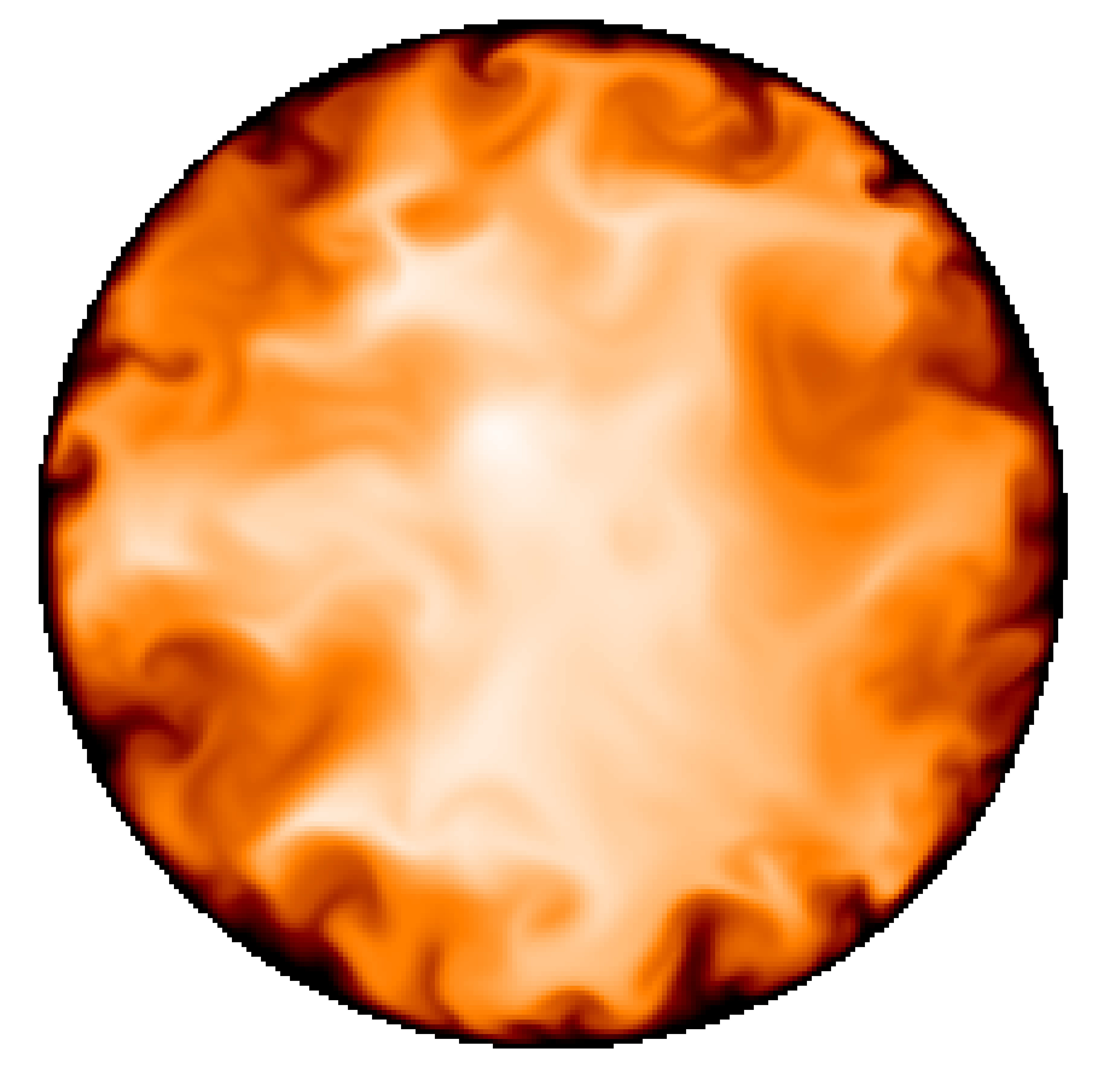}\put(-100,100){$(a)$}
               \includegraphics*[width=3.6cm]{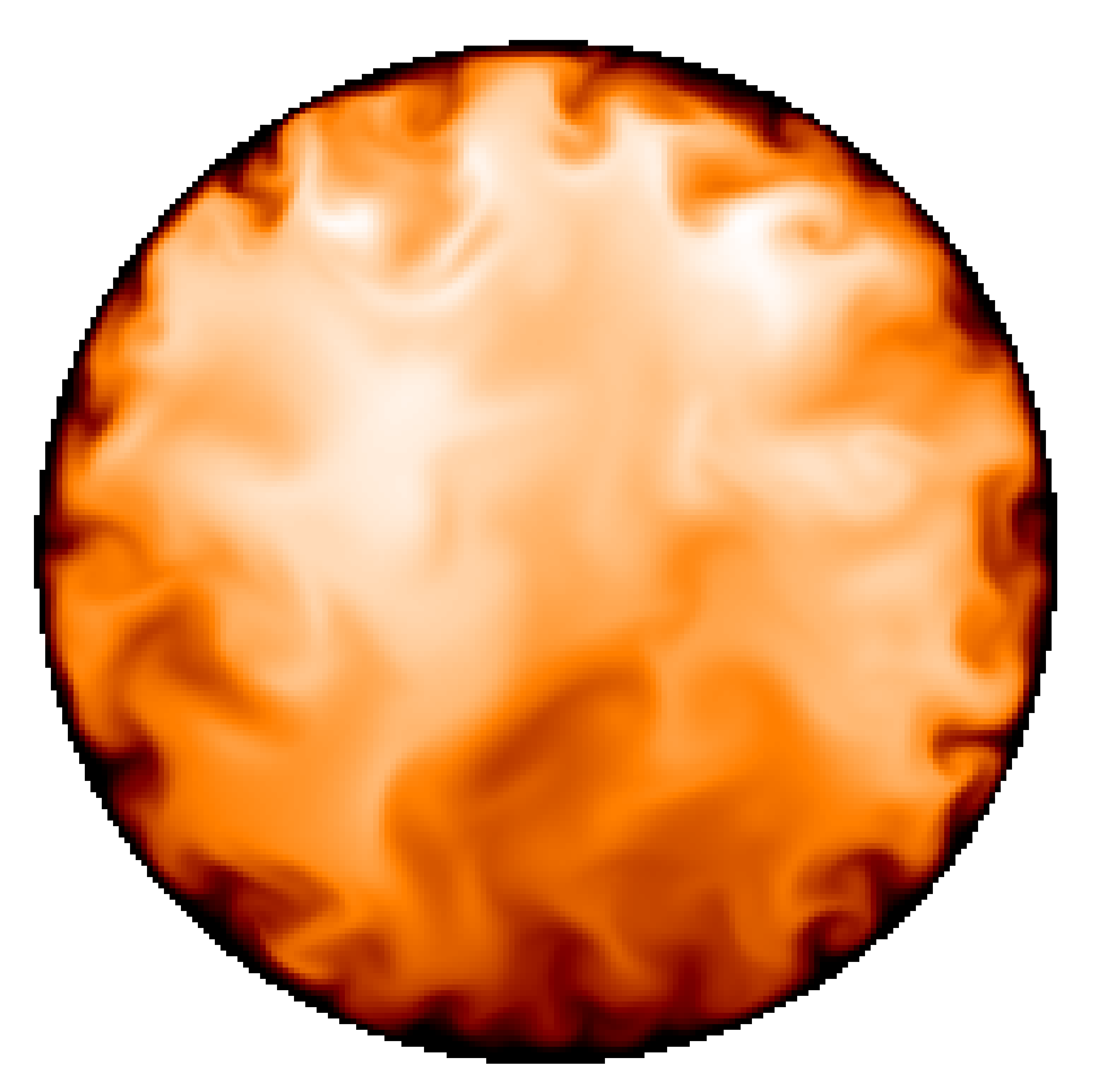}
               \includegraphics*[width=3.6cm]{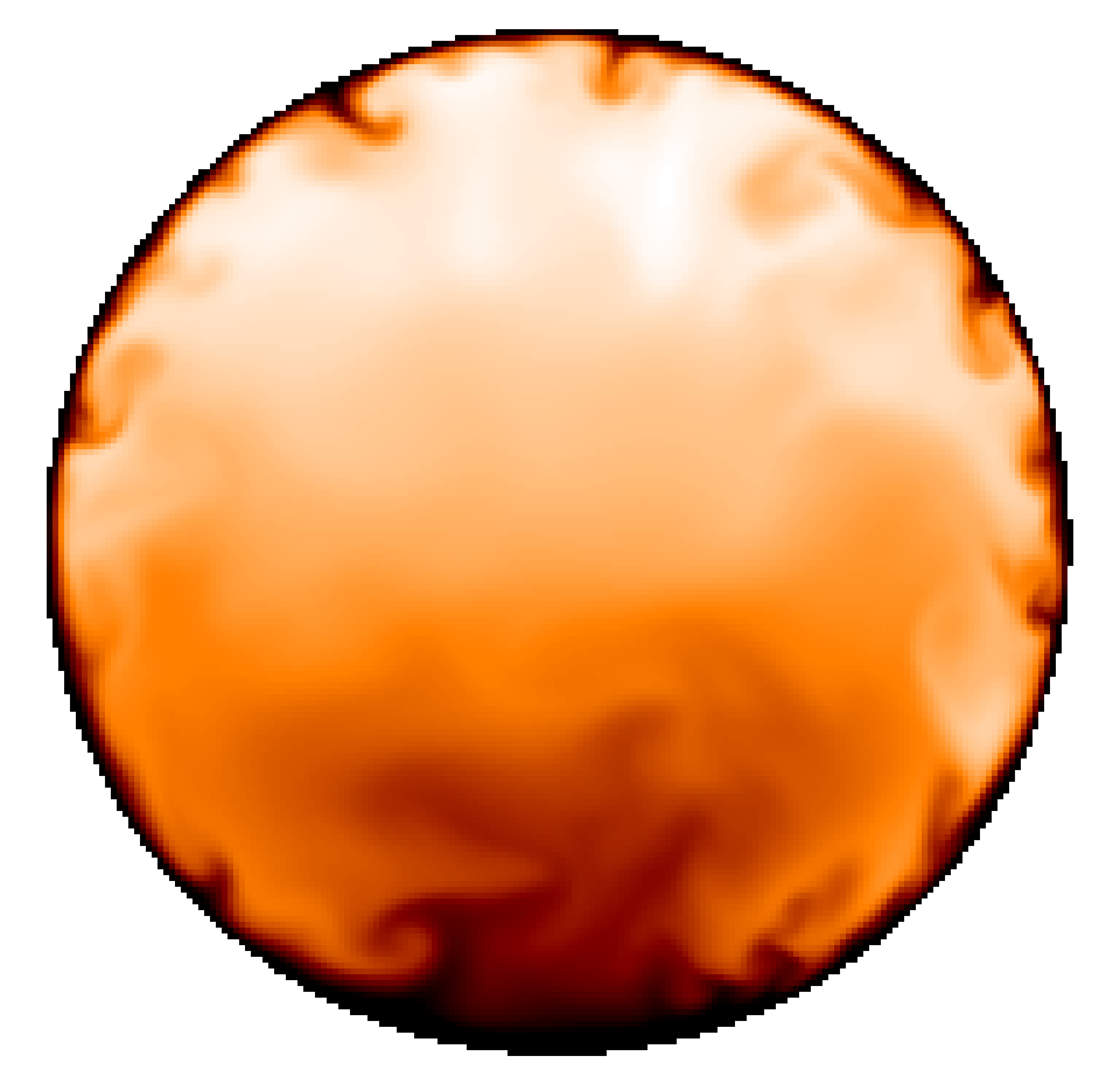}
               \includegraphics*[width=1.35cm]{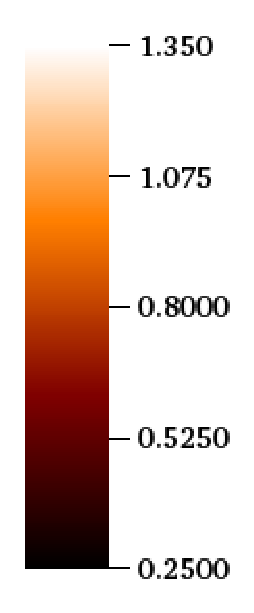}}
   \centerline{\includegraphics*[width=13.6cm]{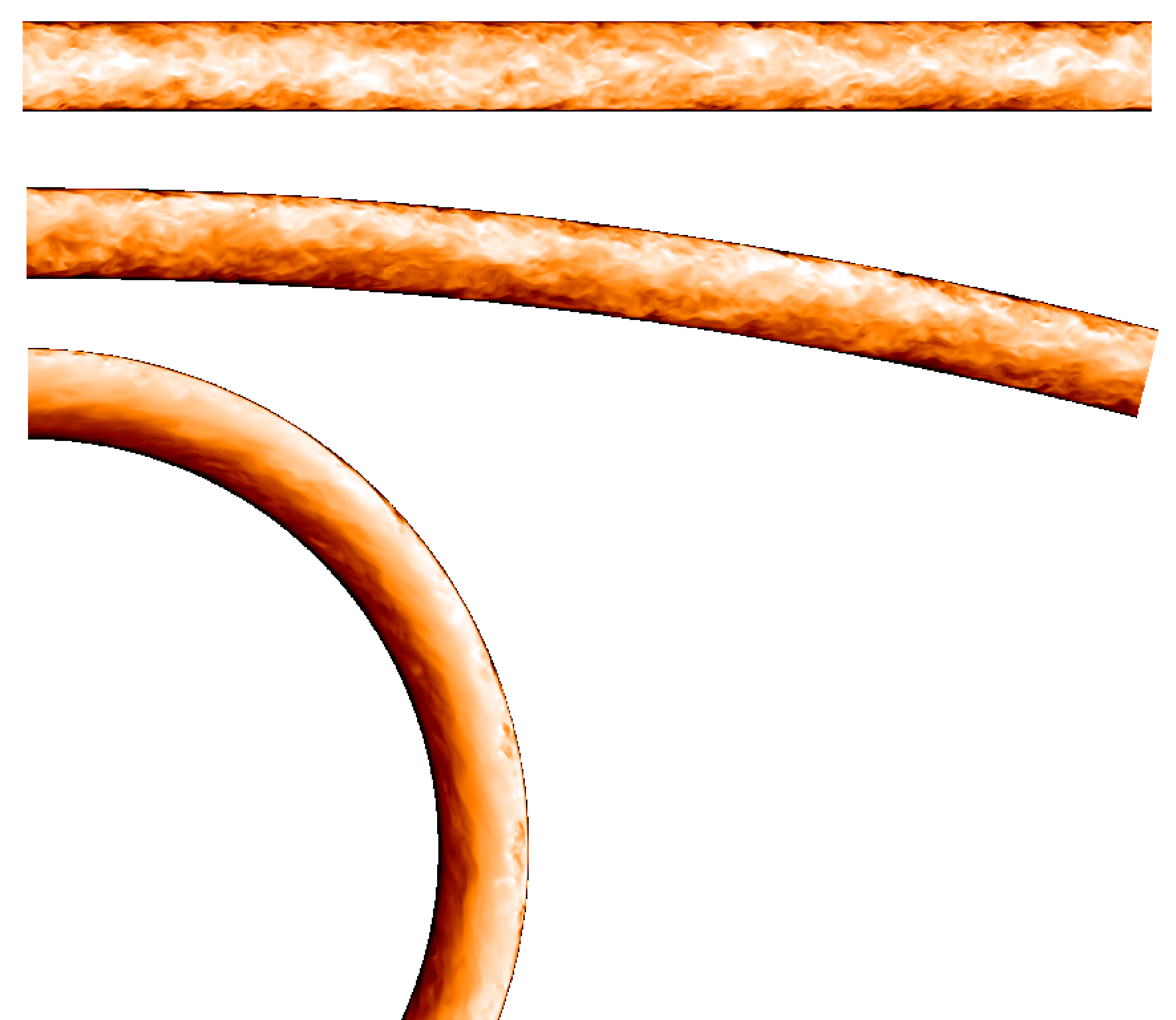}
               \put(-367,340){$(b)$}}
   \caption{Pseudocolours of axial velocity: \emph{(a)} in a cross-section; \emph{(from left:} $\kappa=0$, $0.01$, and $0.1$\emph{)}; \emph{(b)} in an azimuthal section for the straight pipe and the equatorial mid-plane for the mildly and strongly curved pipes.}
\label{fig:fig4}
\end{figure}

The qualitative modifications of the flow  when increasing the curvature parameter are illustrated in figure \ref{fig:fig4} where instantaneous cross-sectional views of the axial velocity are shown in the equatorial mid-plane and at an arbitrary axial position. Evidently, as the curvature parameter increases the  bulk flow is deflected further towards the outer bend, while the maximum of the axial velocity remains in the centreline of the straight pipe. This leads to a reduction in turbulence activity with increasing curvature. In the strongly curved pipe, this turbulence depression is large enough to almost re-laminarise the flow in the inner bend (\emph{c.f.}\ figure \ref{fig:fig8} \emph{c}).
%%%%%%%%%%%%%%%%%%%%%%%%%%%%%%%%%%%%%%%%%%%%%%%%%%%%%%%%%%%%%%%%%%%%%%%%%%%%%%%%%%%%%%
%%%%%%%%%%%%%%%%%%%%%%%%%%%%%%%%%%%%%%%%%%%%%%%%%%%%%%%%%%%%%%%%%%%%%%%%%%%%%%%%%%%%%%

\begin{figure}
  \centerline{\includegraphics*[width=6.6cm]{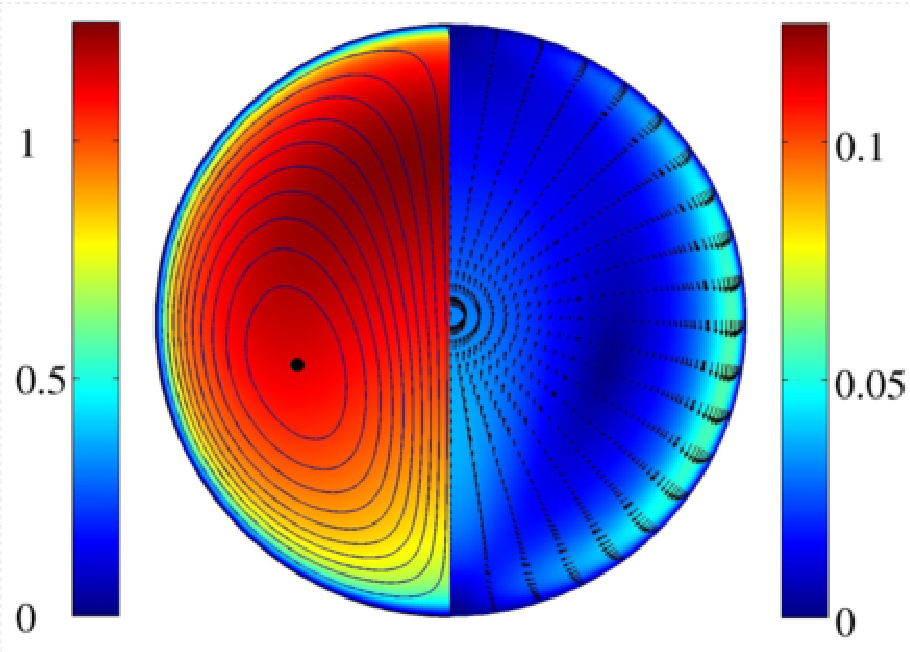} \put(-150,120){$(a)$}
             \includegraphics*[width=6.6cm]{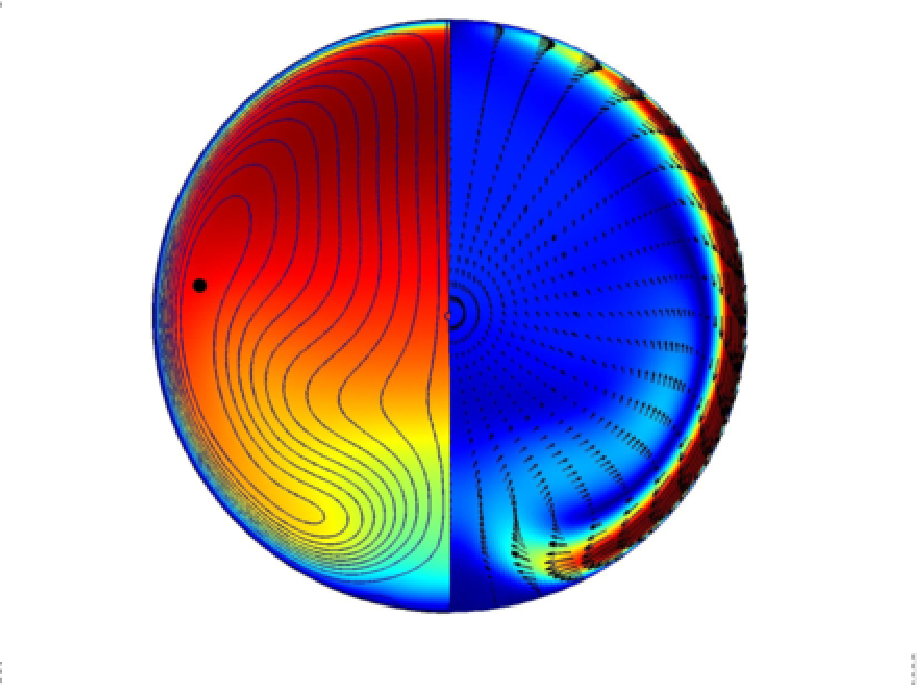} \put(-150,120){$(b)$}}
  \caption{\emph{(Left-half)} Pseudocolours of mean streamwise velocity of the carrier phase with iso-contours of the flow stream function, $\Psi$. The black dot denotes the centres of the Dean cell. \emph{(Right-half)} Pseudocolours of mean in-plane velocity of the flow $\sqrt(\langle u_r \rangle ^2+\langle u_\theta\rangle ^2)$ with the vector plot of the in-plane motion, clearly showing the side-wall boundary layer and its near-separation from the wall; \emph{(a)} the mildly curved pipe, and \emph{(b)} the strongly curved pipe.}
\label{fig:fig5}
\end{figure}

\begin{figure}
  \centerline{\includegraphics*[height=4.3cm]{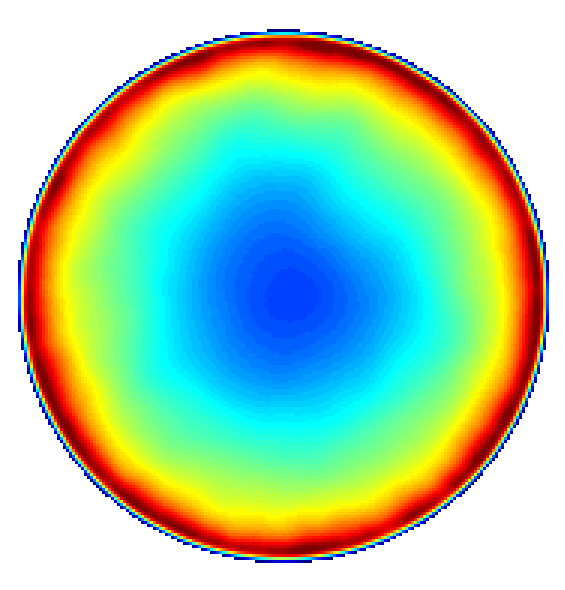}\put(-110,110){$(a)$}
              \includegraphics*[height=4.3cm]{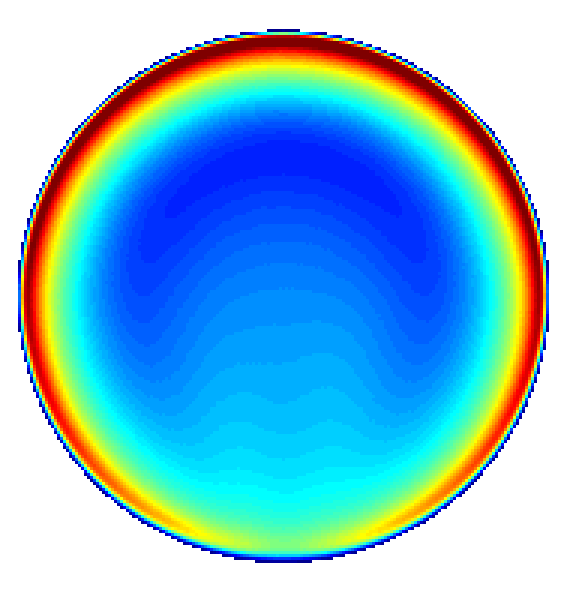}\put(-110,110){$(b)$}
              \includegraphics*[height=4.4cm]{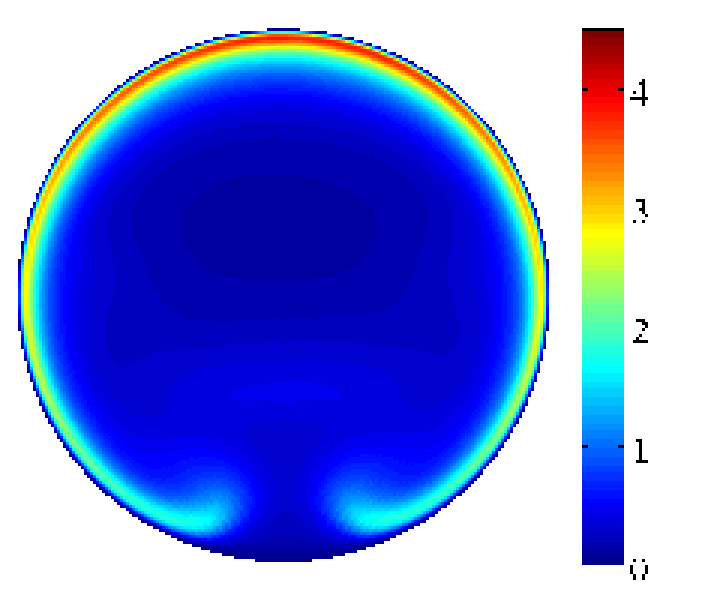}\put(-140,110){$(c)$}}
  \caption{Pseudocolours of turbulent kinetic energy (normalised with the respective $
{\overline{u_{\tau}}}^2$); \emph{(a)} straight pipe; \emph{(b)} curved pipe with $\kappa=0.01$; \emph{(c)} curved pipe with $\kappa=0.1$. Note that, in order to offer a direct comparison no azimuthal averaging has been performed in the straight pipe case.}
\label{fig:fig8}
\end{figure}

With increasing curvature the centrifugal forces increase. From figure \ref{fig:fig5} it is noticeable that the maximum of the mean streamwise velocity deflects more towards the outer bend with increasing $\kappa$. The magnitude of the averaged in-plane velocity $\sqrt(\langle u_r \rangle ^2+\langle u_\theta\rangle ^2)$ and a vector plot of the mean in-plane motion of the flow field in  different configurations are presented in the same figure. A wall-jet like boundary layer forms along the side walls of the curved pipes, which gradually strengthens with increasing the curvature. The secondary motion of the flow, which is merely $5\%$ of the bulk velocity in the mildly curved pipe, reaches up to $15\%$ in the strongly curved configuration at the same bulk Reynolds number. Due to the geometry-induced centrifugal force which acts always outwards, this side-wall boundary layer (SWBL) is subjected to a favourable pressure gradient as it moves towards the inner bend. Correspondingly, the fluid particles at the side walls accelerate to reach their maximum in-plane velocity magnitude close to the horizontal midplane ($\theta=0$). From this point, the in-plane flow decelerates approaching the stagnation point of the Dean vortices in the inner bend to lift up towards the outer bend. This behaviour can be clearly seen in the contours of the mean streamfunction (\emph{i.e.}\ streamlines displayed in figure \ref{fig:fig5}).
%%%%%%%%%%%%%%%%%%%%%%%%%%%%%%%%%%%%%%%%%%%%%%%%%%%%%%%%%%%%%%%%%%%%%%%%%%%%%%%%%%%%%%
%%%%%%%%%%%%%%%%%%%%%%%%%%%%%%%%%%%%%%%%%%%%%%%%%%%%%%%%%%%%%%%%%%%%%%%%%%%%%%%%%%%%%%

The map of the mean secondary motion in the centre region of the pipe section largely differs for the various curvature configurations. In the weakly curved pipe the in-plane flow is accelerated towards the pipe centreline after reaching the stagnation point at the inner bend, whereas the mean in-plane motion accelerates back towards the side-walls in the strongly curved configuration. This generates a bulge region of minimum in-plane velocity magnitude just below the pipe centre of the case with $\kappa=0.1$. The iso-contours of the stream function in figure \ref{fig:fig5} display the resulting antisymmetric mean Dean vortices. The difference among the two curvature configurations is striking and non-trivial to explain \citep{noorani_etal_2013}.

The previous characterisation of the flow in a bent pipe was concerned with mean-flow quantities and as such could be also seen in laminar flow. To provide a clear view of the distribution of turbulent activity in the pipe section the normalised turbulent kinetic energy (TKE) defined as $k=\langle u_i'u_i' \rangle/2$ , with the prime indicating the fluctuating velocity and implied summation over the dummy index $i$, is illustrated in figure \ref{fig:fig8}. Compared to the straight pipe, the value of $k$ in the inner side of the mildly curved pipe is smaller; however, the flow can still be considered  turbulent for all azimuthal positions. On the other hand, the peak in the turbulent kinetic energy diminishes near the inner wall of the strongly curved pipe. The turbulence, hence, is significantly inhibited in this region and the flow is almost quiescent. It is clear from figure \ref{fig:fig8} \emph{(b)} that near the pipe centreline the flow is more turbulent in the weakly curved configuration than in the straight pipe. Although in the majority of the pipe section with $\kappa=0.1$ turbulence is suppressed, it is increased near the region of the bulge visible in the secondary motion (\emph{c.f.} figures \ref{fig:fig5} \emph{b} and \ref{fig:fig8} \emph{c}).
%%%%%%%%%%%%%%%%%%%%%%%%%%%%%%%%%%%%%%%%%%%%%%%%%%%%%%%%%%%%%%%%%%%%%%%%%%%%%%%%%%%%%%
%%%%%%%%%%%%%%%%%%%%%%%%%%%%%%%%%%%%%%%%%%%%%%%%%%%%%%%%%%%%%%%%%%%%%%%%%%%%%%%%%%%%%%

\subsection{Particulate phase} \label{sec:part}
In the current study a total of seven particle populations with different $St_b$ are simulated. The density ratio is fixed to $\rho_p/\rho_f=1000$, and for each particle population, various $St_b$ are obtained by changing the radius of the particles. In order to imitate laboratory experiments the bulk Stokes number of each population is fixed for the various flow cases with different $\kappa$. This corresponds to fixing the particle density and radius (\emph{i.e.}\ fixing the physical particle) when varying the pipe curvature. The letter coding of particle populations and their corresponding parameters are presented in table \ref{tab:tab3}. In the current paper, \textit{heavier} or \textit{larger} particles are referred to as \textbf{Stp25}, \textbf{Stp50}, and the \textit{smaller}, \textit{lighter} or \textit{less inertial} populations are the \textbf{Stp1} and \textbf{Stp5}. The \textbf{Stp100} population is simulated mainly to investigate the asymptotic ballistic behaviour of the heavy inertial particles.
%%%%%%%%%%%%%%%%%%%%%%%%%%%%%%%%%%%%%%%%%%%%%%%%%%%%%%%%%%%%%%%%%%%%%%%%%%%%%%%%%%%%%%
\begin{table}
\begin{center}
\begin{tabular}{lllllll}
%\hline
\emph{Case}     & $d_p/R_a$              & $St_b$     & ${r_p}^+_{(\kappa=0.0)}$    & $St^+_{(\kappa=0.0)}$ & $St^+_{(\kappa=0.01)}$ & $St^+_{(\kappa=0.1)}$  \\ [3pt]
\hline
\textbf{Stp0}   & n/a                   & $0.0$      & n/a    & $0.0$      & $0.0$    & $0.0$\\
\textbf{Stp1}   & $ 3.727\times 10^{-4}$ &  $0.0451$  & $0.0671$& $1.0$      & $1.0503$ & $1.3020$ \\
\textbf{Stp5}   & $ 8.333\times 10^{-4}$ &  $0.2257$  & $0.1500$& $5.0$      & $5.2516$ & $6.5095$ \\
\textbf{Stp10}  & $ 1.179\times 10^{-3}$ &  $0.4514$  & $0.2121$& $10.0$     & $10.503$ & $13.020$ \\
\textbf{Stp25}  & $ 1.863\times 10^{-3}$ &  $1.1284$  & $0.3354$& $25.0$     & $26.258$ & $32.547$ \\
\textbf{Stp50}  & $ 2.635\times 10^{-3}$ &  $2.2569$  & $0.4744$& $50.0$     & $52.516$ & $65.095$ \\
\textbf{Stp100} & $ 3.727\times 10^{-3}$ &  $4.5139$  & $0.6708$& $100.0$    & $105.03$ & $130.20$ \\
%\hline
\end{tabular}
\caption{Parameters for the particle populations: Number of particles per population $N_p=1.28\times10^5$, and the density ration $\rho_p/\rho_f$ is fixed to $1000$ for all populations. The bulk Stokes number $St_b$ and particle diameter $d_p/R_a$ of each population is fixed while the curvature is varied in each flow configuration ($\kappa = 0.0$, $0.01$, and $0.1$).}
\label{tab:tab3}
\end{center}
\end{table}
%%%%%%%%%%%%%%%%%%%%%%%%%%%%%%%%%%%%%%%%%%%%%%%%%%%%%%%%%%%%%%%%%%%%%%%%%%%%%%%%%%%%%%

\subsubsection{Temporal convergence}
\begin{figure}
  \centerline{\includegraphics*[width=6.6cm]{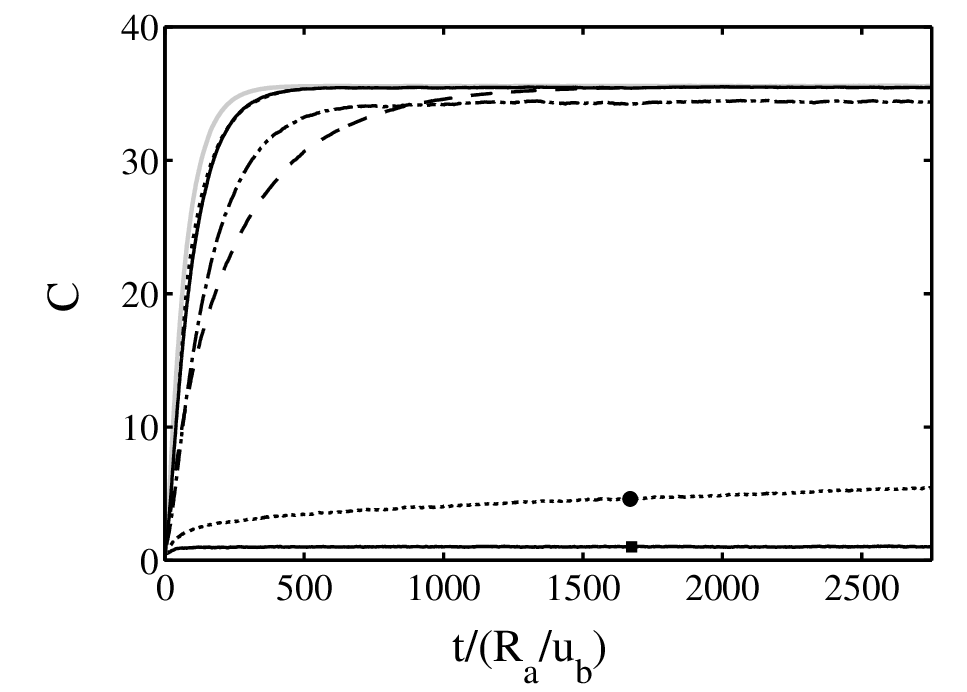}
              \put(-187,117){$(a)$}}
  \centerline{\includegraphics*[width=6.6cm]{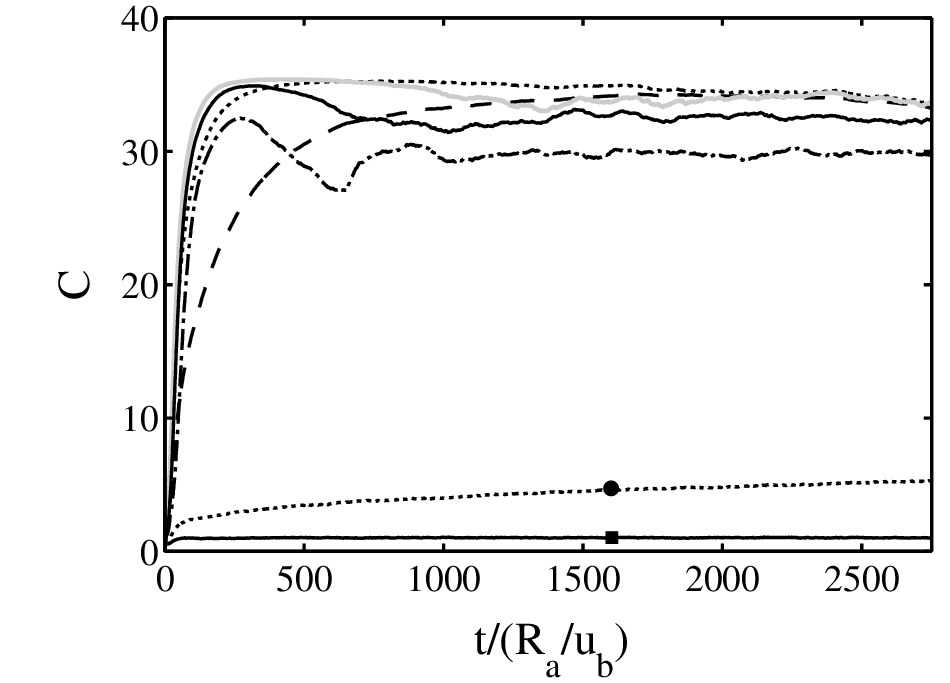}
              \includegraphics*[width=6.6cm]{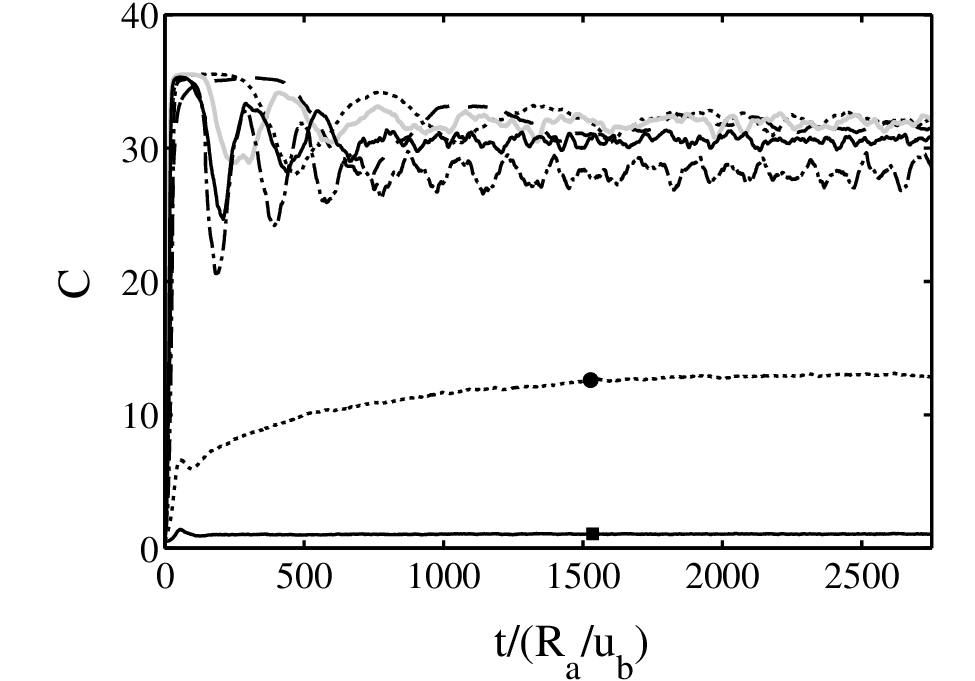}
              \put(-378,120){$(b)$}
              \put(-185,120){$(c)$}}
  \caption{Instantaneous particle concentration $C$ in the viscous sublayer ($y^+\lesssim 5$; $(1-r/R_a)<0.014$) normalised with the mean particle concentration in the domain as a function of time (convective time units): \emph{(a)} $\kappa=0.0$ \emph{(b)} $\kappa=0.01$ ,and \emph{(c)} $\kappa=0.1$. ${\blacksquare}$ \textbf{Stp0}, $\bullet$ \textbf{Stp1}, \dashed \textbf{Stp5}, \dotted \textbf{Stp10}, {\color{light-gray} \solid} \textbf{Stp25}, \solid \textbf{Stp50}, \dotdashed  \textbf{Stp100}.}
\label{fig:fig9}
\end{figure}

The particles are introduced into the velocity field of the fully developed turbulent flow with uniformly random distribution in radial $r$, azimuthal $\theta$ and axial $s$ directions. The initial velocity is set equal to the local flow velocity. To identify when the particle dispersion has reached a statistically stationary stage the instantaneous particle concentration $C$ (normalised with the mean particle concentration in the domain) in the viscous sublayer ($y^+\lesssim 5$; $(1-r/R_a)<0.014$) is monitored. Generally the near-wall concentration is expected to increase due to the turbophoretic motion of the particles until it saturates and the  process reaches a statistically stationary stage. Recent analytical studies of \citet{sikovsky_2014} suggests that a portion of these particles, which are drifting from the outer layer towards the near-wall region, may not possess enough momentum and engages into a slow and diffusional process before completely segregating near the wall. The time for this process is estimated to be inversely proportional to the particle Stokes number. Figure \ref{fig:fig9} shows the evolution of the observable versus the convective time of the simulation. As can be seen, in the straight pipe, after an initial increase, $C$ levels off at $t>500R_a/u_b$ for large particles. For less inertial particles, however, this time is considerably longer as \textbf{Stp5} reaches a steady state at about $1500R_a/u_b$ while \textbf{Stp1} does not reach a statistically steady state within the simulation time.

From figures \ref{fig:fig9} \emph{(b,c)} quasi-periodic oscillations of the value of $C$ can be observed in the curved pipes; these can be explained by the action of the secondary motion on the particles in the initial stage of the simulations. The initial random seeding leads to less particles near the walls and larger initial particle concentration close to the centreline of the pipes. Starting the simulation, the majority of these inertial particles is pushed by the Dean vortices towards the outer bend where they largely accumulate. The same Dean vortices transport the accumulated particles towards the inner bend via the side-walls boundary layers where they are eventually re-entrained back to the region of outward motion in the centre. The whole process is continuously modulated by the carrier phase turbulence until the particle concentration at the wall reaches an equilibrium state. The time period of the oscillations is longer for the lighter particles and shorter for the larger particles. This scaling suggests that the particles are mainly excited by the centrifugal force in this very initial stage. Nevertheless, it is clear that in these configurations the dispersed phase reaches the steady state at about $t=1500R_a/u_b$. Note that in the current plot $C$ is only averaged spatially and no temporal averages were taken. Therefore, even after reaching a statistically stationary value, $C$ still shows some fluctuations in time, due to the instationary near-wall turbulent velocity fluctuations. It is interesting to note that the level of these fluctuations is both dependent on the curvature $\kappa$ and the Stokes number. The largest fluctuations are observed for the heaviest particles in the strongest curvature, \emph{i.e.}\ for those particles that are most ballistic, operating with the highest in-plane velocity. 
%%%%%%%%%%%%%%%%%%%%%%%%%%%%%%%%%%%%%%%%%%%%%%%%%%%%%%%%%%%%%%%%%%%%%%%%%%%%%%%%%%%%%%
%%%%%%%%%%%%%%%%%%%%%%%%%%%%%%%%%%%%%%%%%%%%%%%%%%%%%%%%%%%%%%%%%%%%%%%%%%%%%%%%%%%%%%

\subsubsection{Instantaneous particle distribution}
\begin{figure}
  \centerline{\includegraphics*[width=3.6cm]{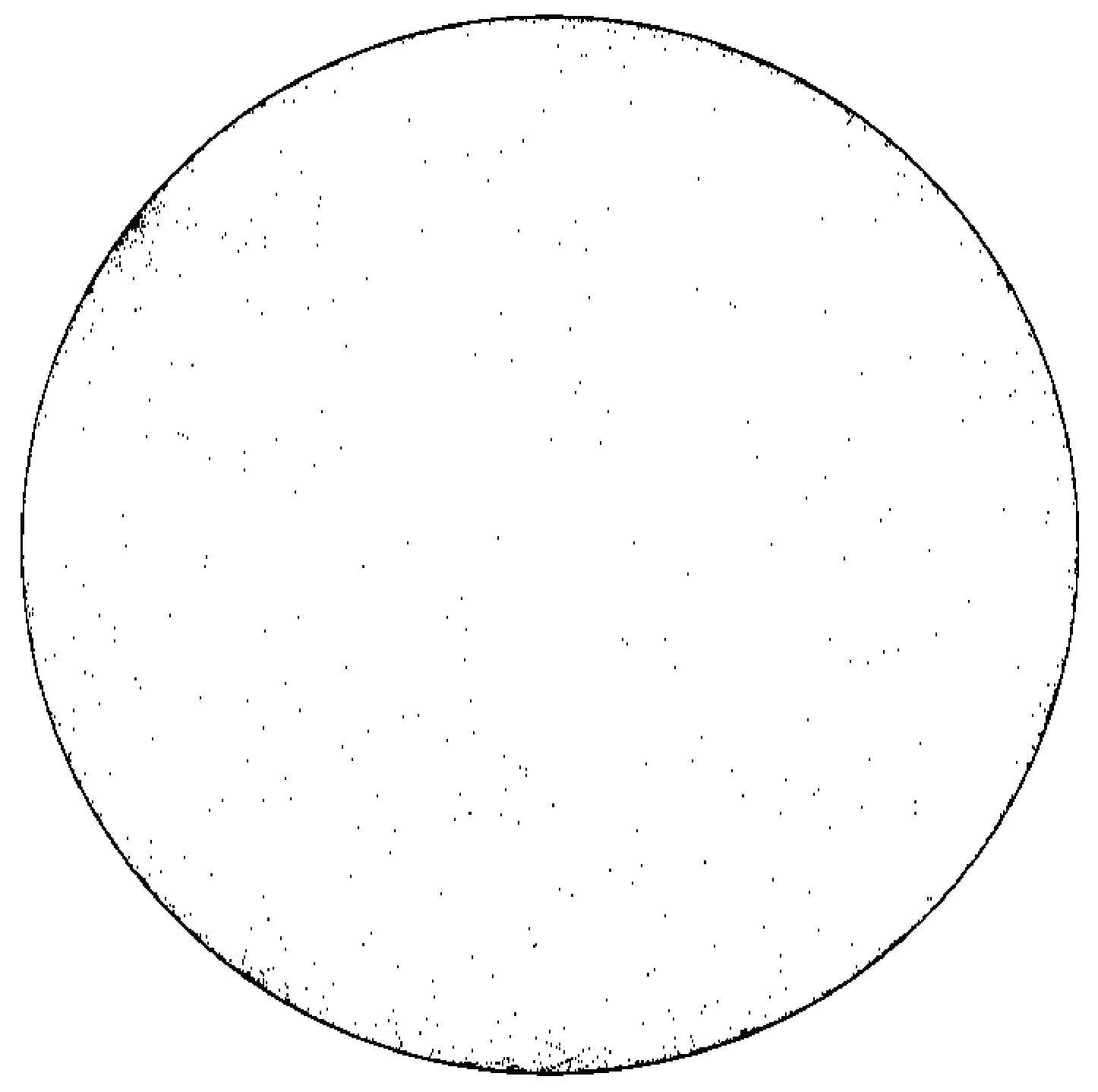}\put(-115,100){$(a)$}
              \includegraphics*[width=3.6cm]{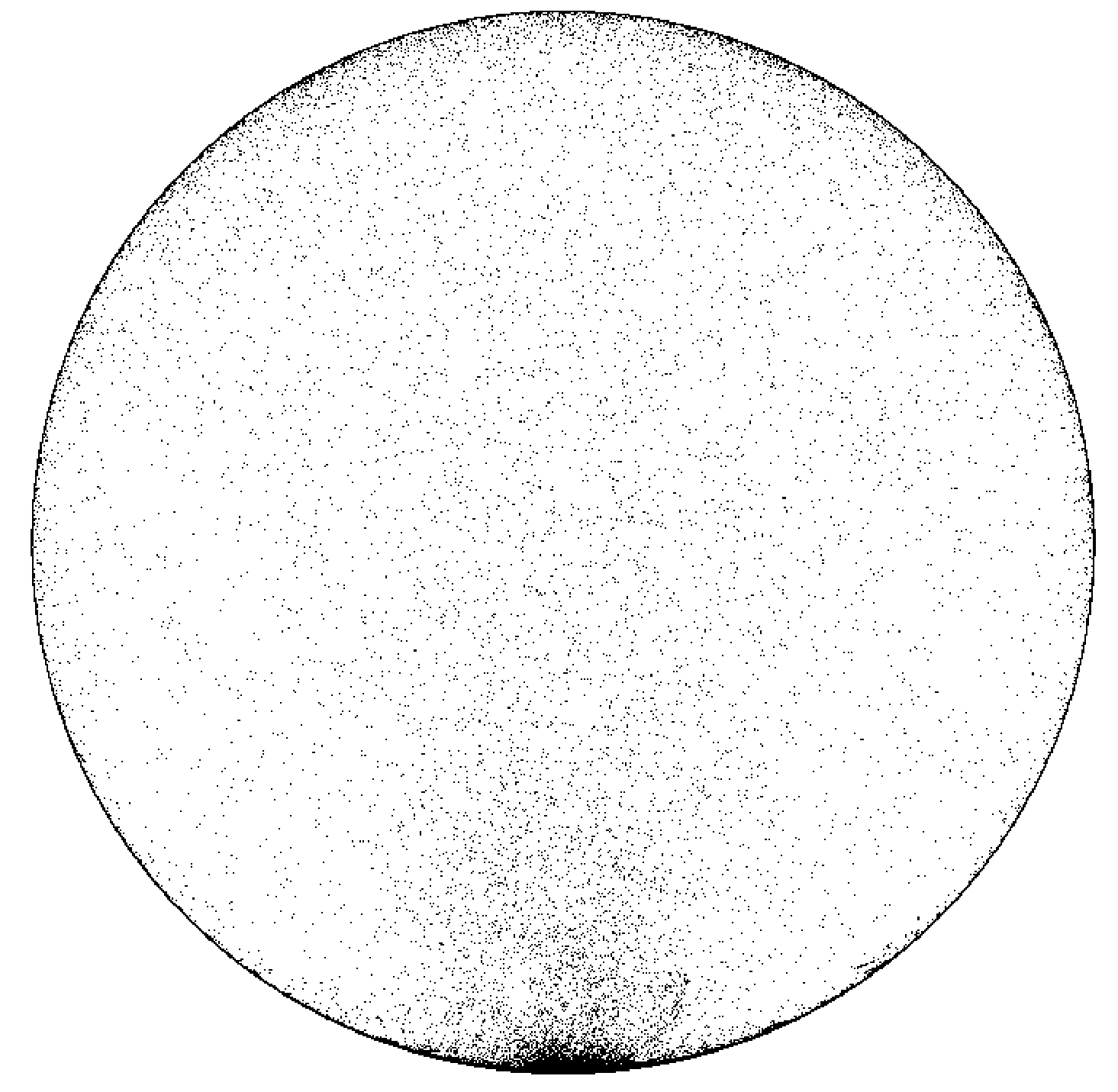}
              \includegraphics*[width=3.6cm]{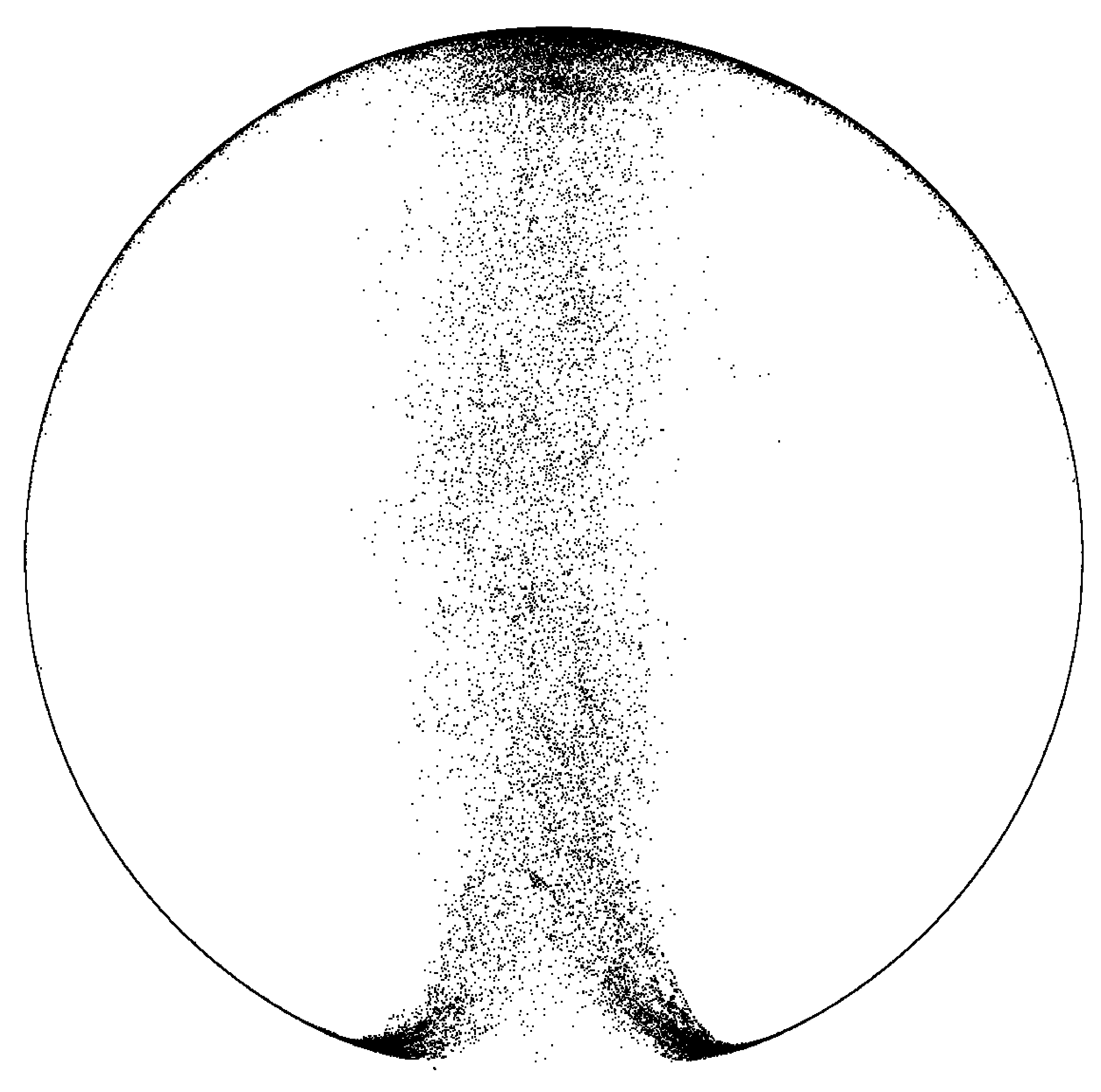}}
   \centerline{\includegraphics*[width=13.6cm]{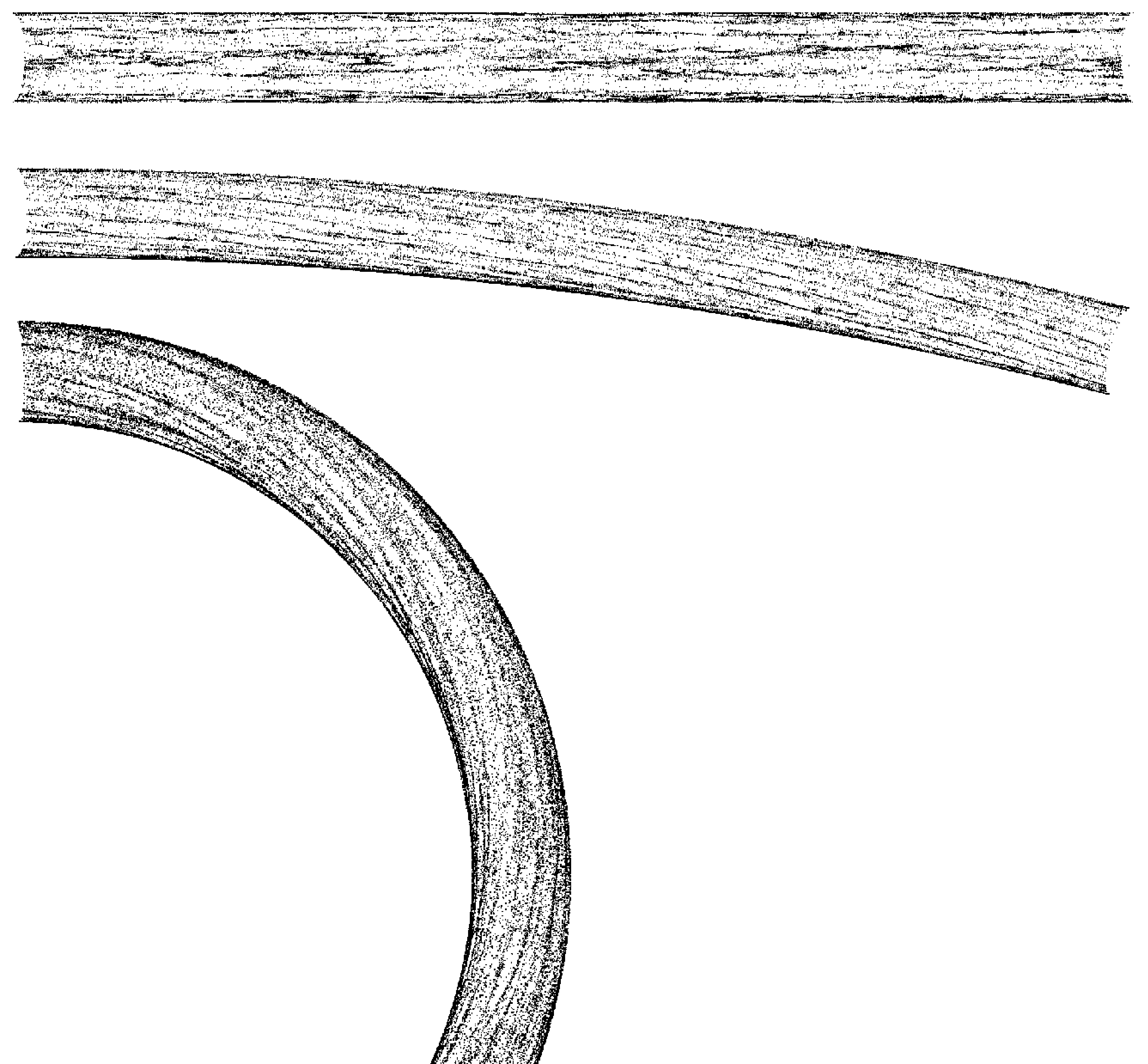}
               \put(-361,365){$(b)$}}
  \caption{\emph{(a)} Projected instantaneous front views of particles (\textbf{Stp50}) distributed in various pipe cofigurations \emph{(from left:} $\kappa=0$, $0.01$, and $0.1$\emph{)}. \emph{(b)} A cut-away view of the particles (\textbf{Stp50}) distributed in the viscous sublayer for pipes with different curvatures. The direction of the flow is from left to right.}
\label{fig:fig10}
\end{figure}

The instantaneous particle distribution for various curvature configurations at the final time of the simulations $t=3000$ is shown in figure \ref{fig:fig10}. The \textbf{Stp50} population has been chosen for illustration purposes here. The projected front view shows that the particles are largely segregated at the wall in the straight pipe, as few particles remain in the outer layer. Some near-wall particle clusters can be seen in this figure, corresponding to the preferential localisation of particles with regards to near-wall dynamics \citep{marchioli_soldati_2002,sardina_schlatter_brandt_picano_casciola_2012}. In comparison to the straight pipe, the amount of particles in the pipe core of the bent configurations is clearly larger. This is also true at the inner and the outer bend locations. At the same time, the core of the mean Dean vortices in the strongly curved configuration is essentially depleted of particles, and a notable plume of particles can be seen rising in the pipe centre towards the outer bend. Overall, the particle configuration marks the Dean vortices that are normally not so clearly distinguishable in instantaneous snapshots of the flow (\emph{c.f.} figure \ref{fig:fig4} \emph{a}). %In the strongly curved pipe no particles are found close to the core of the Dean vortices as opposed to the region nearby the maxima of the flow TKE in the inner bend where there are large particle concentration. This is the place where the secondary motion of the carrier phase lifts up and accelerates towards the outer bend (\emph{c.f.} figures \ref{fig:fig5}, \ref{fig:fig6} \& \ref{fig:fig8}.) 
Furthermore, the intermittent (partially re-laminarised) region in the inner bend is also characterised by harbouring only very few particles; it is clear that the particles follow the SWBL and the region of highest TKE, as shown in the velocity and kinetic energy maps (\emph{c.f.}\ figures \ref{fig:fig5} and \ref{fig:fig8}).
%%%%%%%%%%%%%%%%%%%%%%%%%%%%%%%%%%%%%%%%%%%%%%%%%%%%%%%%%%%%%%%%%%%%%%%%%%%%%%%%%%%%%%
%%%%%%%%%%%%%%%%%%%%%%%%%%%%%%%%%%%%%%%%%%%%%%%%%%%%%%%%%%%%%%%%%%%%%%%%%%%%%%%%%%%%%%

Figure \ref{fig:fig10} \emph{(b)} displays  side views of the particle distribution in the viscous sublayer, again defined as all particles below $y^+=5$, for each flow configuration. The \textbf{Stp50} particles form clear particle streaks due to the preferential concentration of the particles close to the pipe wall. These streaky structures are straight and randomly (in the azimuthal direction) distributed over the wall of the straight pipe. On the contrary, the elongated particle streaks in the bent configuration are curved and of helicoidal shape. This is clearly a consequence of the effect of the secondary motion. In fact, in straight channels and pipes, particles tend to preferentially align in elongated streaks correlated with slow departing fluid motions directed away from the wall and corresponding slow streamwise fluid velocity; namely the ejection events of wall turbulence \citep{rouson_eaton_2001}. \citet{picano_sardina_casciola_2009} showed that this tendency to accumulate in the ejection events is an essential condition for reaching a steady state particle concentration. It is, therefore, not surprising that the particle dynamics in the bent pipe shows the same phenomenology although being influenced and changed by the presence of the Dean vortices.

\begin{figure}
  \centerline{\includegraphics*[width=13.0cm]{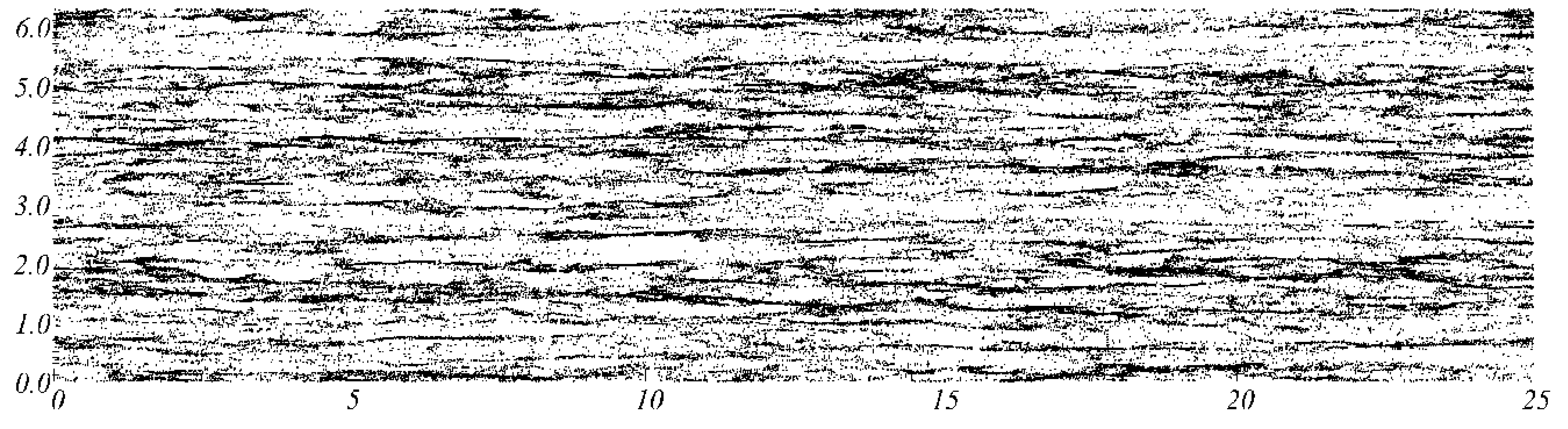}\put(-380,90){$(a)$}}
  \centerline{\includegraphics*[width=13.0cm]{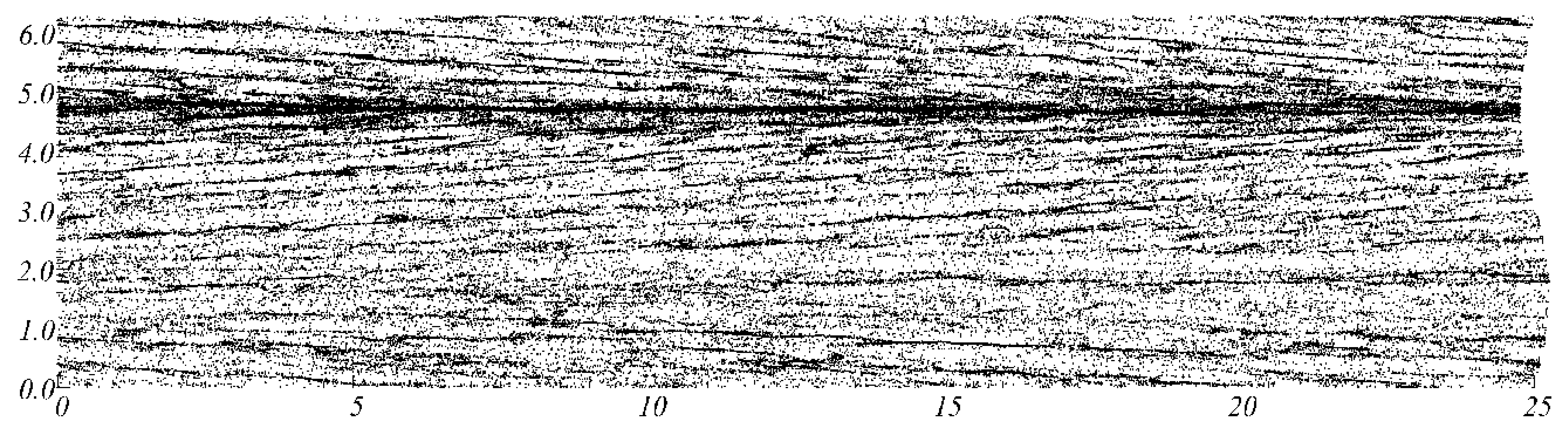}\put(-380,90){$(b)$}}
  \centerline{\includegraphics*[width=13.0cm]{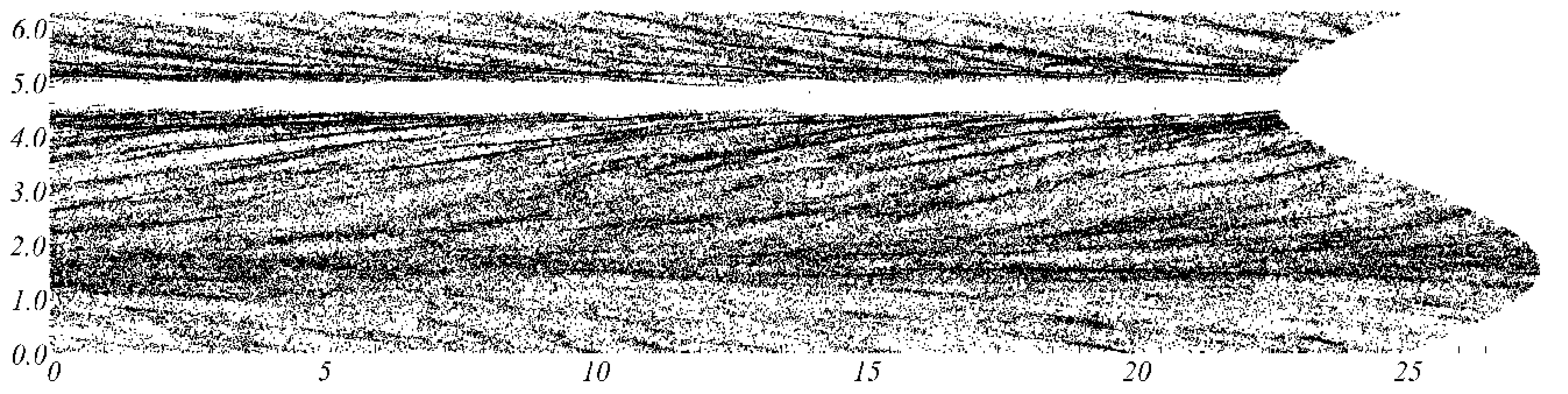}\put(-380,82){$(c)$}}
  \caption{An open-cut view of the particles (\textbf{Stp50}) distributed in the viscous sublayer for pipes with different curvatures; \emph{(a)} straight pipe; \emph{(b)} curved pipe with $\kappa=0.01$; \emph{(c)} curved pipe with $\kappa=0.1$. The ordinates represent the azimuthal direction of the pipe and the abscissae are the pipe lengths with the flow being from left to right.}
\label{fig:fig11}
\end{figure}

In order to provide a clearer view of these near-wall particle streaks, the geometry of each pipe configuration is unwound azimuthally and unfolded. The \textbf{Stp50} particle distribution in the viscous sublayer is  shown in  figure \ref{fig:fig11} in the $s-\theta$ plane. For the straight pipe, $\kappa=0$, the particle streaky structures are homogeneously distributed over the pipe wall circumference. Also, in agreement with previous studies in channel flow \citep{sardina_schlatter_brandt_picano_casciola_2012} a large-scale organisation is clearly visible which modulates the smaller-scale accumulation; these scales are of the order of the pipe radius, $R_a$. However, in the curved configurations the particle streaks are not only elongated in the axial  direction of the pipe, but also inclined in the azimuthal direction to point towards the inner bend. In fact, these streaks are aggregated with a certain angle to resemble \emph{fish-bone} like structures. The inclination of these helices appears to be three times larger in the strongly curved configuration compared to the mildly curved pipe. This can be explained by the fact that the in-plane motion of the carrier phase in pipes with $\kappa=0.1$ is almost three times stronger than that of the mildly curved pipe. From this plane view (figure \ref{fig:fig11} \emph{b}) it is clear now that the particles intensely accumulate in the inner side of the mildly curved pipe ($\theta\approx3\pi/2$) and form a large particle streak all the way though the pipe length. Interestingly the same position in the strongly curved configuration is totally void of these heavy particles as discussed above. A noteworthy increase in the particles segregation in the left half of the pipe with $\kappa=0.1$ is evident, which is changing to the right-half periodically in time at a comparably long time scale (not shown here). This changeover is reminiscent of the swirl-switching phenomena that is normally observed in spatially developing pipe bends and suggested to be a result of up-stream effects or separation in the bends \citep[see][]{kalpakli_orlu_2013}. The fact that the particle concentration carries features of swirl-switching might indicate that these low frequencies are in fact inherent to strongly curved pipes even without upstream effects. A final observation from figure \ref{fig:fig11} is that the large-scale organisation seen for the straight pipe is absent for the bent pipes. The inclined particle streaks are much more uniform. This indicates that the duration a particle spends in the near-wall region is too short to be organised over spatial scales $\mathcal{O}(R_a)$. Similarly, for the pipe with strong curvature even the signature of the near-wall streaks is weaker, which could be associated with the stronger secondary flow.
%%%%%%%%%%%%%%%%%%%%%%%%%%%%%%%%%%%%%%%%%%%%%%%%%%%%%%%%%%%%%%%%%%%%%%%%%%%%%%%%%%%%%%
%%%%%%%%%%%%%%%%%%%%%%%%%%%%%%%%%%%%%%%%%%%%%%%%%%%%%%%%%%%%%%%%%%%%%%%%%%%%%%%%%%%%%%

\subsubsection{Particle trajectories}
\begin{figure}
  \centerline{\includegraphics*[width=4.5cm]{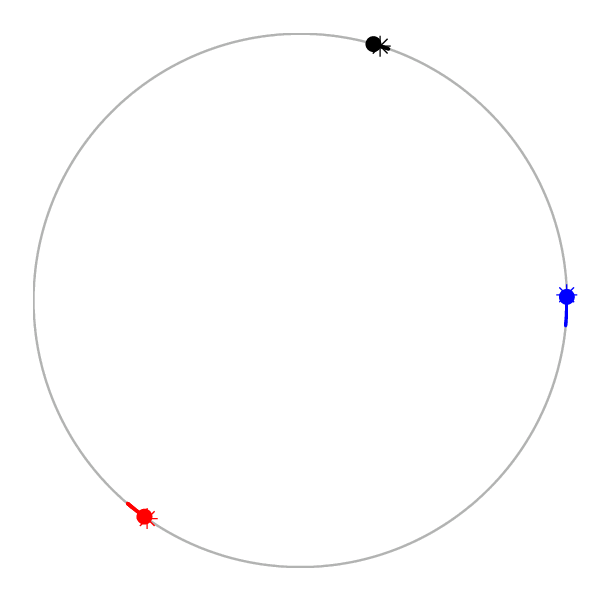}\put(-120,120){$(a)$}
              \includegraphics*[width=4.5cm]{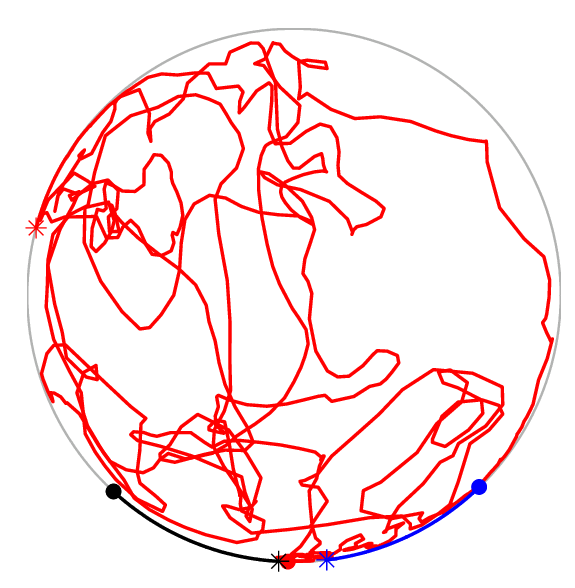}\put(-120,120){$(b)$}
              \includegraphics*[width=4.5cm]{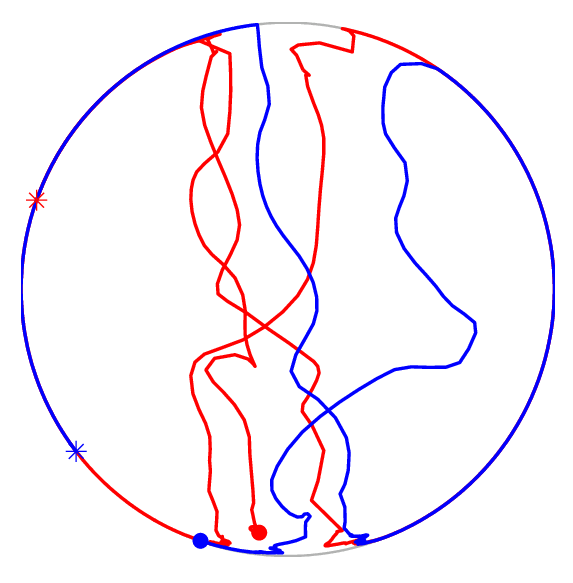}\put(-120,120){$(c)$}}
  \centerline{\includegraphics*[width=4.5cm]{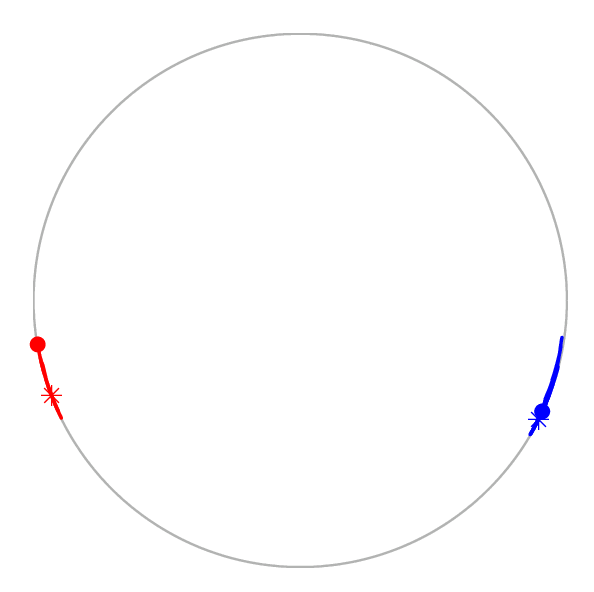}\put(-120,120){$(d)$}
              \includegraphics*[width=4.5cm]{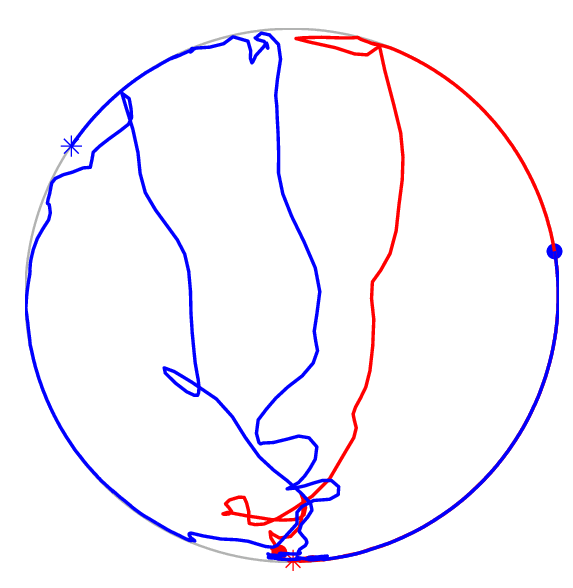}\put(-120,120){$(e)$}
              \includegraphics*[width=4.5cm]{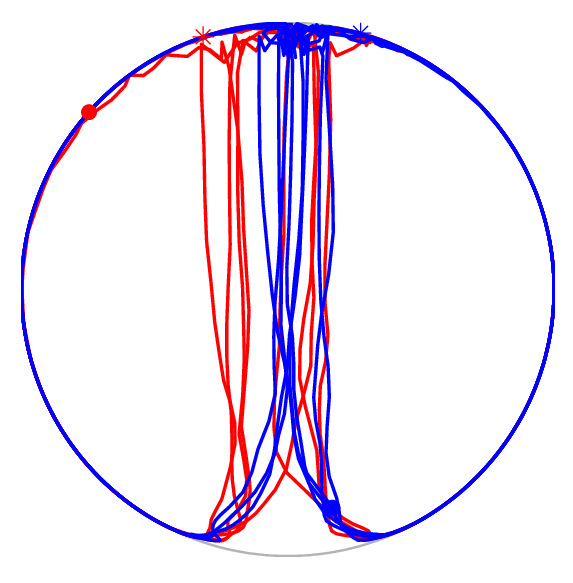}\put(-120,120){$(f)$}}
  \caption{Some typical particle trajectories for \emph{(left)} straight, \emph{(middle)} mildely bent, and \emph{(right)} strongly curved pipe. \emph{(top)} The \textbf{Stp5} and \emph{(bottom)} the \textbf{Stp50} population. The $\bullet$ indicates the beginning of the track and $\ast$ shows the final time of the trajectory. Different colours indicate different particle tracks while the gray line sketches the circumference of the pipe section. All trajectories are 1200 convective units in length.}
\label{fig:fig12}
\end{figure}

Some typical trajectories of \textbf{Stp5} and \textbf{Stp50} particles within the last $1200R_a/u_b$ convection time units of the simulation are shown for different curvatures in figure \ref{fig:fig12}. In the straight pipe these particles are already trapped in the viscous sublayer of the flow and remain there for this long period. Also, the particles hardly move in the azimuthal directions, and thus clearly sample one single low-speed region each. However, considering the same populations the trajectories are completely different in curved configurations. In the case with $\kappa=0.01$ particles of \textbf{Stp5} are erratically modulated by the flow turbulence unless they are trapped in the SWBL where they follow the secondary motion. On the other hand, the heavier particles of \textbf{Stp50} are essentially following the mean Dean vortices performing a spiralling motion through the bent pipe. What is striking is the fact that both particles are spending a long time near the stagnation point of the Dean vortices in the inner bend once they are transferred by the SWBL and before being re-suspended towards the middle. In the case with $\kappa=0.1$ where the secondary motion is strengthened, the trajectories indicate that even the lighter particles are excited mostly by the act of Dean vortices. The trajectories of \textbf{Stp50} suggest that these heavy particles are hardly modulated by the flow turbulence for the majority of the pipe section. Once being trapped in the SWBL and descending towards the inner side, these particles are lifted up towards the middle section of the pipe in almost straight lines ignoring the carrier phase in-plane streamlines; in particular the bulge feature discussed in figure \ref{fig:fig5}(b). This event is followed by a series of wall collisions in the outer bend. These particle-wall collisions occur at an angle of incidence almost straight which might indicate a very large erosive effect \citep[see][]{edwards_mclaury_shirazi_2001}. 
%%%%%%%%%%%%%%%%%%%%%%%%%%%%%%%%%%%%%%%%%%%%%%%%%%%%%%%%%%%%%%%%%%%%%%%%%%%%%%%%%%%%%%
%%%%%%%%%%%%%%%%%%%%%%%%%%%%%%%%%%%%%%%%%%%%%%%%%%%%%%%%%%%%%%%%%%%%%%%%%%%%%%%%%%%%%%

\subsubsection{Concentration statistics}
In order to obtain  Eulerian statistics from the Lagrangian particle data in the straight pipe the whole computational domain is divided into wall-parallel axisymmetric slabs. The slabs in the viscous sublayer are distributed equidistantly with size of one particle diameter away from each other. Above this region ($y^+>5$) the slabs are distributed non-uniformly according to the function: $r=R_a(1-(\sinh \gamma \eta)/(\sinh \gamma)) $ with $\gamma=3$ being the stretching factor. The vector $\eta$ consists of equally distributed grid points spanning the wall-normal direction. The resulting grid is more resolved near the wall than in the middle of the pipe. The curved pipe data is treated by an additional grid in the azimuthal direction. This mesh consists of cells (in the in-plane poloidal space) extended in the axial direction. A total of $2821$ cells is used for the two-dimensional representation of the Eulerian statistics pertaining to the particle data in curved geometries: $N_{\theta}=91$ and $N_r=39$, with $\gamma=2$; $N_{\theta}$ and $N_r$ are the number of cells in the azimuthal and wall-normal directions, respectively. Similarly, the statistical analysis at the equatorial mid-plane of the curved pipes is carried out with 100 wall-parallel slabs where $\gamma=2$ in the distribution function above and a constant cell width of 8  viscous units.

\begin{figure}
  \centerline{\includegraphics*[width=7.0cm]{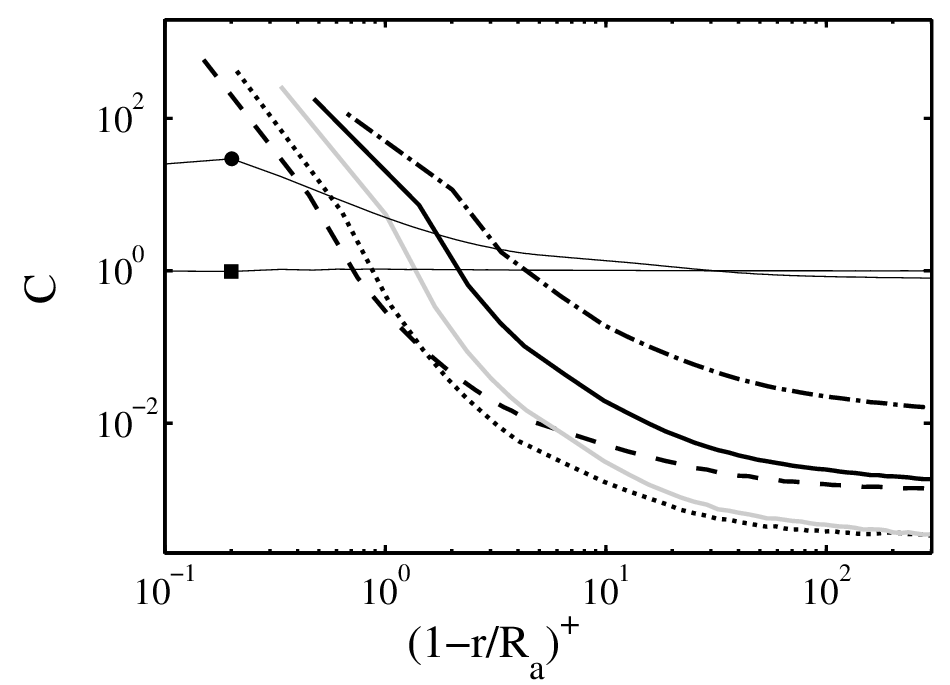}\put(-25,120){$(a)$}
              \includegraphics*[width=7.0cm]{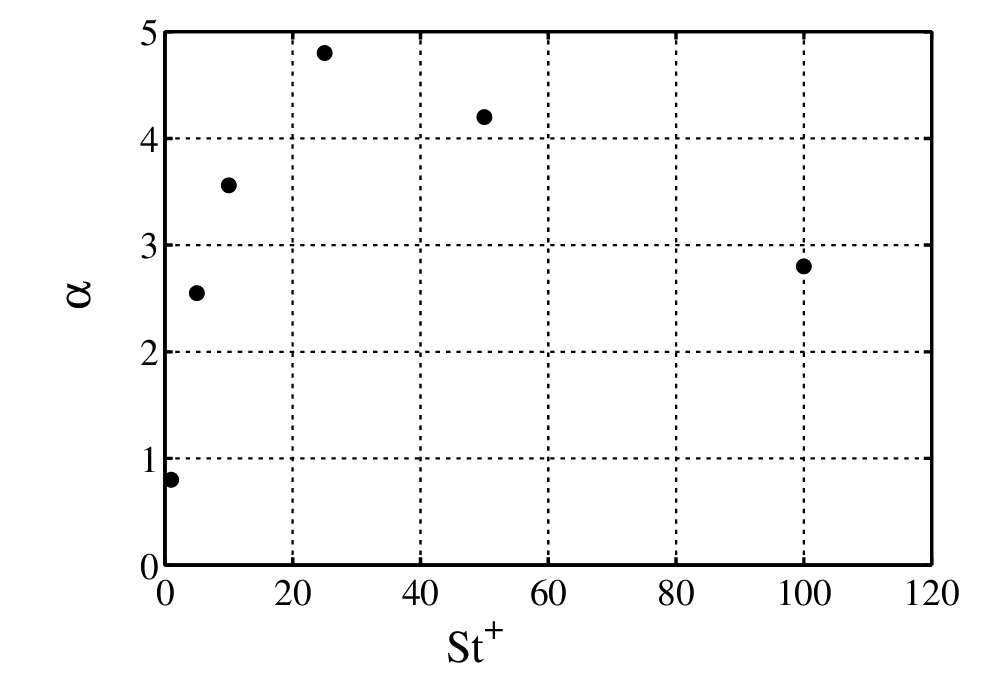}\put(-30,120){$(a)$}}
  \caption{\emph{(a)} Wall-normal mean particle concentration profiles for various populations (normalised by the mean particle concentration in the pipe); ${\blacksquare}$ \textbf{Stp0}, $\bullet$ \textbf{Stp1}, \dashed \textbf{Stp5}, \dotted \textbf{Stp10}, {\color{light-gray} \solid} \textbf{Stp25}, \solid \textbf{Stp50}, \dotdashed  \textbf{Stp100}. \emph{(b)} The exponent of the power-law dependence $C \sim (y^+)^\alpha$ computed in the range of $1<y^+<5$ (straight pipe).}
\label{fig:fig13}
\end{figure}

The particle accumulation in various regions of the pipe section is quantified by means of the particle concentration $C$ that is defined as the number of the particles per unit volume, normalised with the mean (bulk) concentration. Figure \ref{fig:fig13} \emph{(a)} displays this normalised mean particle concentration in logarithmic scale for each population in the straight pipe. While the Lagrangian tracers (\textbf{Stp0}) are uniformly distributed ($C=1$), the inertial heavy particles accumulate close to the wall and the maximum concentration appears at the wall due to the turbophoresis \citep[see][]{caporaloni_tampieri_trombetti_vittori_1975, reeks_1983, young_leeming_1997}. Naturally, the particle radius serves as a cut-off of the profile in the vicinity of the wall. Despite averaging over a long period it is hard to say that these log-log concentration profiles show a linear behaviour. Though following \citet{sikovsky_2014}, however, the exponent $\alpha$ of the power-law dependence $C \sim (y^+)^\alpha$ is computed in the range $1<y^+<5$ by a least-square fit. Figure \ref{fig:fig13} \emph{(b)} clearly shows that with increasing  particle relaxation time this power-law exponent increases to reach a maximum  at $St^+=25$. This is of course consistent with the the previous results of turbulent particle-laden channel and pipe flows \citep[see][]{young_leeming_1997, sardina_schlatter_brandt_picano_casciola_2012, picano_sardina_casciola_2009}, which show that \textbf{Stp25} particles are the most efficient in accumulating near the wall. Note that in the concentration plot, the lowest possible distance between particle centre and the wall is the particle radius, which increases with the Stokes number. No particle can be located below this minimum distance, which necessarily leads to  zero concentration.
%%%%%%%%%%%%%%%%%%%%%%%%%%%%%%%%%%%%%%%%%%%%%%%%%%%%%%%%%%%%%%%%%%%%%%%%%%%%%%%%%%%%%%
%%%%%%%%%%%%%%%%%%%%%%%%%%%%%%%%%%%%%%%%%%%%%%%%%%%%%%%%%%%%%%%%%%%%%%%%%%%%%%%%%%%%%%

\begin{figure}
  \centerline{\includegraphics*[width=5.6cm]{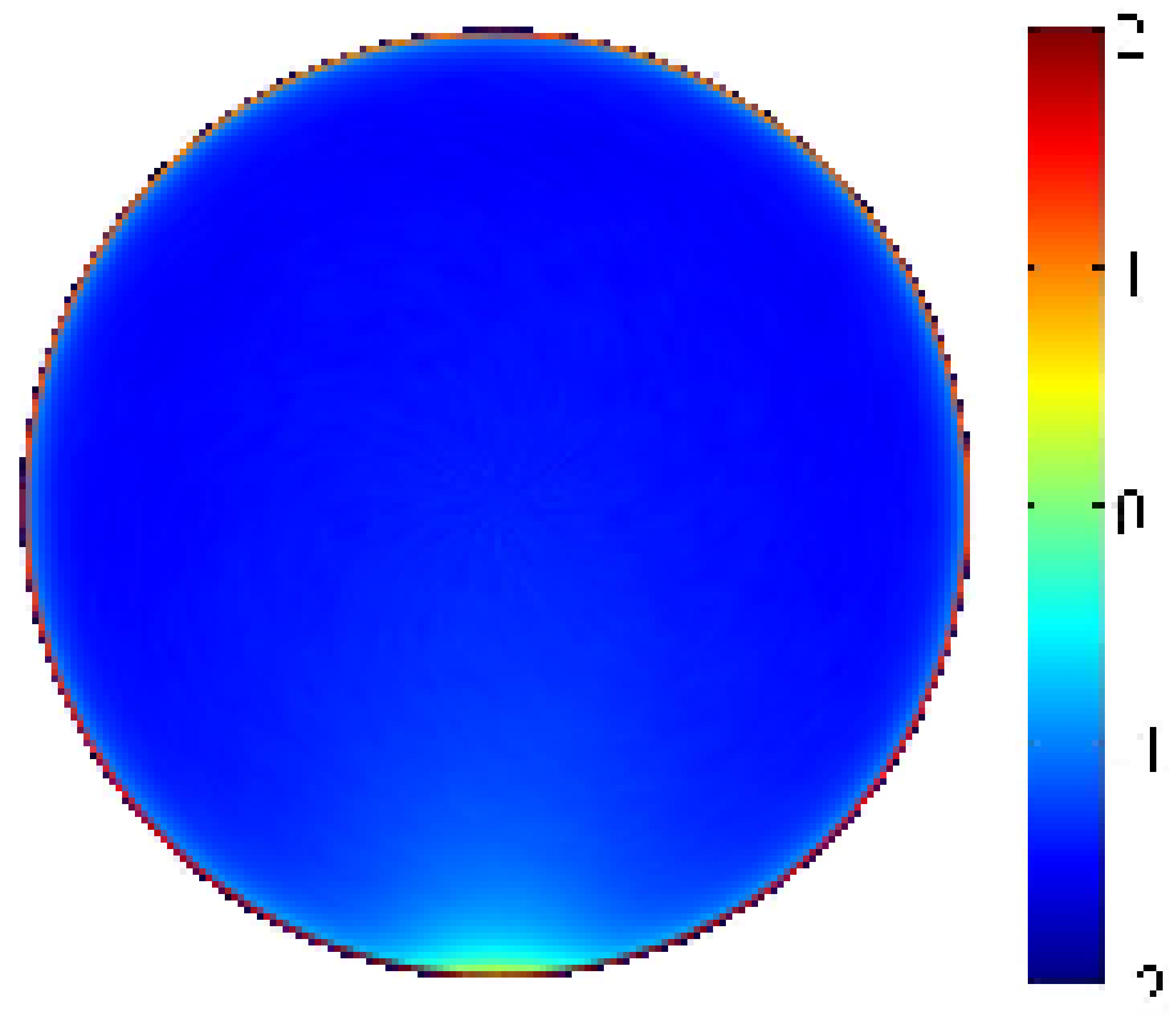}\put(-160,120){$(a)$}
              \includegraphics*[width=5.6cm]{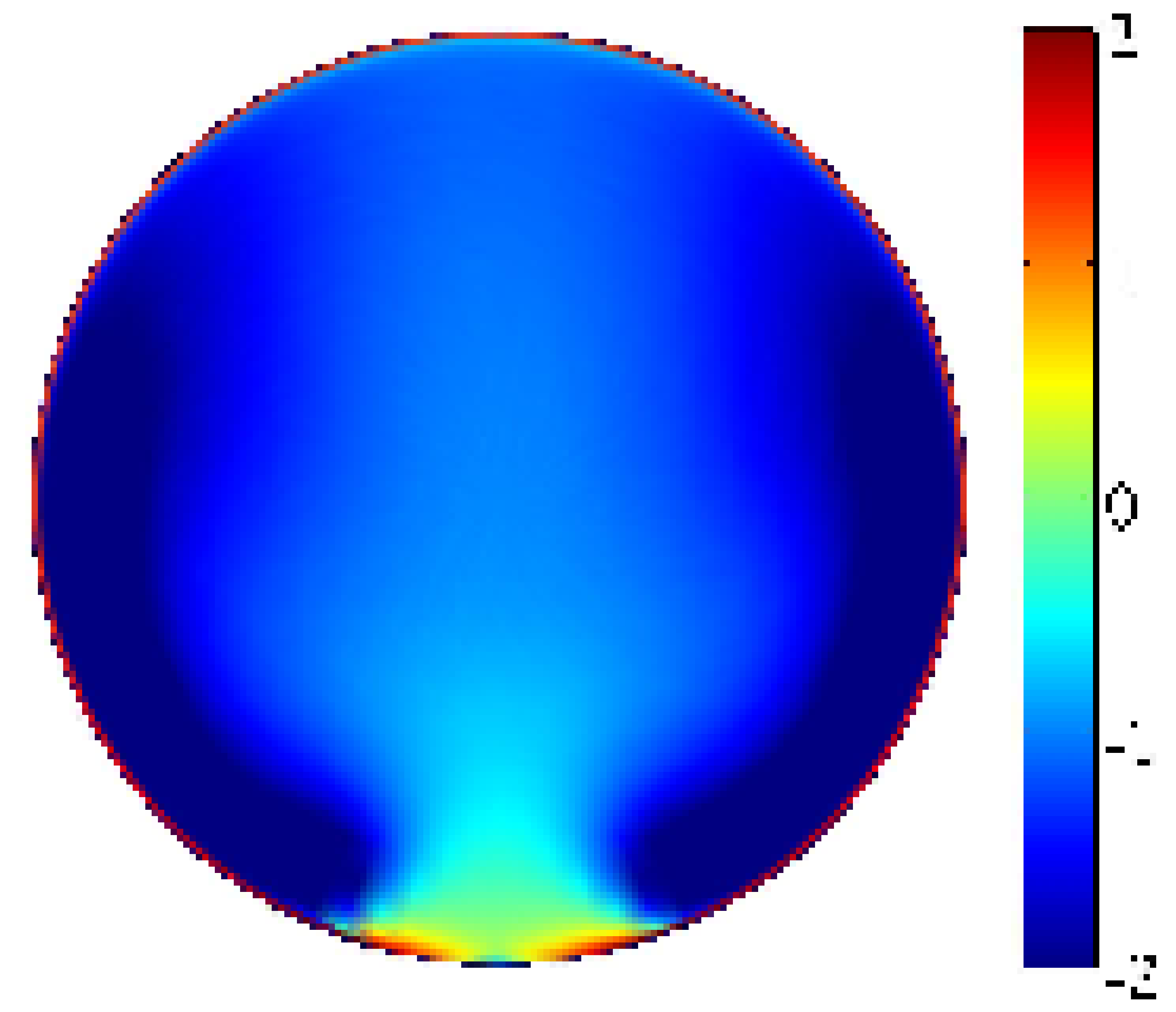}\put(-158,120){$(d)$}}
  \centerline{\includegraphics*[width=5.6cm]{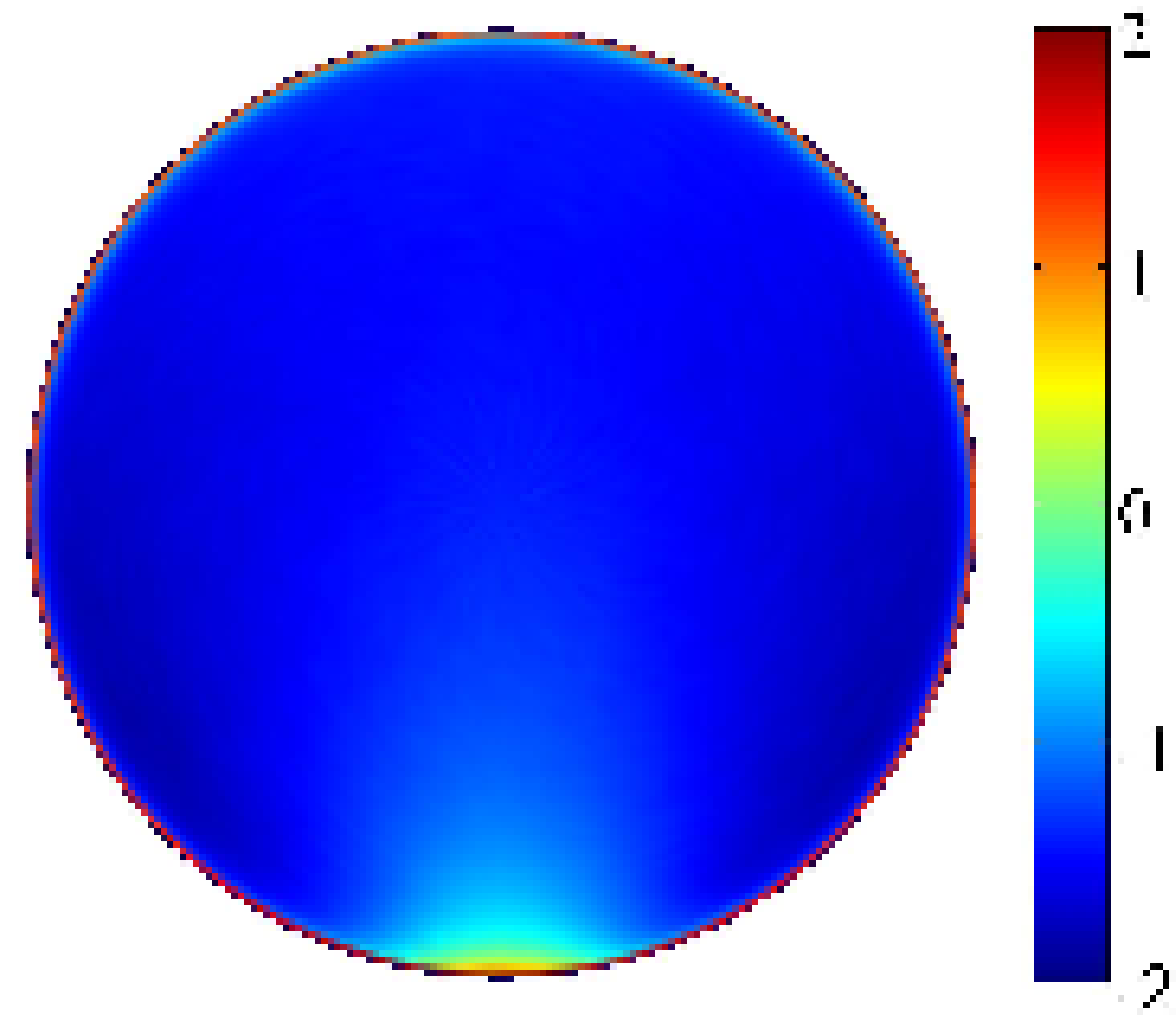}\put(-160,120){$(b)$}
              \includegraphics*[width=5.6cm]{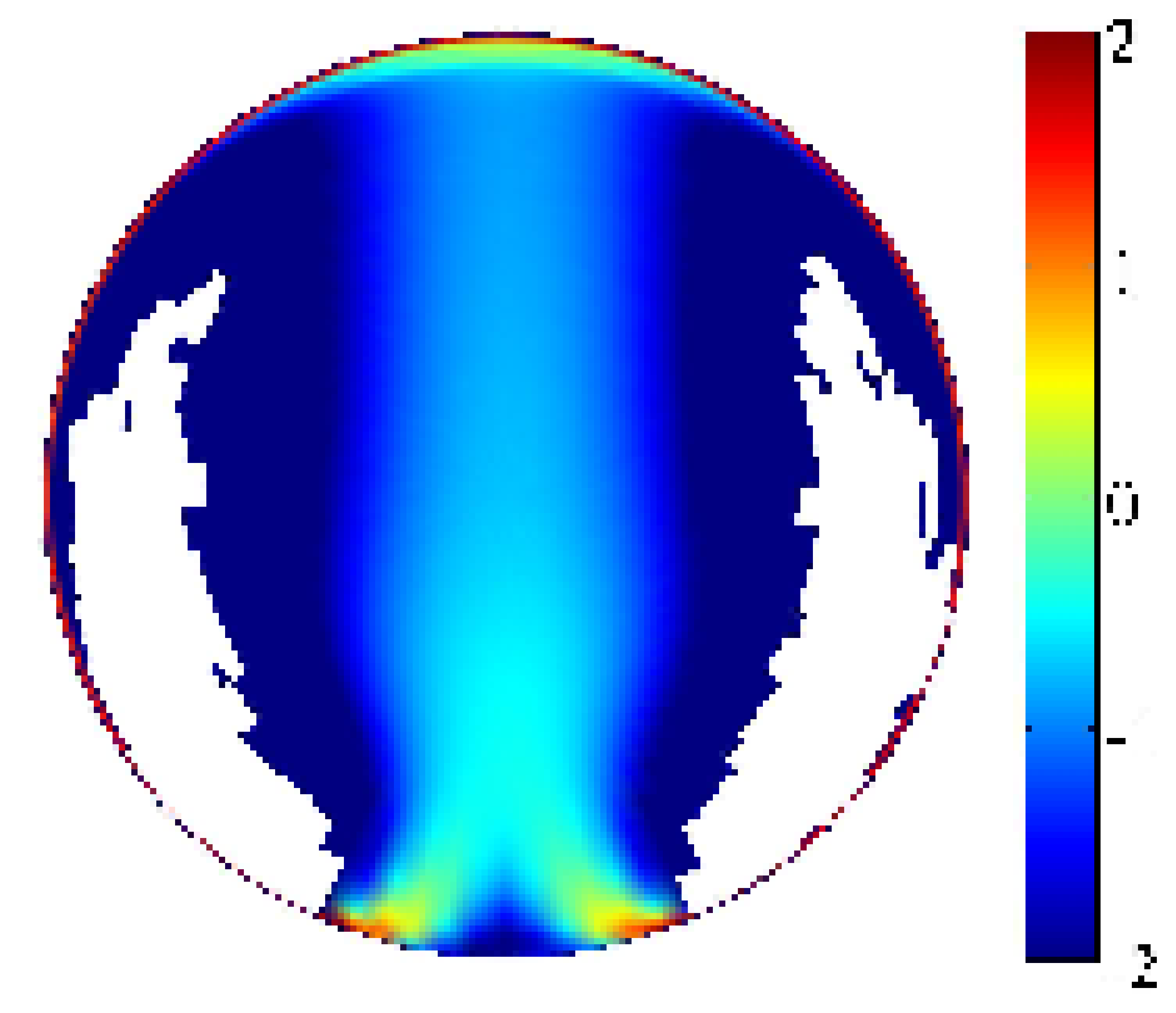}\put(-158,120){$(e)$}}
  \centerline{\includegraphics*[width=5.6cm]{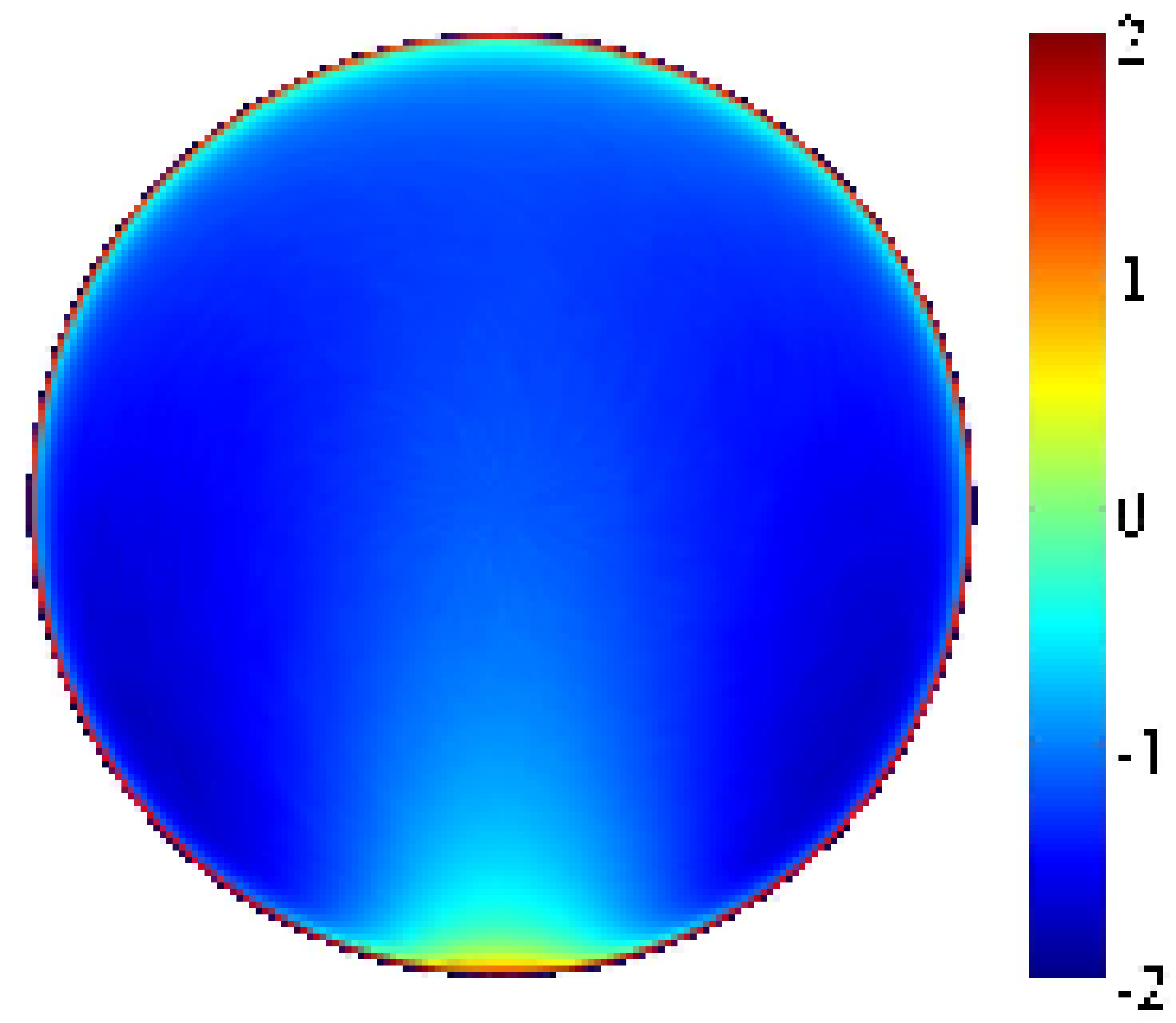}\put(-160,120){$(c)$}
              \includegraphics*[width=5.6cm]{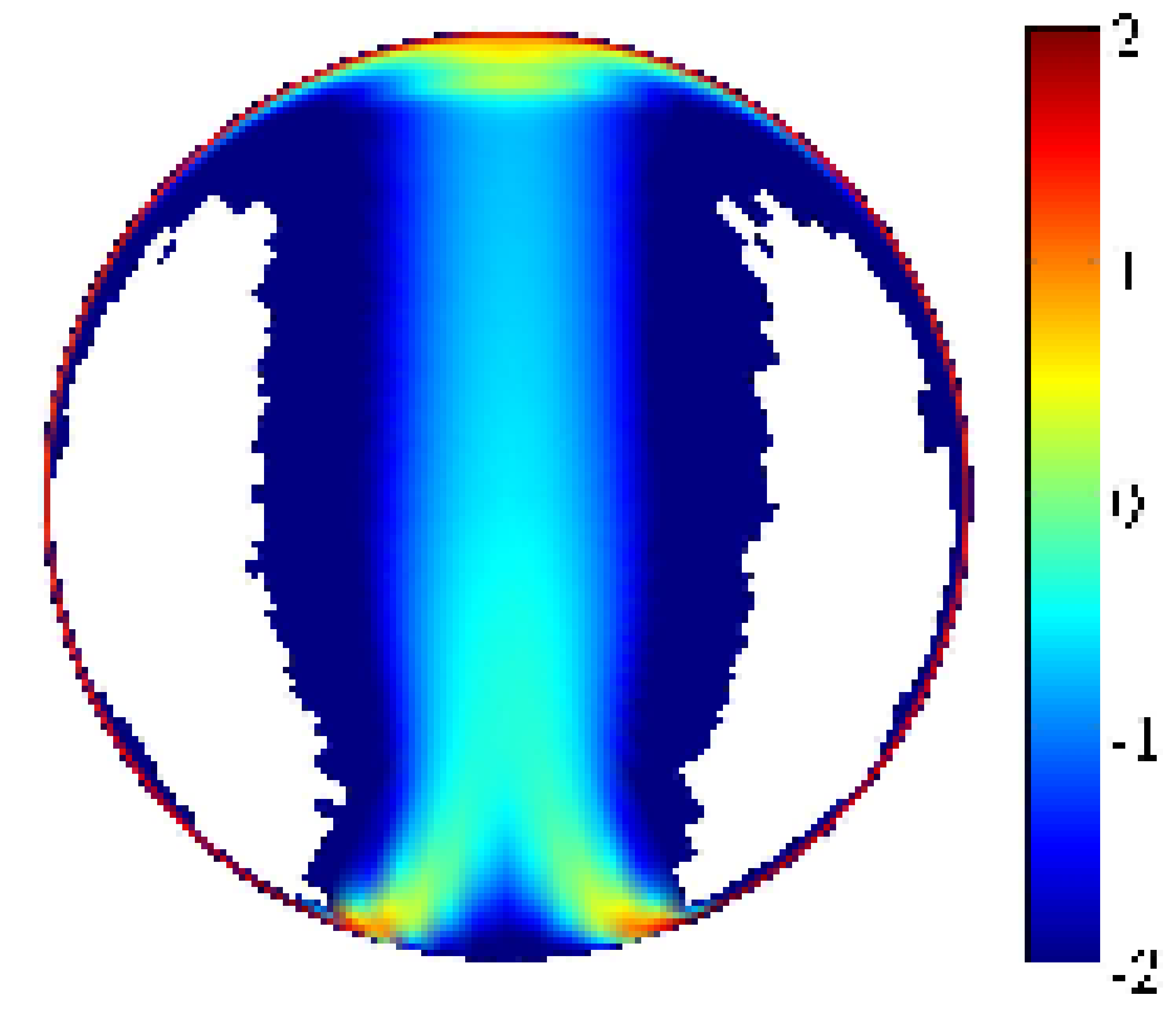}\put(-158,120){$(f)$}}
  \caption{Logarithmic representation ($log_{10}(C)$) of mean normalised concentration map of the particles dispersion in bent pipes with $\kappa=0.01$ \emph{(left)} and $\kappa=0.1$ \emph{(right)}. From \emph{top:} \textbf{Stp5}, \textbf{Stp25}, and \textbf{Stp50}. The white cells indicate regions completely void of particles in the statistical time frame of the simulation. }
\label{fig:fig14}
\end{figure}

Figure \ref{fig:fig14} displays the normalised mean particle concentration in logarithmic scale for mildly and strongly curved pipes for three different populations. Although particles in curved configurations  also largely accumulate at the wall, the mean particle concentration is quite different from that of the straight configuration. As previously inferred from instantaneous snapshots (\emph{c.f.}\ figure \ref{fig:fig10} \emph{a}), the statistics clarify that heavy particles cluster densely outside the core of the mean Dean vortices,  avoiding these vortical regions. As the strength of the Dean cells increases (in the case with $\kappa=0.1$) the concentration decreases as much as there are areas close to the vortex core where it is unlikely to find a particle at all. The cells in which no particles enter during a simulation interval of $1500R_a/u_b$ (the statistical time frame) are illustrated in white in figure \ref{fig:fig14} \emph{(e,f)}. The particle concentration in the inner bend stagnation point differs greatly for different curvatures. While the mildly curved configuration exhibits a high concentration in this region, in fact including the concentration maximum, whereas in the case with $\kappa=0.1$, $C$ is very low (almost zero) in the same area. For both cases, this region grows in size with increasing the Stokes number of the particles. The strongly curved pipe also displays two high-concentration regions in the inner bend near the SWBL's separation (\emph{c.f.}\ figures \ref{fig:fig5} \emph{b} and \ref{fig:fig8} \emph{c}). These high concentration regions extend towards the pipe centreline and merge in the equatorial mid-plane to generate a particle confluence region. The bulge region visible in the velocity streamlines is not clearly represented in the concentration map of the heavier particles; however, smaller particles, \emph{e.g.}\ \textbf{Stp5}, display an enlargement of concentration in this region. Both curved configurations present a relatively larger concentration near the outer side where the near-wall turbulent intensity of the carrier phase is relatively stronger. It is worth noting that the almost perfect left-right symmetry of these plots is clearly a sign of the statistical convergence of the present simulations. 
%%%%%%%%%%%%%%%%%%%%%%%%%%%%%%%%%%%%%%%%%%%%%%%%%%%%%%%%%%%%%%%%%%%%%%%%%%%%%%%%%%%%%%
%%%%%%%%%%%%%%%%%%%%%%%%%%%%%%%%%%%%%%%%%%%%%%%%%%%%%%%%%%%%%%%%%%%%%%%%%%%%%%%%%%%%%%

\begin{figure}
  \centerline{\includegraphics*[width=10.0cm]{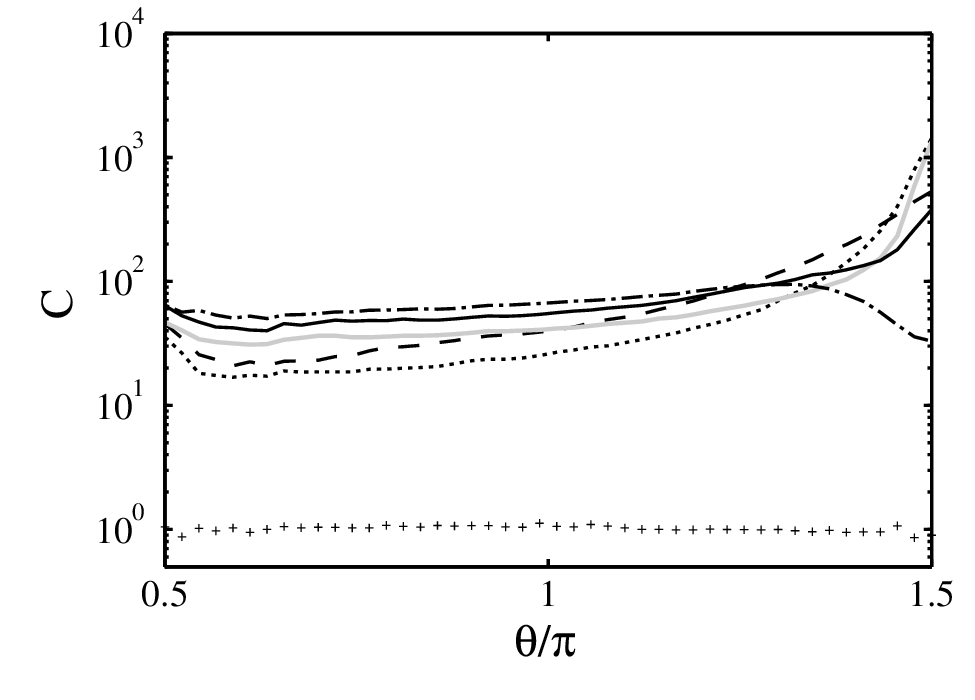}\put(-280,190){$(a)$}}
  \centerline{\includegraphics*[width=10.0cm]{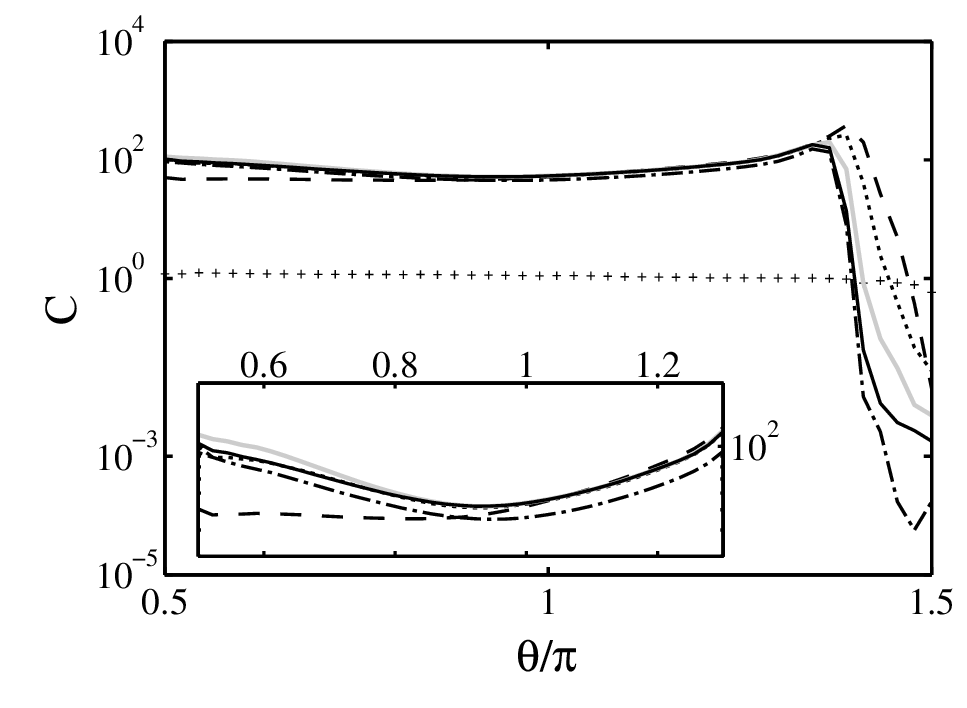}\put(-280,190){$(b)$}}
  \caption{Azimuthal distribution of mean normalised particle concentration close to the wall ($y^+ \lesssim 5$) of bent pipes with $\kappa=0.01$ \emph{(a)} and $\kappa=0.1$ \emph{(b)}. $+$ \textbf{Stp0}, \dashed \textbf{Stp5}, \dotted \textbf{Stp10}, {\color{light-gray} \solid} \textbf{Stp25}, \solid \textbf{Stp50}, and \dotdashed \textbf{Stp100}. The inner side and the outer bend of the curved pipes are located at $\theta/\pi=3/2$ and $\theta/\pi=1/2$, respectively.}
\label{fig:fig15}
\end{figure}

Figure \ref{fig:fig15} shows the azimuthal distribution of particle concentration in the viscous sublayer of the curved pipes. In the mildly curved pipe, although the turbulence intensity of the carrier phase is higher in the outer half ($\pi/2<\theta<\pi$), the near-wall concentration is higher in the inner half for almost all the populations except the smallest and largest particles (\textbf{Stp1} and \textbf{Stp100}). Closer to the inner bend, with increasing Stokes number the accumulation increases dramatically for \textbf{Stp5} to \textbf{Stp50} particles. The largest growth of the near-wall particle concentration from the outer bend towards the inner bend belongs to \textbf{Stp10} population. 

The same analysis in the strongly curved configuration (figure \ref{fig:fig15} \emph{b}) shows that for heavier populations than \textbf{Stp5}, traversing from the outer bend towards the inner side, the concentration at the wall first slightly decreases and then increases again with the minimum around $\theta\approx\pi$ (see the inset of figure \ref{fig:fig15} \emph{b}).  Approaching the inner bend farther the accumulation increases for smaller populations more dramatically. This behaviour continues up to a maximum in the particle concentration around $\theta=4\pi/3$. Further down towards the inner bend the concentrations decrease sharply, clearly linked to the strong outward motion of the carrier phase and the attenuation of turbulence in this part. The rate of this drop in the concentration value has a direct relation with the inertia of the particles.
%%%%%%%%%%%%%%%%%%%%%%%%%%%%%%%%%%%%%%%%%%%%%%%%%%%%%%%%%%%%%%%%%%%%%%%%%%%%%%%%%%%%%%
%%%%%%%%%%%%%%%%%%%%%%%%%%%%%%%%%%%%%%%%%%%%%%%%%%%%%%%%%%%%%%%%%%%%%%%%%%%%%%%%%%%%%%

\begin{figure}
  \centerline{\includegraphics*[width=10.cm]{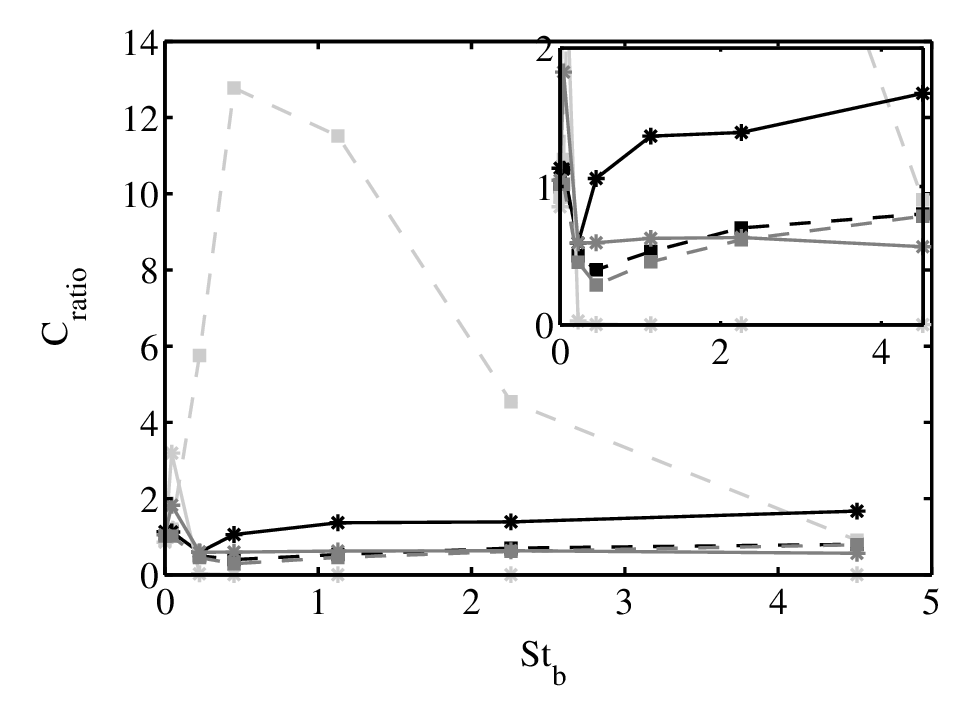}}
  \caption{The ratio of concentration in the viscous sublayer of the bent pipes ($\kappa=0.01$ \dashed$\Box$; $\kappa=0.1$ \solid$\ast$) to that in the straight pipe, $C_{ratio}$, for different populations. \emph{Black} represents the ratio for the outer bend, \emph{light-gray} for the inner bend and \emph{dark-gray} shows $C_{ratio}$ at the horizontal plane. The close-up view is shown in the inset.}
\label{fig:NEWfig1}
\end{figure}

The data points in figure \ref{fig:NEWfig1} shows the ratio of concentration in the viscous sublayer of the bent pipes (at horizontal and vertical cuts) to that in the straight pipe ($C_{ratio}$) for various particle populations. The data for the bent pipes are obtained by averaging the particle data in near-wall sectors at horizontal and vertical cuts with $(1-r/R_a)^+\leq5$ and $(R_a\theta)^+\approx30.6$. For both bent pipes, in the stagnation point of the Dean cells at the outer bend (black lines),  $C_{ratio}$ of \textbf{Stp5} particles is decreased by almost $50\%$ compared to the \textbf{Stp1} population. Increasing particle inertia further, the $C_{ratio}$ continues to grow for both cases in the outer bend. However, the growth is much faster for the strongly curved pipe as $C_{ratio}\approx1$ for \textbf{Stp10} and $C_{ratio}\approx1.7$ for \textbf{Stp100} particles while in the mildly bent pipe the $C_{ratio}$ of the \textbf{Stp100} population only reaches up to $0.8$. The ratio of the concentration in the inner bend position, denoted by light gray, of mildly curved pipe to that in the straight pipe rises very rapidly with increasing particle inertia to reach a maximum for \textbf{Stp10} particles where $C_{ratio}\approx13$ after which the value decreases again. On the contrary, the $C_{ratio}$ in this position for the strongly curved pipe is nearly zero for heavy populations. Notwithstanding, for \textbf{Stp1} particles the concentration in the inner bend for the strongly curved pipe is three times larger than that in the straight pipe. It is worthwhile to note that for the mildly bent pipe the inner bend concentration for \textbf{Stp5}, \textbf{Stp10} and \textbf{Stp25} particles is approximately $12$ times, $32$ times and $22$ times larger that in the outer bend, respectively. Interestingly, in the near-wall horizontal cut (dark-gray lines) of the strongly curved pipe (where the magnitude of the secondary motion is maximum) the $C_{ratio}$ is almost constant at $0.6$ for all the populations except \textbf{Stp1}. This is not the case for the mildly curved pipe as the trend of $C_{ratio}$ in the near-wall horizontal cut is identical to the outer bend profile, but with an almost $20\%$ deficit.
%%%%%%%%%%%%%%%%%%%%%%%%%%%%%%%%%%%%%%%%%%%%%%%%%%%%%%%%%%%%%%%%%%%%%%%%%%%%%%%%%%%%
%%%%%%%%%%%%%%%%%%%%%%%%%%%%%%%%%%%%%%%%%%%%%%%%%%%%%%%%%%%%%%%%%%%%%%%%%%%%%%%%%%%%%

\begin{figure}
  \centerline{\includegraphics*[width=13.5cm]{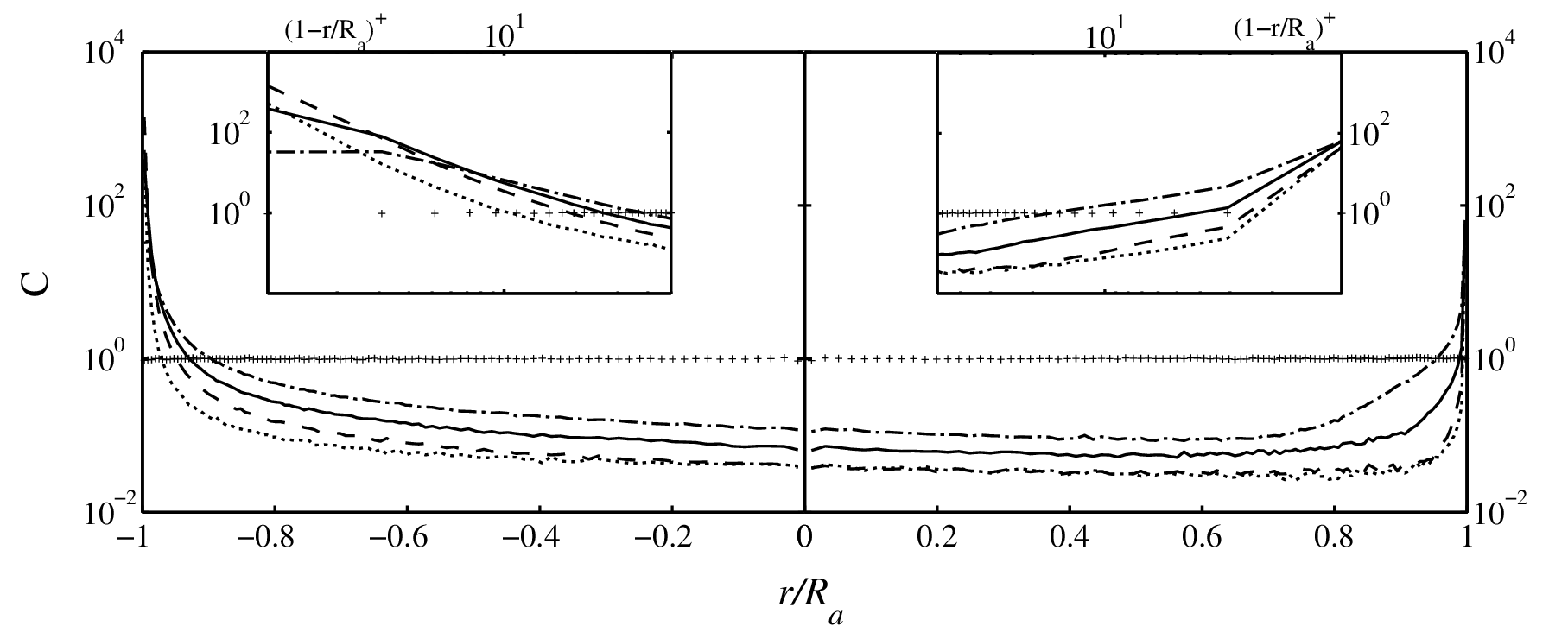}\put(-380,135){$(a)$}}
  \centerline{\includegraphics*[width=13.5cm]{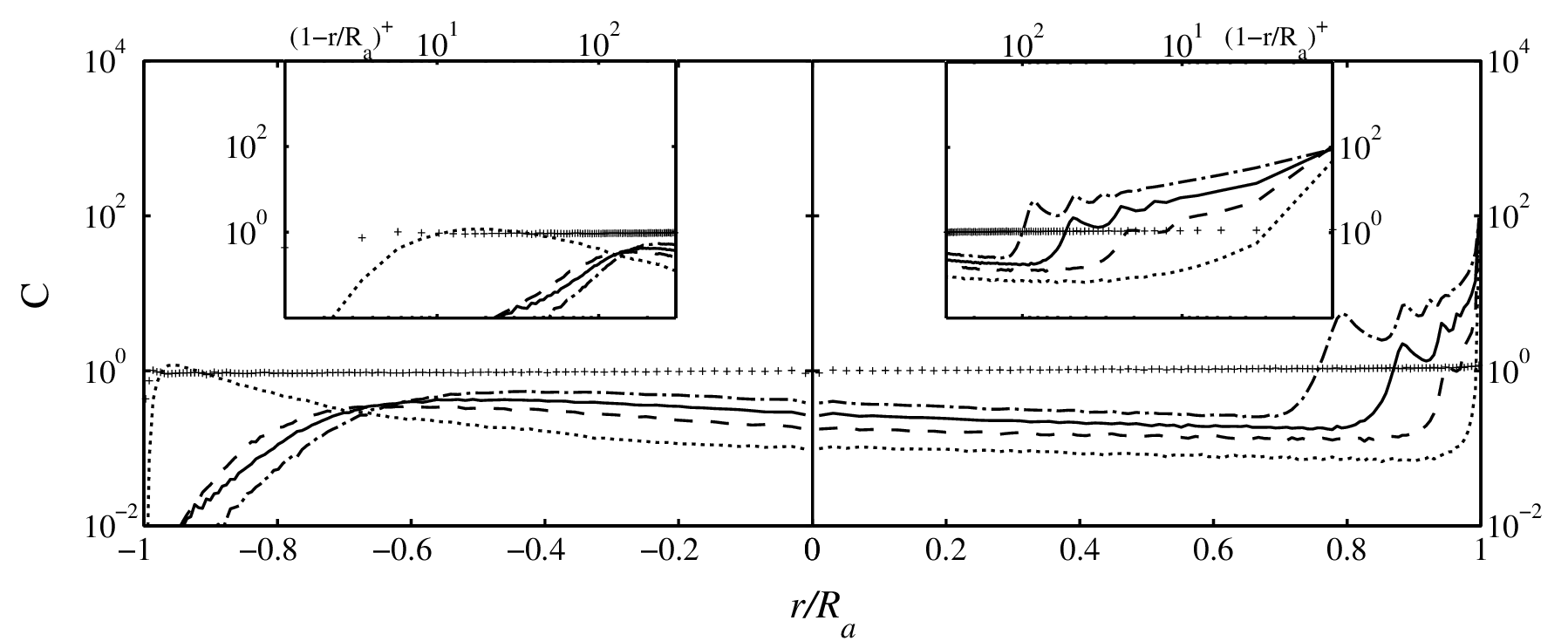}\put(-380,135){$(b)$}}
  \caption{Distribution of the mean normalised particle concentration along the equatorial mid-plane (vertical cut) of the curved pipe with $\kappa=0.01$ \emph{(a)} and $\kappa=0.1$ \emph{(b)}. $+$ \textbf{Stp0}, \dotted \textbf{Stp5}, \dashed \textbf{Stp25}, \solid \textbf{Stp50}, and \dotdashed \textbf{Stp100}. The inner bend is located at $r/R_a=-1$ and the outer side of the curved pipe is at $r/R_a=1$.}
\label{fig:fig16}
\end{figure}

Further insight can be gained by measuring the wall-normal distribution of the particle concentration at the equatorial mid-plane of the curved pipes in polar coordinates (figure \ref{fig:fig16}). The behaviour in the  strongly and mildly curved configurations is similar in the middle of the pipe section ($-0.7<r/R_a<0.7$); namely the concentrations of the inertial particles are below the uniform distribution, but increase with larger particle inertia (except for lighter particles (\textbf{Stp5-Stp10}) in the mildly curved pipe). Traversing towards the outer bend, the concentration of the particles generally reduces slightly. The accumulation of the particles in the inner bend is consistent with the observations in figure \ref{fig:fig15}. However, an interesting feature of the particle concentration appears in the outer bend of the strongly curved pipe where there are series of peaks in the concentration map; these peaks are not a consequence of insufficient temporal averaging but rather a robust statistical feature of this flow. This can also be seen in figure \ref{fig:fig14} \emph{(f)}. It is apparent that the magnitude of $C$ and the radial position of these peaks is directly connected to the particle Stokes number. The reason for these concentration peaks is that particles hit the outer wall with considerable speed, and are thus bouncing from the wall (elastic collision). The incoming speed essentially determines how far the re-bounce will reach into the domain. If the incoming particles have a large-enough and quite uniform speed, such concentration peaks will be visible. The results show that the re-bounce is stronger for heavier particles. This phenomenon has also previously been observed by \citet{huang_durbin_2012} close to the outer bend of highly curved S-shaped channel laden with heavy particles and denoted as \emph{reflection layer}. These authors explained the oscillations and the multiple peaks of the concentration in their study via a simplified model which shows that for large enough particle inertia the re-bounced ballistic individuals could have multiple velocities in the wall-normal direction.
%%%%%%%%%%%%%%%%%%%%%%%%%%%%%%%%%%%%%%%%%%%%%%%%%%%%%%%%%%%%%%%%%%%%%%%%%%%%%%%%%%%%
%%%%%%%%%%%%%%%%%%%%%%%%%%%%%%%%%%%%%%%%%%%%%%%%%%%%%%%%%%%%%%%%%%%%%%%%%%%%%%%%%%%%%

% \clearpage

\subsubsection{Force statistics}
In this section we try to elucidate the asymmetry in the concentration profiles in the equatorial midplane of bent pipes (observed in the previous figure) providing a comparison between two important forces contributing to the Stokes drag acting on particles in the vertical plane; namely, the centrifugal and turbophoretic force. It is worthwhile to note that the centrifugation of the mean Dean cells is absent in this plane due to symmetry.  

\begin{figure}
  \centerline{\includegraphics*[width=12.cm]{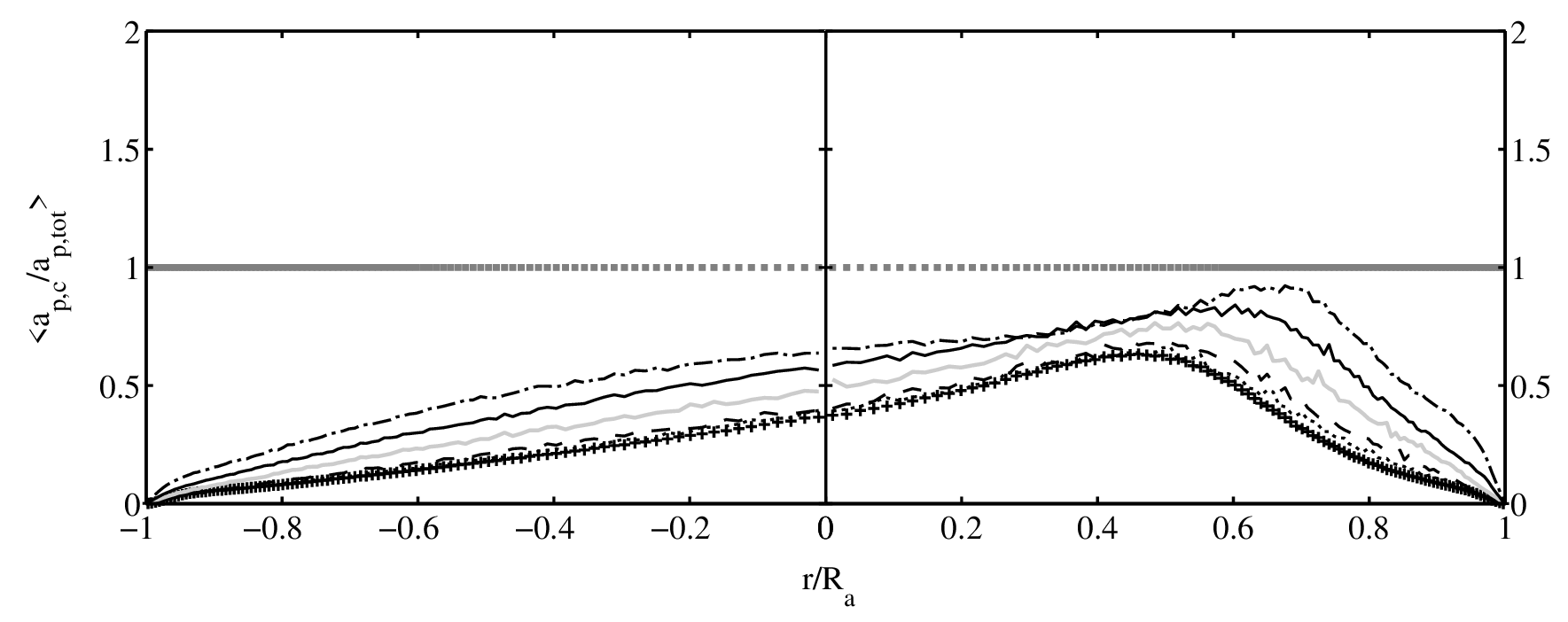}\put(-340,130){$(a)$}}
  \centerline{\includegraphics*[width=12.cm]{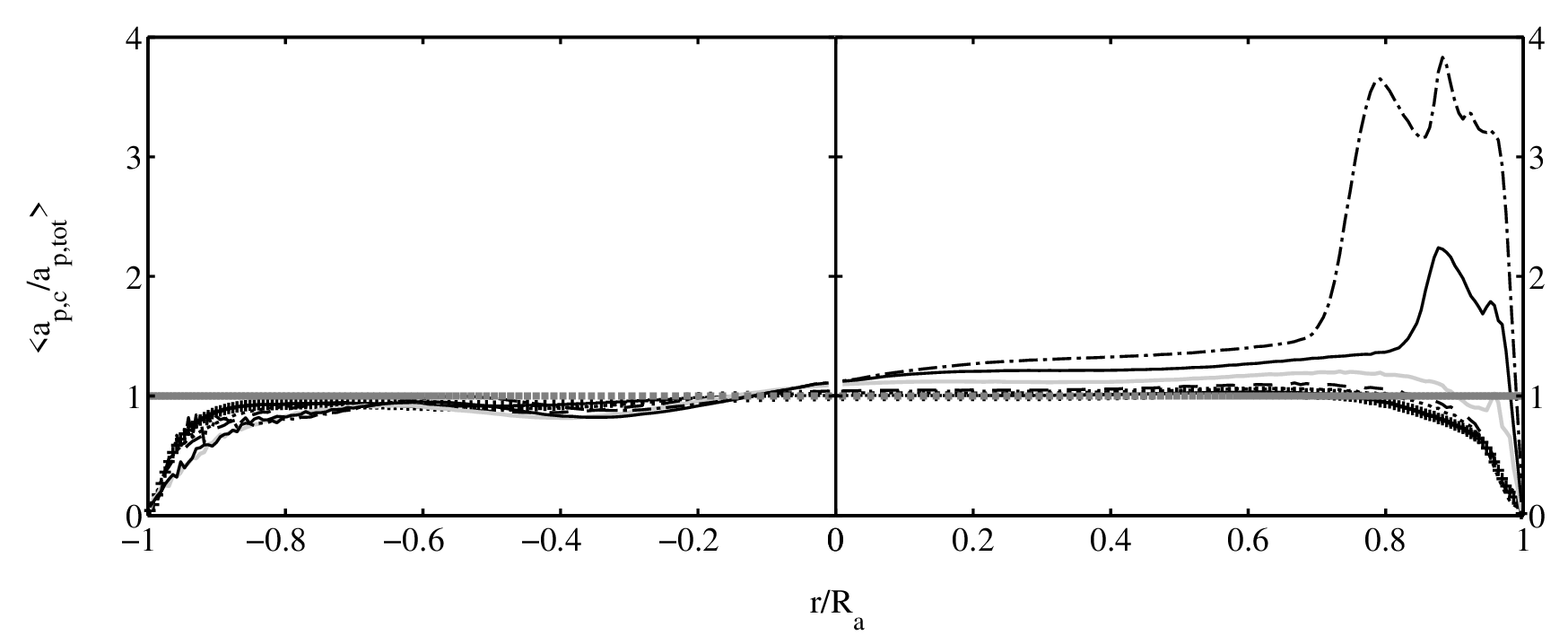}\put(-340,130){$(b)$}}
  \caption{Wall-normal distribution of the ratio of particle centrifugal acceleration $a_{p,c}$ to total particle acceleration magnitude $a_{p,tot}$ averaged in time and streamwise direction in the equatorial midplane of \emph{(a)} the mildely bent pipe and \emph{(b)} the strongly curved pipe. $+$ \textbf{Stp1}, \dotted \textbf{Stp5}, \dashed \textbf{Stp10}, {\color{light-gray} \solid} \textbf{Stp25}, \solid \textbf{Stp50}, and \dotdashed \textbf{Stp100}. The value of $1$ is indicated by {\color{light-gray} $\blacksquare$}. The inner bend is located at $r/R_a=-1$ and the outer side of the curved pipe is at $r/R_a=1$.}
\label{fig:NEWfig2}
\end{figure}

The distribution of the ratio of particle centrifugal acceleration $a_{p,c}=v_{s_p}^2/R$ to the magnitude of the total particle acceleration $a_{p,tot}=\surd(a_{r_p}^2+a_{\theta_p}^2+a_{s_p}^2)$ averaged in time and streamwise direction in the equatorial midplane of both bent pipes is shown in the figure \ref{fig:NEWfig2}. Here, $v_{s_p}$ indicates the streamwise particle velocity while $a_{r_p}$, $a_{\theta_p}$ and $a_{s_p}$ denote particle acceleration in radial, azimuthal and streamwise direction of the pipe section. In the mildly bent pipe, the importance of the centrifugal force is increasing directly with particle inertia. Nevertheless, the maximum value for all the populations is in the outer-bend half of the pipe section between $r/R_a\approx0.5$ and $0.7$. These extrema traverse farther towards the outer bend with $St_b$. For the heaviest population in the maximum point of the profile, $a_{p,c}$ is almost $90\%$ of the total acceleration although $\langle a_{p,c}/a_{p,tot} \rangle$ remains smaller than unity for all the positions in the vertical cut of the mildly bent pipe.   
%%%%%%%%%%%%%%%%%%%%%%%%%%%%%%%%%%%%%%%%%%%%%%%%%%%%%%%%%%%%%%%%%%%%%%%%%%%%%%%%%%%%
%%%%%%%%%%%%%%%%%%%%%%%%%%%%%%%%%%%%%%%%%%%%%%%%%%%%%%%%%%%%%%%%%%%%%%%%%%%%%%%%%%%%%

The importance of centrifugal acceleration is much more obvious in the strongly bent pipe (figure \ref{fig:NEWfig2} \emph{(b)}) as the ratio of $\langle a_{p,c}/a_{p,tot} \rangle$ is almost at unity in the majority of the pipe section. This value slightly reduces near the bulge ($-0.6<r/R_a<-0.2$) due to the fanning effect of turbulence in this region. The large peaks at the outer bend are purely due to reflection layers observed in the concentration plots of heavy particles. As shown by \citet{noorani_sardina_brandt_schlatter_FTC_2015}, these specular reflections change the direction of the radial velocity at the outer bend and expel particles from the near-wall region. The consecutive peaks of radial velocity near the outer bend (figure $6$ $(c)$ in \citet{noorani_sardina_brandt_schlatter_FTC_2015}) show how the radial velocity cancels due to the superposition of reflected particles with negative radial velocity. Consequently, in these positions the magnitude of the in-plane velocities is drastically reduced and the only effective component is the streamwise particle velocity which will be directly translated into the centrifugal acceleration. The peaks in the vicinity of the outer bend wall in figure \ref{fig:NEWfig2} \emph{(b)} are directly corresponding to the troughs at the outer bend of radial velocity profile shown by \citet{noorani_sardina_brandt_schlatter_FTC_2015}. It is also important to note that $\langle a_{p,c}/a_{p,tot} \rangle$ is larger than one in $0<r/R_a<0.6$ in figure \ref{fig:NEWfig2} \emph{(b)}. This is only possible if a negative acceleration exists in this vertical cut to balance the particle centrifugation. Such decelerations are depicted also by Noorani \emph{et al.}\ (2015) computing the radial acceleration in similar positions (\emph{c.f.}\ figure $12$ \emph{(d)} therein). These negative radial accelerations are directly related to the lag of the particle centrifugal force behind the fluid centrifugal force which stems from different streamwise velocities of the particles and fluids. \citet{noorani_sardina_brandt_schlatter_FTC_2015} further show that the streamwise velocity of the particle phase is smaller than that of the fluid (\emph{c.f.}\ figure $6$ \emph{(a)} therein).  Therefore it takes some time for the inertial particles to reach the same local streamwise velocity as the fluid $u_s$, leading to  the phase lag and strong deceleration in $R$. 

\begin{figure}
  \centerline{\includegraphics*[width=7cm]{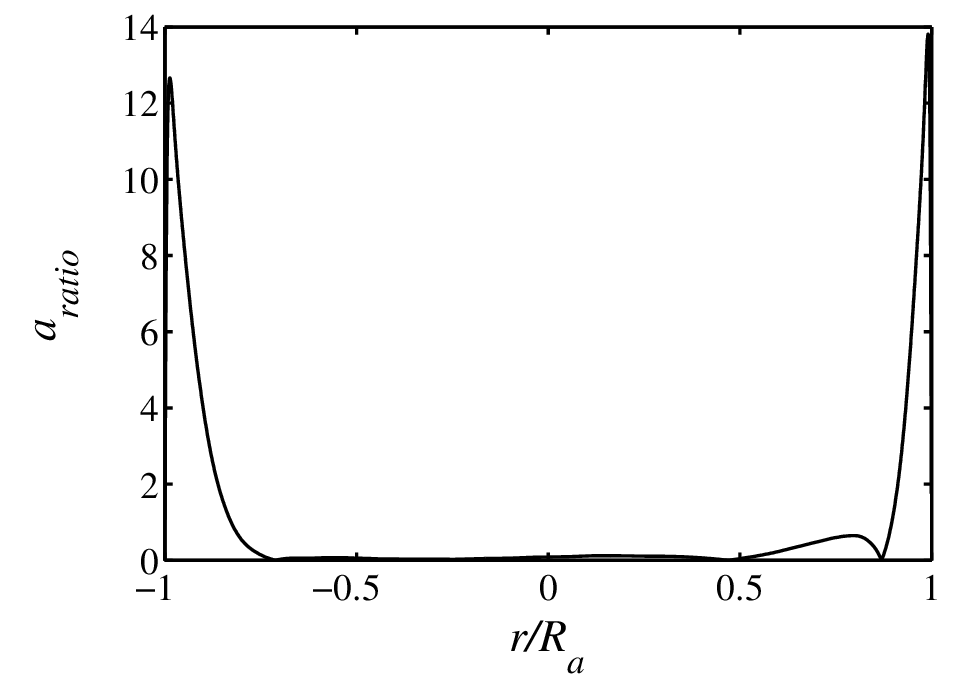}\put(-160,130){$(a)$}
              \includegraphics*[width=7cm]{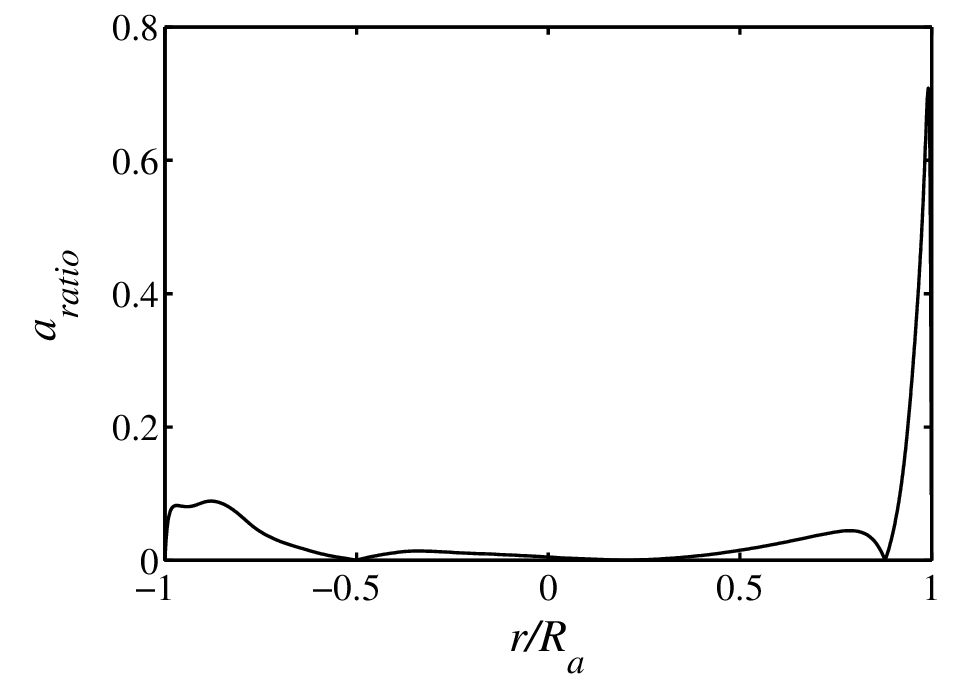}\put(-160,130){$(b)$}}
  \centerline{\includegraphics*[width=7cm]{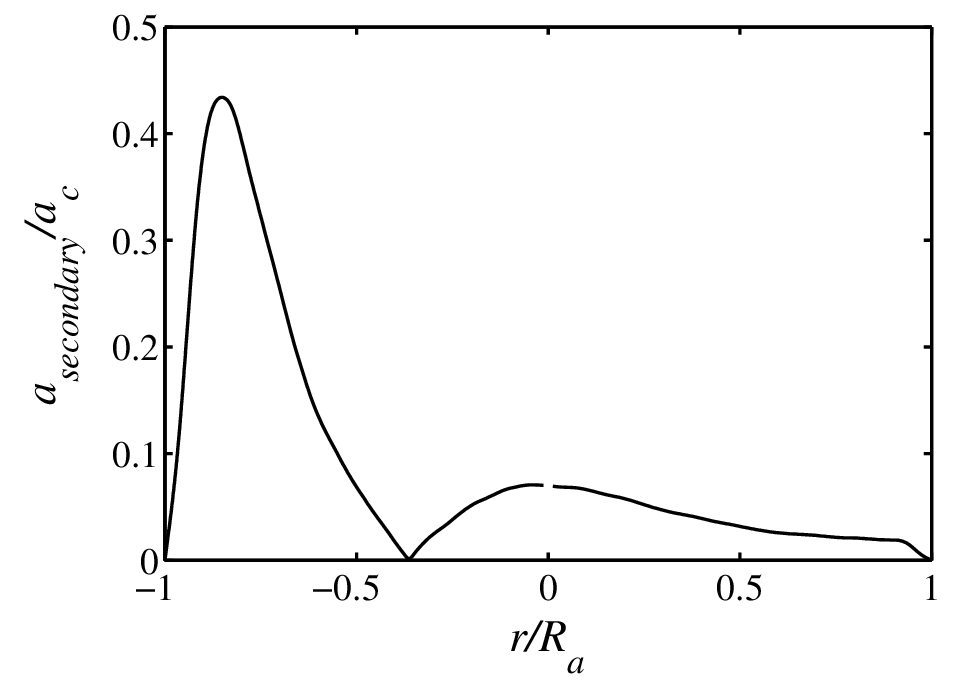}\put(-160,130){$(c)$}
              \includegraphics*[width=7cm]{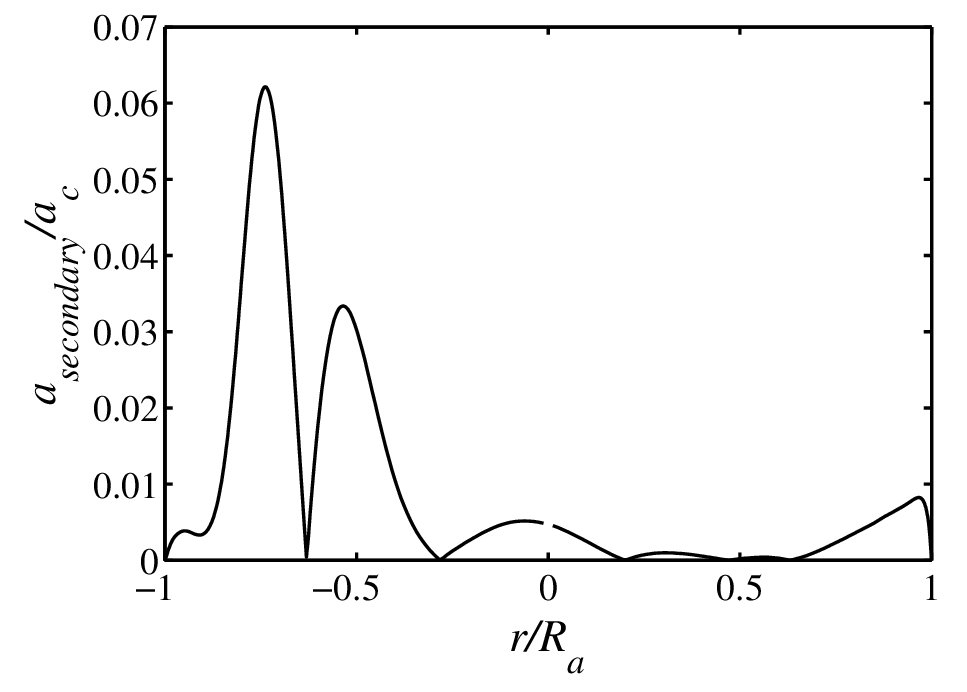}\put(-160,130){$(d)$}}
  \caption{\emph{(Top)} Distribution of the ratio of the main component of turbophoretic acceleration to the mean centrifugal acceleration $a_{ratio}$ along the equatorial mid-plane (vertical cut) of the curved pipe with $\kappa=0.01$ \emph{(a)} and $\kappa=0.1$ \emph{(b)}. The absolute value is plotted here. \emph{(Bottom)} The ratio of the absolute value of the mean secondary-motion acceleration to the mean centrifugal acceleration in the plane of symmetry of the curved pipe with $\kappa=0.01$ \emph{(c)} and $\kappa=0.1$ \emph{(d)}. The inner bend is located at $r/R_a=-1$ and the outer side of the curved pipe is at $r/R_a=1$.}
\label{fig:NEWfig3}
\end{figure}

Assuming $\tau_p$ is small, the contribution of various forces to the in-plane particle dynamics in turbulent bent pipes is derived from the particle equation of motion in Appendix \ref{appA}. Each term in the equations (\ref{equation9}) and (\ref{equation10}) can be computed from the carrier-phase statistics. Among the contributions to the turbophoretic drift of the particles (the second and third terms on the right-hand side of equations (\ref{equation9}) and (\ref{equation10})) the only non-negligible term in the equatorial midplane is 
$$a_{turb}=-{\frac{\partial {\langle u_r'u_r' \rangle}}{\partial r}} \, .$$  
Figure \ref{fig:NEWfig3} shows the absolute value of the ratio of this acceleration term to the mean centrifugal acceleration ($a_{c}=\langle  u_s^2/R\rangle$) denoted as $a_{ratio}$. The ratio of the absolute value of the secondary-motion contribution to in-plane acceleration ($a_{secondary}$) to $a_{c}$ in the plane of symmetry of the pipes is also shown in the same figure \emph{(bottom)}. In the equatorial midplane, $a_{secondary}$ is the summation of the $6{th}$ to $9{th}$ terms of the right-hand-side of the equation (\ref{equation9}) in the Appendix.     
%%%%%%%%%%%%%%%%%%%%%%%%%%%%%%%%%%%%%%%%%%%%%%%%%%%%%%%%%%%%%%%%%%%%%%%%%%%%%%%%%%%%
%%%%%%%%%%%%%%%%%%%%%%%%%%%%%%%%%%%%%%%%%%%%%%%%%%%%%%%%%%%%%%%%%%%%%%%%%%%%%%%%%%%%%

For both cases $a_{turb}$ is only significant in the vicinity of the walls. However, $a_{ratio}$ in the inner and outer bend position of the mildly curved pipe is much larger than that of the strongly curved pipe. This is mainly due the reduction of turbulence activity in the highly curved configuration. For the strongly curved pipe, $a_{c}$ is larger than $a_{turb}$ even at the outer bend where the flow is fully turbulent. The peaks of $a_{ratio}$ near the inner bend of this case ($\kappa=0.1$) are related to stronger turbulence in the bulge region. Moving to figures \ref{fig:NEWfig3} \emph{(c)} and \emph{(d)}, the secondary motions have only a negligible impact with respect to the centrifugal acceleration for the high curvature case (figure \ref{fig:NEWfig3} \emph{d}). For the mildly bent pipe, however, the effect is of the order of $10\%$ compared to centrifugal acceleration, reaching a peak at $40\%$ in the inner part of the bent pipe. 

Comparing the two previous figures, it appears that in the outer layer of the mildly bent pipe in the vertical cut the centrifugal force is the dominant force, and this dominancy increases with particle inertia. However, at the walls the turbophoresis is still the dominant cause of the particle in-plane motion. In the strongly curved pipe, on the other hand, the centrifugal force dominates the particle transport even close to the walls.

%\clearpage
\section{Concluding remarks} \label{sec:conclusions}

In this paper, a series of new direct numerical simulations (DNSs) of the turbulent flow in straight and bent pipes have been discussed. The curvature parameter $\kappa$ varies from 0, 0.01 to 0.1. Together with the velocity and pressure  fields, populations of inertial point-particles were advected. All simulations were performed at moderate Reynolds number, corresponding to $Re_\tau=360$ in the straight pipe configuration; this $Re_\tau$ is large enough for the flow to be not completely dominated by low-$Re$-effects, but yet the study is computationally feasible. A total of 120 million grid points was used for each simulation, and 0.9 million Lagrangian particles were advected in each case. The employed numerical method is a high-order spectral-element method coupled with spectrally accurate interpolation employed for the particle solver. The implementation and set-up were validated against existing channel-flow data.

The DNS results show that even a slight curvature of the pipe geometry drastically changes the map of particle concentration due the cross-stream motion. The inner bend of the mildly curved pipe exhibits larger concentration than the outer side for a range of particle Stokes number. Enhancing the curvature alters the dynamics of the particles even further as the change in the concentration map is more pronounced. While the core of the secondary-flow Dean vortices and their stagnation points in the inner bend are almost void of any particles, considerable accumulations appear in the outer bend. Generally the heavier (\emph{i.e.}\ larger) particles do not exactly follow the in-plane flow streamlines leading to a large collision rate at the wall at the outer bend. This generates a reflection layer with large concentration adjacent to the outer bend. This finding is consistent with the study of particulate dispersion in S-shaped turbulent channel at very strong curvature conducted by \citet{huang_durbin_2012}. 

In addition to concentration map, the instantaneous flow structures were examined. It turns out that the particle streaks, commonly observed in canonical wall-bounded particle-laden flows, exist also in the near-wall region of the present curved geometries. However, due to the general tendency of the particles to follow the secondary flow, these particle streaks are inclined from the outer towards the inner bend, and as a consequence, their length and strength is weaker. In general, due to the presence of secondary flow, the particles are more active, and cover larger portions of the cross-stream plane. Particles are not only found in near-wall regions as in the canonical case, but pass through the pipe centre region. However, as discussed above, if the strength of the secondary flow becomes large, certain regions of the flow are depleted of particles, whereas others are oversampled. 

Another important feature is that the particle dispersion reaches a statistically stationary state inside the curved configuration and there is no dependency of the inlet/outlet condition. The present case thus allows us to study particle dispersion in the presence of secondary flow without temporal transients. It is worth noting that the Dean vortices are locked in their position unlike the situation appearing \emph{e.g.}\ in a curved channel or curved boundary layer where vortical Dean cells or G\"{o}rtler vortices appear. In those cases, the spanwise homogeneity leads to uniform particles concentrations in that direction. The present case of bent pipe, as a first of its kind, provides a suitable configuration to study the effect of geometry induced centrifugal forces, the Dean cell centrifuges and the turbophoretic motion of the particles on the dispersion map in the pipe cross-section. The current study may provide a reliable base to conduct further theoretical studies on the effect of the curvature on turbulent wall-bounded particle-laden flows. The data can also be employed as a bench-mark to facilitate modelling advancement for dispersed multiphase turbulent flow in wall-bounded complex geometries.

\section*{Acknowledgment}
The authors gratefully acknowledge computer-time allocation from the Swedish National Infrastructure for Computing (SNIC). This research is also supported by the ERC grant '2013-CoG-616186, TRITOS' to L.B.

\appendix
\section{}\label{appA}
This appendix contains the derivation of different force contributions to the in-plane particle motion in turbulent bent pipes. In order to distinguish between the main three effects determining the particle motion, namely turbophoresis, preferential concentration, and centrifugal effects of the Dean vortices we start from the particle-dynamics equation in the approximation of small relaxation time $\tau_p$. In this case the particle acceleration can be approximated by the fluid acceleration $\mathrm{d}{\bf v_p}/\mathrm{d}t\simeq \mathrm{D}{\bf u}/\mathrm{D}t$ where ${\bf v_p}$ is the particle velocity, and ${\bf u}$ is the fluid velocity \citep{nowbahar_sardina_picano_brandt_2013}:
\begin{equation}
{\bf v_p}={\bf u_p}-\tau_p\frac{\mathrm{D}{\bf u}}{\mathrm{D}t} \, .
\label{equation1}
\end{equation}
Here,  ${\bf u_p}$ is the fluid velocity at particle positions and $\mathrm{D}/{\mathrm{D}t}={\partial}/{\partial t}+{\bf u}\cdot\nabla$ is the material derivative following the fluid.

Let us first briefly recall the implications of the previous equation in turbulent plane channel and straight pipe flows. In the case of channel flow, we indicate the wall-normal direction with $y$ and  apply averages to equation (\ref{equation1}) in time and in the streamwise and spanwise directions, denoted by an \emph{overbar}. Projecting this mean equation in the non-homogeneous wall-normal direction we obtain
\begin{equation}
\overline{v_{y_p}}=\overline{u_{y_p}}-\tau_p\frac{\mathrm{d} \overline{ u_y^2}}{\mathrm{d}y};
\label{equation2}
\end{equation}
applying Reynolds decomposition, $f=\overline{f}+f'$, where  $f$ is a generic scalar or vectorial field and the prime indicates the fluctuations around the mean \citep{pope_2000} gives
\begin{equation}
\overline{v_{y_p}}=\overline{ u_{y_p}}-\tau_p\frac{\mathrm{d} \overline{u_y'u_y'}}{\mathrm{d}y} ,
\label{equation3}
\end{equation}
where the mean particle velocity is due to two different contributions:  $\overline{u_{y_p}}$ that represents the preferential concentration, sampling of specific fluid events, and $\tau_p{\mathrm{d} \overline{u_y'u_y'}}/{\mathrm{d}y}$ the turbophoretic drift. 
For channel flow at steady state, the mean particle velocity on the left-hand side is zero, which implies a net balance between the preferential concentration and the turbophoretic drift, namely, $\overline{u_{y_p}}=\tau_p{\mathrm{d} \overline{ u_y'u_y'}}/{\mathrm{d}y}$. Since the sign of the derivative is negative, the the particles tend to preferentially accumulate in the low-speed regions as shown among others in \citet{sardina_schlatter_brandt_picano_casciola_2012}.

For the straight pipe flow, proceeding as for the case of the channel, \emph{i.e.}\ averaging in time, streamwise and azimuthal ($\theta$) direction and projecting the resulting equation (\ref{equation1}) in the radial direction $r$ we obtain after Reynolds decomposition
\begin{equation}
\overline{v_{r_p}}=\overline{u_{r_p}}-\tau_p\frac{\mathrm{d} \overline{u_r'u_r'}}{\mathrm{d}r}-\tau_p\frac{\overline{u_r'u_r'}-\overline{ u_\theta'u_\theta'}}{r}
\label{equation4}
\end{equation}
where the last term is negligible in the buffer layer (as shown in \citet{picano_sardina_casciola_2009}); $\overline{ u_{r_p}}$  represents preferential concentration and $\tau_p{\mathrm{d} \overline{u_r'u_r'}}/{\mathrm{d}r}$ the turbophoretic drift. The equation has the same structure as in channel flow, as the last term is almost zero, therefore the same conclusions as for the channel can be drawn. In particular, the particles tend to oversample the negative flow events once the statistical steady state has been reached.

For the bent pipe, we can make use of the orthogonal thoroidal coordinate system as introduced in \citet{germano_1982}. We average only in time and streamwise direction ($s$) denoting this average with $\langle \cdot \rangle$ and project the equation in the radial ($r$) and spanwise coordinate ($\theta$), obtaining
\begin{equation}
{\langle v_{r_p}\rangle}={\langle u_{r_p}\rangle}-\tau_p{\frac{1}{h_sr}\frac{\partial h_s r {\langle u_ru_r\rangle}}{\partial r}}-\tau_p{\frac{1}{h_sr}\frac{\partial h_s  {\langle u_ru_\theta \rangle}}{\partial \theta}}+\tau_p\frac{\kappa \sin(\theta)}{h_s}{\langle u_su_s \rangle}+\tau_p\frac{{\langle u_\theta u_\theta \rangle}}{r}
\label{equation5}
\end{equation}
\begin{equation}
{\langle v_{\theta_p}\rangle}={\langle u_{\theta_p}\rangle}-\tau_p{\frac{1}{h_sr}\frac{\partial h_s r {\langle u_ru_\theta \rangle}}{\partial r}}-\tau_p{\frac{1}{h_sr}\frac{\partial h_s  {\langle u_\theta u_\theta \rangle}}{\partial \theta}}+\tau_p\frac{\kappa \cos(\theta)}{h_s}{\langle u_su_s \rangle}-\tau_p\frac{{\langle u_r u_\theta\rangle}}{r}
\label{equation6}
\end{equation}
where $\kappa$ is the curvature of the bent pipe and the metric term $h_s$ is defined as $1+\kappa r \sin(\theta)$. The previous equations can be rewritten as
\begin{eqnarray}
{\langle v_{r_p}\rangle}&=&{\langle u_{r_p}\rangle}-\tau_p{\frac{\partial {\langle u_ru_r \rangle}}{\partial r}}-\tau_p\frac{2h_s-1}{h_sr}{\langle u_ru_r \rangle}-\tau_p{\frac{1}{r}\frac{\partial  {\langle u_ru_\theta \rangle}}{\partial \theta}} \nonumber \\
&& -\tau_p\frac{\kappa \cos(\theta)}{h_s}{\langle u_ru_\theta \rangle}+\tau_p\frac{\kappa \sin(\theta)}{h_s}{\langle u_su_s \rangle}+\tau_p\frac{{\langle u_\theta u_\theta \rangle}}{r}
\label{equation7}
\end{eqnarray}
\begin{eqnarray}
{\langle v_{\theta_p}\rangle}&=&{\langle u_{\theta_p}\rangle}-\tau_p{\frac{\partial {\langle u_ru_\theta \rangle}}{\partial r}}-\tau_p\frac{2h_s-1}{h_sr}{\langle u_ru_\theta \rangle}-\tau_p{\frac{1}{r}\frac{\partial  {\langle u_\theta u_\theta \rangle}}{\partial \theta}} \nonumber \\
&& -\tau_p\frac{\kappa \cos(\theta)}{h_s}{\langle u_\theta u_\theta \rangle}+\tau_p\frac{\kappa \cos(\theta)}{h_s}{\langle u_su_s \rangle}-\tau_p\frac{{\langle u_r u_\theta \rangle}}{r}.
\label{equation8}
\end{eqnarray}
Applying Reynolds decomposition and rearranging we finally obtain
\begin{eqnarray}
{\langle v_{r_p}\rangle}&=&{\langle u_{r_p}\rangle}-\tau_p{\frac{\partial {\langle u_r'u_r' \rangle}}{\partial r}}-\tau_p{\frac{1}{r}\frac{\partial  {\langle u_r'u_\theta' \rangle}}{\partial \theta}}\nonumber\\
&& +\tau_p\frac{\kappa \sin(\theta)}{h_s}({\langle u_s\rangle}^2+{\langle u_s'u_s'\rangle})+\tau_p \left (\frac{{\langle u_\theta \rangle}^2+{\langle u_\theta'u_\theta' \rangle}}{r} \right )+\nonumber\\
&& -\tau_p{\frac{\partial {\langle u_r\rangle}^2}{\partial r}}-\tau_p\frac{2h_s-1}{h_sr}{\langle u_r\rangle}^2-\tau_p{\frac{1}{r}\frac{\partial  {\langle u_r \rangle}\, {\langle u_\theta\rangle}}{\partial \theta}}-\tau_p\frac{\kappa \cos(\theta)}{h_s}{\langle u_r\rangle}\, {\langle u_\theta\rangle}\nonumber\\
&& -\tau_p\frac{2h_s-1}{h_sr}{\langle u_r'u_r'\rangle}-\tau_p\frac{\kappa \cos(\theta)}{h_s}{\langle u_r'u_\theta'\rangle}
\label{equation9}
\end{eqnarray}
\begin{eqnarray}
{\langle v_{\theta_p}\rangle}&=&{\langle u_{\theta_p}\rangle}-\tau_p{\frac{\partial {\langle u_r'u_\theta'\rangle}}{\partial r}}-\tau_p{\frac{1}{r}\frac{\partial  {\langle u_\theta' u_\theta'\rangle}}{\partial \theta}} \nonumber \\
&&+\tau_p\frac{\kappa \cos(\theta)}{h_s}({\langle u_s\rangle}^2+{\langle u_s'u_s'\rangle}) \nonumber \\
&& -\tau_p{\frac{\partial {\langle u_r\rangle}\,{\langle u_\theta\rangle}}{\partial r}}-\tau_p\frac{3h_s-1}{h_sr}{\langle u_r\rangle}\,{\langle u_\theta\rangle}-\tau_p{\frac{1}{r}\frac{\partial  {\langle u_\theta\rangle}^2}{\partial \theta}}-\tau_p\frac{\kappa \cos(\theta)}{h_s}{\langle u_\theta\rangle}^2 \nonumber \\
&&-\tau_p\frac{3h_s-1}{h_sr}{\langle u_r'u_\theta'\rangle}-\tau_p\frac{\kappa \cos(\theta)}{h_s}{\langle u_\theta' u_\theta'\rangle}
\label{equation10}
\end{eqnarray}
Here we assume the effect of the secondary vortex is only due to the mean flow.
On the right-hand side of equations (\ref{equation9}) and (\ref{equation10}), the first terms are considered as preferential concentration, and the second and third terms as the contributions to turbophoretic drift. The forth and fifth term in equation (\ref{equation9}) and the forth term in equation (\ref{equation10}) are also considered as the contribution to centrifugal forces. Except for the last two terms in those equations which are negligibly small, the rest of the terms on the right-hand sides is directly related to the secondary-motion contributions to the in-plane particle motion.

\bibliographystyle{jfm}
% Note the spaces between the initials

\bibliography{bpPart}

\end{document}